\documentclass[sn-basic]{sn-jnl}

\usepackage{graphicx}
\usepackage{natbib}
\usepackage{amssymb}
\usepackage{gensymb}
\usepackage[dvipsnames]{xcolor}
\usepackage{soul}
\usepackage{amsmath,amssymb,amsfonts}%
\usepackage{amsthm}%

\usepackage{multirow}%
\usepackage{mathrsfs}%
\usepackage[title]{appendix}%
\usepackage{xcolor}%
\usepackage{textcomp}%
\usepackage{manyfoot}%
\usepackage{booktabs}%
\usepackage{algorithm}%
\usepackage{algorithmicx}%
\usepackage{algpseudocode}%
\usepackage{listings}%

\usepackage[normalem]{ulem}%

\usepackage[T1]{fontenc}

\usepackage{lmodern}
\usepackage{anyfontsize}
\usepackage{array}

\newcommand{\rearth}{R$_\mathrm{Earth}$}
\newcommand{\mearth}{M$_\mathrm{Earth}$}

\newcommand{\mj}[1]{\textcolor{cyan}{#1}}

\begin{document}

\title[Status end of Phase B]{The PLATO Mission}


\author[1,2]{\fnm{Heike} \sur{Rauer}}
\author[3]{\fnm{Conny} \sur{Aerts}}
\author[1]{\fnm{Juan} \sur{Cabrera}}
\author[4]{\fnm{Magali} \sur{Deleuil}}
\author[1]{\fnm{Anders} \sur{Erikson}}
\author[5]{\fnm{Laurent} \sur{Gizon}}
\author[6]{\fnm{Mariejo} \sur{Goupil}}
\author[7]{\fnm{Ana} \sur{Heras}}
\author[7]{\fnm{Jose} \sur{Lorenzo-Alvarez}}
\author[7]{\fnm{Filippo} \sur{Marliani}}
% ...

\affil[1]{\orgdiv{Institute of Planetary Research,}\orgname{German Aerospace Center,}\orgaddress{  \city{Berlin,} \country{Germany}}}
\affil[2]{\orgname{Free University of Berlin,}\orgaddress{  \city{Berlin,} \country{Germany}}}
\affil[3]{\orgdiv{Institute of Astronomy,}\orgname{KU Leuven,}\orgaddress{  \city{Leuven,} \country{Belgium}}}
\affil[4]{\orgdiv{CNRS, CNES, LAM,}\orgname{Aix Marseille University,}\orgaddress{  \city{Marseille,} \country{France}}}
\affil[5]{\orgname{Max Planck Institute for Solar System Research,}\orgaddress{  \city{G\"ottingen,} \country{Germany}}}
\affil[6]{\orgdiv{LESIA, Paris Observatory,}\orgname{PSL University, Sorbonne University, Paris Cit\'e University, CNRS,}\orgaddress{  \city{Meudon,} \country{France}}}
\affil[7]{\orgdiv{ESTEC,}\orgname{European Space Agency,}\orgaddress{  \city{Noordwijk,} \country{The Netherlands}}}
% ...

\abstract{
PLATO (PLAnetary Transits and Oscillations of stars) is ESA’s M3 mission designed to detect and characterise extrasolar planets and perform asteroseismic monitoring of a large number of stars. PLATO will detect small planets (down to $<$2\rearth) around bright stars ($<$11 mag), including terrestrial planets in the habitable zone of solar-like stars. With the complement of radial velocity observations from the ground, planets will be characterised for their radius, mass, and age with high accuracy (5\%, 10\%, 10\% for an Earth-Sun combination respectively). PLATO will provide us with a large-scale catalogue of well-characterised small planets up to intermediate orbital periods, relevant for a meaningful comparison to planet formation theories and to better understand planet evolution. It will make possible comparative exoplanetology to place our Solar System planets in a broader context. In parallel, PLATO will study (host) stars using asteroseismology, allowing us to determine the stellar properties with high accuracy, substantially enhancing our knowledge of stellar structure and evolution. 

The payload instrument consists of 26 cameras with 12cm aperture each. For at least four years, the mission will perform high-precision photometric measurements. Here we review
the science objectives, present PLATO´s target samples and fields, provide an overview of expected core science performance as well as a description of the instrument and the mission profile towards the end of the serial production of the flight cameras. PLATO 
is scheduled for a launch date end 2026. This overview therefore provides a summary of the mission to the community in preparation of the upcoming operational phases.}


\maketitle

\tableofcontents

\section{Introduction}
\label{sec:intro}
The PLATO mission (PLAnetary Transits and Oscillations of stars) is designed to detect and characterise a large number of exoplanetary systems, including terrestrial exoplanets orbiting bright solar-type stars in their habitable zone. PLATO was selected as ESA's M3 mission in the Cosmic Vision 2015-2025 Programme in 2014~\citep{rauer2014}. PLATO satellite data will provide accurate and precise planetary radii and architectures of a large number of planetary systems via the photometric transit method. The light curve data will also enable accurate stellar parameters, including evolutionary ages, to be derived via asteroseismic analysis. In combination with high-precision spectroscopic ground-based follow-up observations, accurate planetary masses, and hence mean planetary densities, will be determined. 

\medskip
The overall scientific objectives that PLATO will investigate are:
\begin{itemize}
    \item How do planets and planetary systems form and evolve?
    \item Is our Solar System special or are there other systems like ours?
    \item Are there potentially habitable planets?
\end{itemize}

The photometric precision and observing mode of PLATO will help determine the frequency of Earth-like planets. In general, analysis of PLATO data is expected to give new insights into the formation and evolution of planets and planetary systems as well as the evolution of stars. 

Benchmark cases will be used to improve stellar models to further minimise the impact of poorly understood internal
physics on stellar parameters. With these benchmark cases, classical methods for stellar modelling
can also be improved. Furthermore, light curves are used to determine the magnetic variability and activity of planet host stars. The combined analysis of improved stellar physics and well-characterised host stars with high photometric precision light curves and follow-up data will provide accurate planetary parameters. Such parameters are key ingredients for the analysis of the planetary 
radius, mass, density, and age with important implications for models of interior structure and planetary evolution. 

The PLATO satellite consists of an ESA-provided satellite platform and a payload including 26 cameras (see Section \ref{sec:payload}). Of these, 24 cameras will operate in white light and obtain high-precision photometric light curves of thousands of stars, and 2 cameras will provide colour information by observing in the blue (505-700 nm) and in the red (665-1000 nm)
wavelength range respectively. Adding to this are the 
on-board computers and power supply units. The PLATO payload is developed by the international PLATO Mission Consortium (PMC) including contributions from ESA. Launch is foreseen for end 2026.

The initial
PLATO mission layout and its science objectives are described in \cite{rauer2014}, presenting the status just before mission selection. In the present paper we summarise the status of the consolidated mission design  towards the end of flight model production and show the mission in the context of current exoplanetary research.

However, let us first have a look at the major updates on the mission since 2014. The scientific goals and research questions described in \cite{rauer2014}, and which can be addressed with PLATO, are still valid today. After the successful launch of NASA´s TESS mission in 2018, PLATO puts, however, less emphasize on short-period planets and short pointing target fields, although the option to do so is kept in the mission requirements. The focus of PLATO´s nominal mission is clearly on long orbital period planets now.
The major design change since 2014 concerns the move from 32 "Normal"-cameras to 24 N-Cams in the current design. See Section \ref{sec:TargetFieldsandObsStrat} for the  expected science performances with the current design. There have been multiple further updates, refinements and adaptations of the payload design made since 2014, which are however too detailed to be listed here. The overall design of the cameras with their support units remained, however, similar to the concept presented in 2014. Concerning the mission itself, the final satellite design and choice of industry prime of ESA was made from three competing studies during the Phase B study phase.

In this review we provide an overview of the expected impact of PLATO in exoplanet science  including planet yields and performance for planet characterisation (Section \ref{sec:ExoplanetScience}). Section \ref{sec:StellarScience} discusses PLATO's impact on stellar science, including a discussion on accuracy versus precision and the prospects for detecting solar-like oscillations in PLATO samples. The potential of PLATO concerning complementary science 
topics beyond its core programme is discussed in Section \ref{sec:ComplScience}. Section \ref{sec:StellarSamples} defines the PLATO P1 to P5 stellar samples and presents options for the distribution of on-board processed light curves versus imagettes for detailed analysis on ground. The status of target field and observing mode selection is summarised
in Section \ref{sec:TargetFieldsandObsStrat}. The science ground segment, PLATO data products and data releases are 
presented in Section \ref{sec:SGS}. The organised ground-based follow-up to determine masses for planets in PLATO's prime sample is a key element of the mission (Section \ref{sec:GOP}). In addition to its core science goals, PLATO will provide a wealth of complementary science  from its observed samples P1 to P5 and its science calibration stars. Moreover, PLATO will also offer an extensive
Guest Observer programme, which will be part of the legacy of the mission, see Section \ref{sec:GOProgram}. The expected instrument signal-to-noise levels as well as the mission and payload designs are highlighted
in Sections \ref{sec:InstrPerformance}, \ref{sec:mission} and \ref{sec:payload}. 
Finally, the synergies with other missions are presented in \ref{sec:synergy-missions}.

\section{Overview of PLATO Science Objectives}
\label{sec:ScienceProg}


The scientific programme of PLATO addresses the following science goals in the fields of exoplanet and stellar sciences:

\begin{itemize}
\item {\it Determine the bulk properties (radius, mass, and mean density) of planets in a wide range of systems, including terrestrial planets in the habitable zone (HZ) of solar-like stars. }\\

Among the key goals of PLATO are high accuracy parameters for planets orbiting F5-K7 dwarf and sub-giant stars. Note that we use the term ``accurate'' planet parameters for those cases where stellar parameters can be derived with highest precision and stellar models are well constrained (see Section\,\ref{sec:StellarScience}). For an Earth-sized planet (Earth radius: \rearth) orbiting a 10~mag G0V star an accuracy of 3\% in planetary radius (5\% for a 11~mag star) and 10\% in stellar age shall be derived (see also Table \ref{tab:NoiseLevels}). 
This configuration (for planets up to 1 AU) is used as the benchmark for the mission. Better performance is expected for larger planets or brighter host stars (see Section \ref{sec:PlanetCharact}). In addition, PLATO will observe stars that are sufficiently bright to enable the
determination of planetary masses (Earth mass: \mearth) with an accuracy of 10\% for a terrestrial planet (defined as $\le$2 \rearth\, and $\le$10 \mearth) 
orbiting a G0V star via the radial velocity method from ground-based follow-up observations. 
The resulting planetary bulk properties allow us to explore the diversity of planets and identify typical planet populations, which in turn help better constrain planet formation models. Small planets with high mean densities in the HZ will be prime candidates for potentially habitable planets and allow an estimation of $\eta_\textrm{Earth}$,
here defined as the fraction of Earth-like planets in the HZ per host star.
Furthermore, precise measurements for close-in high-density planets will show how many have a Mercury-like interior structure which will allow a study of their refractory (e.g. Fe and Si) content and help test formation theories for this class of planets. Extending our knowledge towards longer orbital periods will show whether extreme-density small planets also exist on such orbits. Other questions which can be addressed with PLATO data (and combined with e.g. stellar properties) include, ``What is a typical internal structure and composition of terrestrial and mini-gas planets?
How does it evolve due to stellar interactions (losses)? What is the core size of gaseous planets?''.
\\

\item {\it Study how planets and planetary systems evolve with age.}\\

The age of planetary systems is a key parameter to explore how planetary system properties evolve and provides observational constraints to formation models as well as to how gravitational instabilities deplete planetary systems over time \citep{veras2015MNRAS.453...67V}. Furthermore, details of the evolutionary stages of planets are only known from one example so far, our Solar System. In particular for terrestrial planets and their complex evolution as well as for the new and poorly known class of mini-Neptunes better constraints can be provided once a sample of planets observed at different evolution stages becomes available. Open questions are, e.g., ``When does the magma ocean phase for terrestrial planets end?'' and ``Are our model timescales for the evolution of gaseous planets correct?''. To further our understanding of these issues, homogeneous and precise age determination is required, an issue we are still facing today with the various age indicators. The seismic characterisation of a large sample  of bright stars across the Hertzsprung-Russell (HR) diagram will lead to significantly improved stellar models, allowing for substantially more reliable age characterisation of stars in general.
Gyrochronology, magneto-gyrochronology \citep{mathur2023} and  age-activity relationships (see Sect. 4.6) will be used to provide estimates for the ages of targets where p-mode oscillations cannot be detected,
 notably those beyond spectral type K2-K3. This will considerably increase the statistical samples upon which the evolution of planetary systems can be 
 investigated. \\

\item {\it Study the typical architectures of planetary systems.}\\

The architecture of planetary systems includes parameters such as the distribution of planet
masses and types (terrestrial or gaseous) over orbital separation, the co-planarity of systems,
and orbital parameters such as, e.g., orbital eccentricities (e.g. \citealt{vaneylen2015ApJ...808..126V, vaneylen2019AJ....157...61V}) and inclinations.  Key questions to address include, e.g., ``What fraction of planetary systems have a structure similar to the Solar System? How many have multiple/no gas giants? Was migration caused by planet-planet interaction or disk migration?''. First analyses of {\it Kepler} mission results, but also comparisons with planet synthesis models (e.g. \citealt{borucki2010a,lissauer2011a,weiss2018AJ....156..254W,millholland2017ApJ...849L..33M,wang2017AJ....154..236W,mishra2021}) suggest adjacent planets having similar masses and sizes with a trend towards increasing masses for outer planets 
\citep[e.g.,][]{ciardi2013ApJ...763...41C}.
Dynamical interactions between planet embryos can lead to typical spacing between planets with small planets being more densely packed than large/massive ones. Today, these apparent trends are based on only a few tens of well-characterised planetary systems in a limited parameter space \citep[e.g.,][]{mishra2021} and there are many exceptions to the trends. Furthermore, the dynamical evolution and development of, e.g., orbital resonances during early ages of planetary systems need to be better understood and linked to system properties. Nevertheless, the available early studies already show the large potential for well-characterised samples of planet system architectures to constrain planet formation theories. PLATO will significantly extend the number of well-characterised planetary systems, in particular for orbital periods longer than 80 days. These longer periods are underrepresented in past and ongoing mission data (see Section \ref{sec:PlanetYield} for a comparison to the Transiting Exoplanet Survey Satellite (TESS, \citealt{ricker2015}).
\\

\item {\it Analyse the correlation of planet properties and their frequencies with stellar parameters (e.g., stellar metallicity, stellar type).}\\

Occurrence rates for different types of planets around different types of stars have been provided by both transit and radial velocity surveys (e.g. \citealt{howard2012,fressin2013,christiansen2015,deleuil2018,kunimoto2022}, see Section \ref{sec:PlanetYield} for more). A proper estimate of an occurrence rate requires however good knowledge of some key parameters such as the survey detection sensitivity to estimate its completeness, the planet parameters, and a good  overview of the underlying stellar population. The latter is a serious limitation that prevents direct comparisons of published occurrence rates and limits the reliability of the estimates. PLATO will benefit from the in-depth characterisation of the stellar fields the instrument will observe as well as an analysis with the most up-to-date stellar models. This will allow for the exploration of possible correlations of planet parameters with stellar properties.  For example, the chemical element ratios of a stellar photosphere are thought to be a good first proxy of its planets' initial compositions  alleviating the degeneracies that occur when deriving the planet's composition from its radius and mass only \citep{dorn2015,brugger2017,acuna2021}. This hypothesis has been recently questioned by various  studies \citep{plotnykov2020,schulze2021PSJ.....2..113S,adibekyan2021Sci...374..330A} which highlight discrepancies between the actual planetary composition and what is expected from a primordial origin as reflected by the chemical ratios of the stars. In addition, stellar properties such as activity are key parameters for our understanding of potential planet habitability
 because it impacts, among other, on the evaporation and the chemical evolution of planetary atmospheres.
In particular and as discussed in Sect.4.6, stellar rotational evolution is a key ingredient to understand planetary migration through tidal and magnetic interaction (e.g. \citealt{lai2012MNRAS.423..486L, mathis2018A&A...620A..22M, strugarek2018ApJ...863...35S, strugarek2019ApJ...881..136S, damiani2018A&A...618A..90D, ahuir2021A&A...650A.126A}).
\\

\item {\it Analyse the dependence of the frequency of terrestrial planets on the environment in which they formed.}\\

Planets form in different regions of our Galaxy, in clusters and around field stars.
Correlations of planet occurrence frequency with their environment will provide constraints on planetary formation processes.\\

\item {\it Study the internal structure of stars and how it evolves with age.}\\

Determining planet host star parameters requires improving today’s stellar models and
stellar evolution theory in general. PLATO light curve data will be used to measure the oscillation frequencies of stars, which will be interpreted via asteroseismic modelling \citep[e.g.,][for recent reviews]{Garcia2019, Aerts2021} to test evolution theory. Stellar models constrained by asteroseismology and cross-calibration with classical methods of stellar modelling will be key in obtaining planet host star parameters to address the overall PLATO science goals. 

In the absence of seismic age determination, measurements of rotation periods from the PLATO light curves will allow us to  estimate the age of a given target through gyrochronology \citep{skumanich1972ApJ...171..565S, barnes2003ApJ...586..464B, 
barnes2007} for stars like solar-type stars which are losing angular momentum through the stellar wind due to their magnetic activity. The PLATO rotation period catalogue will complete and extend the catalogues already available for {\it Kepler\/}/K2/TESS (e.g. \citealt{nielsen2013, mcquillan2014, santos2019, santos2021, holcomb2022ApJ...936..138H}). It is also expected to detect and characterise activity cycles which are shorter than the solar cycle, such as the Rieger-like cycle \citep{rieder2016, gurgenashvili2021A&A...653A.146G}. For stars with planets it is furthermore important to understand how activity and rotation is affected by close-in planets, a task for which PLATO will expand our currently available data base.
\\

\item  {\it Identify good targets for spectroscopic follow-up measurements to investigate planetary atmospheres.}\\

Planets identified around the brightest stars will likely be ``Rosetta Stones'' 
for spectroscopic follow-up to study their atmospheric structure and composition. Apart from the gaseous planets, those low-mass PLATO planets accessible to e.g. JWST (James-Web-Space-Telescope), or its successors, have the potential to provide new insights into the link between mantle composition, iron-to-silicate ratio, redox state and atmospheric compositions. They would allow, e.g., the identification of secondary atmospheres (water, CO$_2$) on such low-mass planets (e.g., \citealt{ortenzi2020, kabath2007}), which are barely distinguishable from just mass and radius modelling alone. 
\\
\end{itemize}

To reach these objectives, PLATO data will encompass a greater parameter space - well beyond those accessible by previous and ongoing exoplanet missions. In addition, PLATO will address a large number of complementary and legacy science topics, e.g., asteroseismology of young massive single and binary stars, of evolved stars nearing the end of their lives,
and of compact objects. PLATO's asteroseismic characterisation of stellar ensembles, binaries, clusters and populations will be a significant addition to the Gaia 
and {\it Kepler\/} data, to mention some examples of the huge legacy of PLATO. 

The following sections detail PLATO's science objectives and provide analysis of the respective accuracy and precision that the instrument can achieve.

\section{Exoplanet Science}
\label{sec:ExoplanetScience}
\normalsize

\subsection{Exoplanet Detection}

\subsubsection{State of the Art}
The pioneering early discoveries made by ground-based telescopes  (e.g. WASP, \citealt{pollacco2006}; HATNet, \citealt{bakos2004}) and the CoRoT space mission \citep{baglin2006,deleuil2018} provided a first glimpse on the diversity of extrasolar giant and close-in planet properties. It was, however, the {\it Kepler\/} mission \citep{koch2010a, borucki2011ApJ...736...19B} which enlarged the sample of small-sized planets and enabled the first studies on exoplanet population properties~\citep{batalha2014}. For ensemble properties to become apparent, a sufficiently large number of planets with well-known parameters are needed. Today, TESS data provide a wealth of information on the population of planets with orbital periods up to about 100 days \citep{Guerrero2021}.

\begin{figure}[!htbp]
	\centering
 	\includegraphics[width=0.35\textheight]{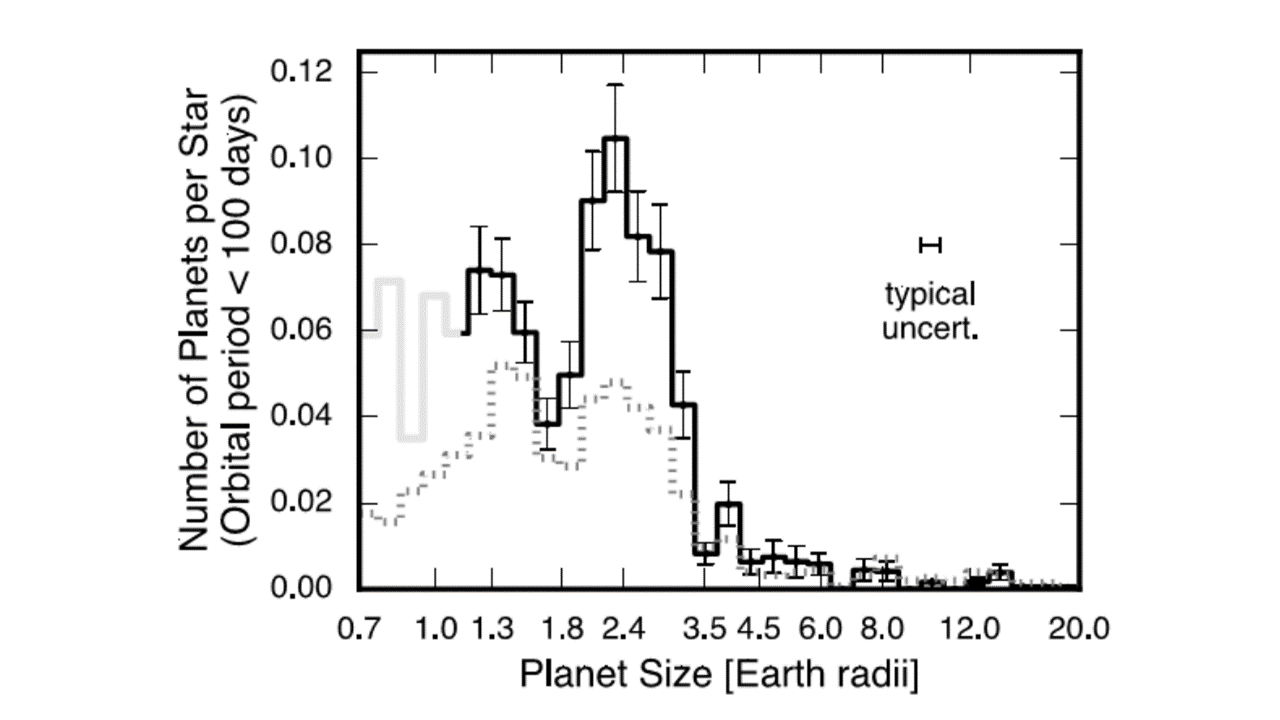}\hspace{-0.67cm}
 	\includegraphics[width=0.25\textheight]{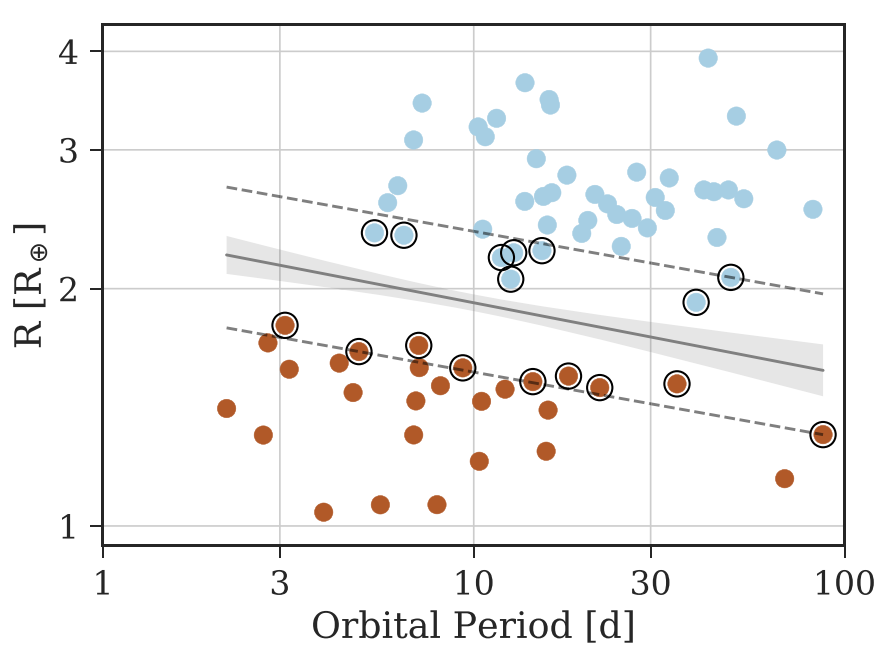}
	\caption{Planetary populations found from the accurately characterised {\it Kepler\/} host stars. The left panel shows the
 bi-modal distribution found in planetary radii (\citealt{fulton2017,fulton2018}, their Fig. 5 shown). The right panel shows the slope of the radii gap  with increasing orbital period, derived using asteroseismology for improved stellar and planetary parameters (\citealt{vaneylen2018}, their Fig. 7 shown). The cause of the gap is not completely understood but is thought to represent a break point where larger planets can retain their primordial atmospheres while the smaller objects lose their atmospheres.
	}
    \label{fig:exoDetections}
\end{figure}

Analyses of the planets with the most accurately determined radii (i.e. those with the best characterised stars) reveal a non-homogeneous distribution in bulk planet properties. Whereas in our Solar System the distribution of planetary radii follow a power law, {\it Kepler\/} data show that  there is structure in the distribution of exoplanet systems
when evaluating the planet population at large. This is particularly interesting for small planets. While there are still questions over completeness at small radii, a deficit of planets at $\sim$1.8\,\rearth\,(Fig. \ref{fig:exoDetections}) suggests physical processes (photo evaporation or core-powered mass loss) that separate the super-earth and sub-neptune populations - the so-called radius valley \citep{lundkvist2016, fulton2017, 2017ApJ...847...29O, fulton2018, venturini2020A&A...643L...1V, Izidoro2022, Ho2023MNRAS.519.4056H}. The benefit of well-characterised host stars from asteroseismology when investigating the slope of the radius valley with increasing orbital distance was shown by \cite{vaneylen2018}. 
However, although our current knowledge on exoplanet populations already provides exciting results challenging theories of planetary system formation and evolution, it is still heavily affected by observational biases. For example, the population of small planets represented in Figure \ref{fig:exoDetections} is dominated by planets with orbital periods of less than 80-100 days. There is a clear need to explore whether and how the radius valley extends to larger orbits. PLATO data will be excellent to analyze such trends at intermediate orbital distances. 

To understand the complexity of processes affecting the evolution of planets further we need to assess planet (system) properties and correlate them to, e.g.,
stellar type, activity, and age, to name just a few of the relevant effects. The large number of correlations to investigate requires not only to increase the sample of planetary systems known, but also to determine their parameters with the best accuracy. As outlined above, it is essential thereby to close observational gaps, in particular for temperate and cool planets. It is the scope of PLATO to address these unknown population parameter ranges.

\begin{figure}[h]
\begin{minipage}{1\textwidth}
\includegraphics[width=0.5\textwidth]{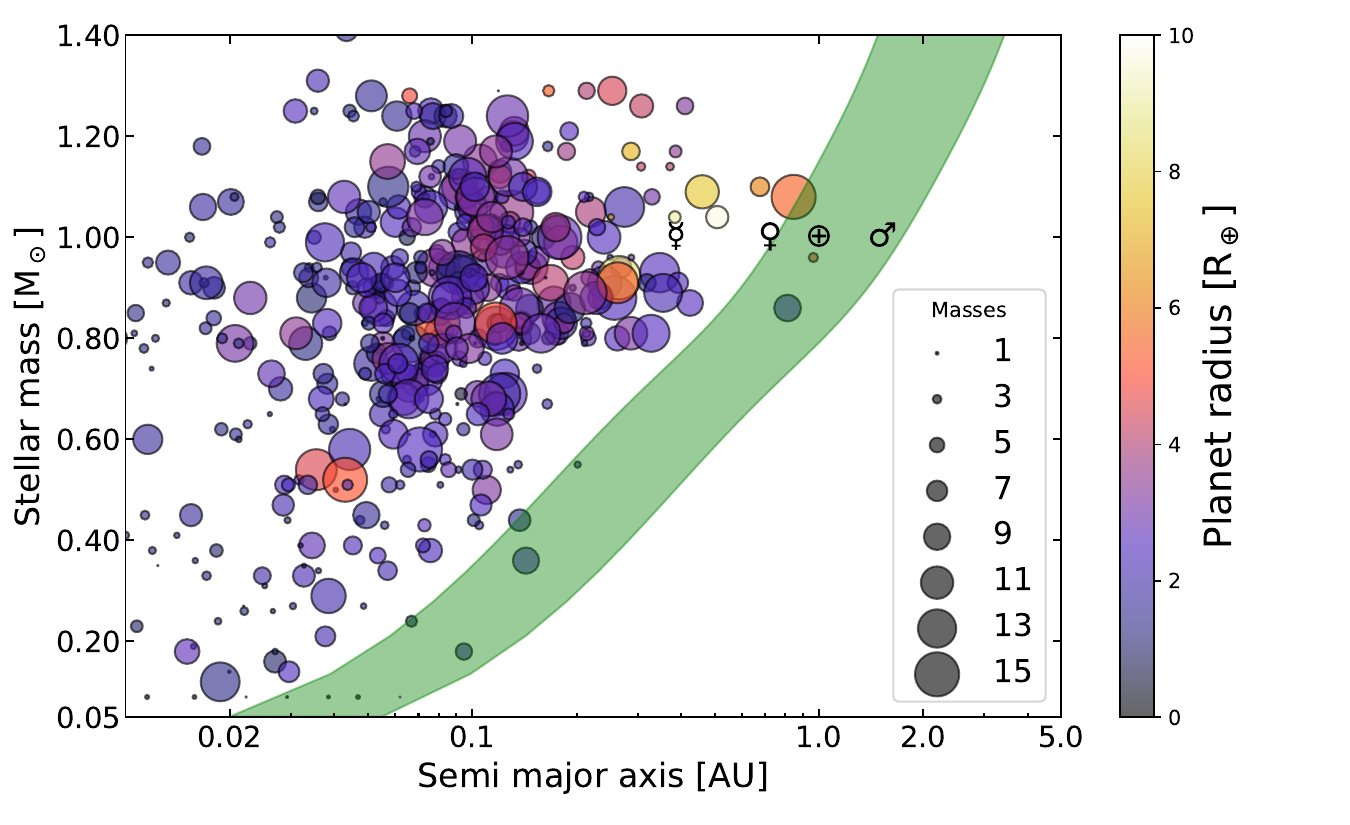}
\includegraphics[width=0.5\textwidth]
{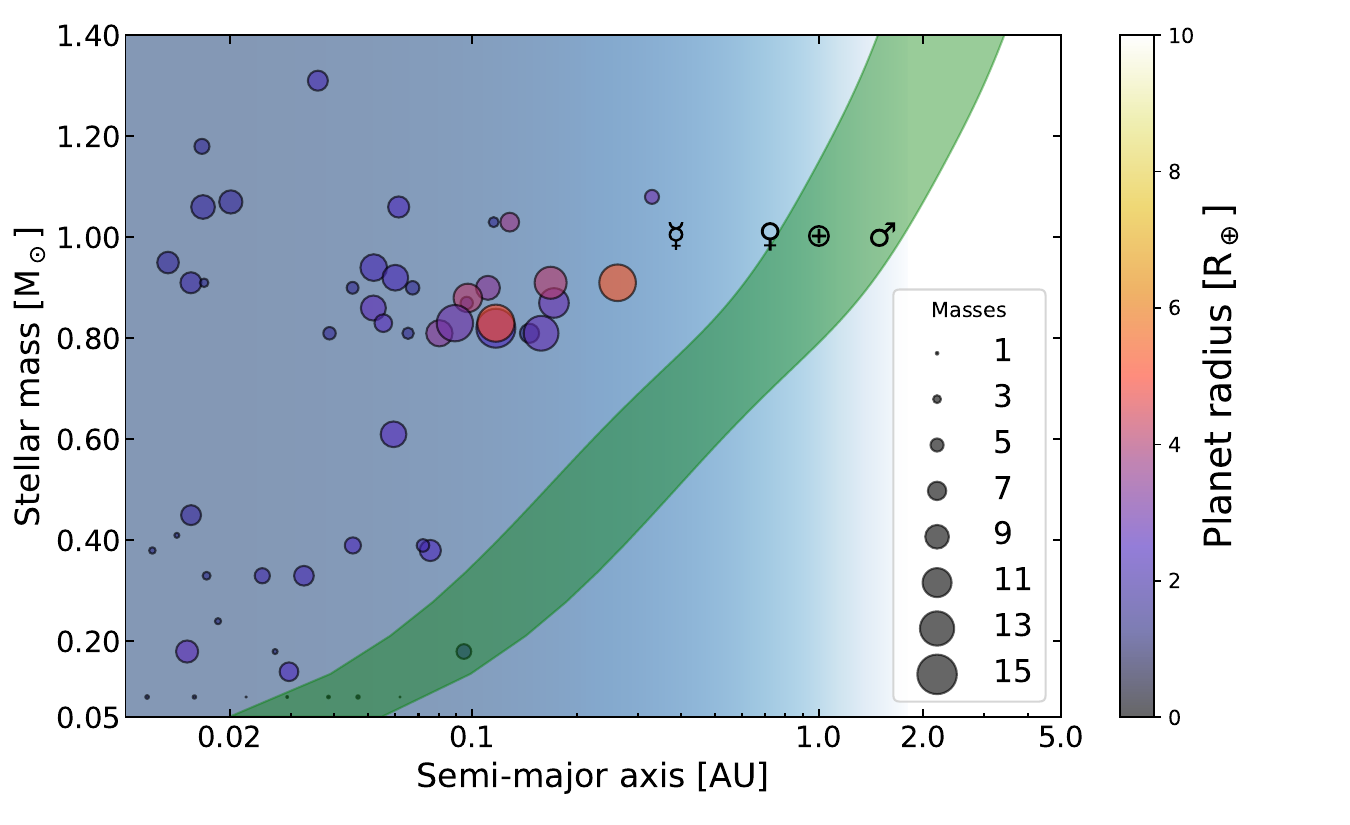}
\end{minipage}
\caption{Illustration of the PLATO objectives: populating the still 
quite empty HZ. Known small-low mass planets are shown in a stellar mass -- orbital distance diagram, with symbol size and colour scaled with their masses and radii, respectively. The green band indicates the approximate position of the HZ, accounting for a potential early Mars or Venus in our Solar System, based on \cite{kopparapu2013b}. Telluric planets in the Solar System are labelled according to their symbol. Left panel: planets \mbox{$\le$ 10 \rearth}\, and $\le$ 15 \mearth; Right panel: same, but only showing planets with precision better than 5\% and 10\% in radius and mass, respectively. The blue shaded area emphasises PLATO targeted detection domain. Data retrieved (on Sept. 2024) from the Planetary System Composite Data of the NASA Exoplanet Archive, excluding planets with mass or radius estimates from calculations.}


\label{fig:PLATOHZ2}       
\end{figure}

Figure \ref{fig:PLATOHZ2} illustrates the anticipated detection range of PLATO with respect to the location of the HZ of solar-like and M dwarf host stars. Only planets with measured radii and masses are shown in the figure. A few characterised small planets in the HZ of cool M dwarf host stars are already known (e.g. the TRAPPIST-1 system \citealt{gillon2017a}, but see also LHS-1140, \citealt{2017Natur.544..333D,2020A&A...642A.121L,2024ApJ...960L...3C}) and more detections around cool stars are expected in the near future from NASA's TESS mission, including planets in the habitable zone, \citep{2021AJ....161..233K,2023A&A...670A..84K}. In addition, ESA's CHEOPS (CHaracterising ExOPlanet Satellite) mission \citep{benz2020} can provide 
higher-precision radii for known exoplanets and therefore improve our knowledge of well-characterised bodies. However, these missions cover orbital distance ranges well below 1\,AU. 

\begin{table}[]
    \centering
    \begin{tabular}{l|l|l}
Occurrence rates & Host stellar type & Reference \\ \hline
1\% - 3\%             & Sun-like stars  & \cite{catanzarite2011} \\
20\% - 58\% (34\%)    & FGK stars       & \cite{traub2012} \\
31\% - 64\% (46\%)    & dwarf stars     & \cite{gaidos2013} \\
7\% - 15\% (11\%)     & GK stars        & \cite{petigura2013PNAS..11019273P}\\
11\% - 22\%           & GK stars        & \cite{batalha2014} \\
0.8\% - 2.5\% (1.7\%) & G stars         & \cite{foreman-mackey2014ApJ...795...64F} \\
5.3\% - 9.8\% (6.4\%) & FGK stars       & \cite{silburt2015}\\
20\% - 30\%           & Sun-like stars  & \cite{kopparapu2018}\\
16\% - 85\% (37\%)    & Sun-like stars  & \cite{bryson2021}\\
11\% - 21\% (14\%)    & FGK             & \cite{bergsten2022AJ....164..190B} \\ 
$<$14.1\%           & FGK             & \cite{kunimoto2020} \\ \hline

28\% - 95\% (41\%)    & M dwarfs        & \cite{bonfils2013} \\
9\% - 28\% (15\%)     & M dwarfs        & \cite{dressing2013} \\
24\% - 60\% (48\%)    & M dwarfs        & \cite{kopparapu2013a} \\ 
\hline
    \end{tabular}
    \caption{Examples of published values for $\eta_\textrm{Earth}$ occurrence rates from various radial velocity and transit surveys. 
    The most likely value given in each respective publication is given in brackets.
    Note that the different studies are not completely consistent with each other regarding the definitions of habitable zone, stellar types, and Earth-like planets. Actually, the diversity of definitions shows the importance of having the consistent approach of PLATO, fully characterizing planets and stars.}
    \label{tab:etaEarth}
\end{table}

Our current best knowledge on planet occurrence rates results from homogeneous re-analyses of {\it Kepler}/K2 data \citep{kunimoto2020,bryson2021}. 
Derived rates  (number of planets per star) are in the range of a few percent down to less than one percent, depending on size and orbital period. However, for small, long-period planets only upper limits can be given and uncertainties are large. Table \ref{tab:etaEarth} provides an overview of published values on the average number of Earth-like planets in the HZ. Uncertainties are large, whether they address solar-like or M dwarf host stars. For example, \citet{kunimoto2020} provide upper limits from {\it Kepler\/} data reaching from few percent up to $>$40\%  for Earth-analogues (see their Figures 4 to 7). 
It is one of the prime goals of PLATO to derive an improved estimate of the occurrence rate of  well-characterised small planets in the HZ of solar-like stars. 

For cool host stars the exploration of orbital periods longer than 100 days allows us to characterise planets beyond the HZ and  even beyond the respective ice lines in such systems. Access to these planets with well-known parameters will open a new door to comparative exo-planetology.

\medskip 
\subsubsection{Expected PLATO Planet Detection Yield}
\label{sec:PlanetYield}

To minimise the impact of biases on planet occurrence rates and derive precise analysis of planet population ensemble properties, we point out how important it is to use homogeneously analysed data sets. These data
sets do not only concern planet parameters but also those of the underlying stellar populations.
The PLATO Catalogue generated by the PLATO Consortium pipeline will provide such a data set for the community. The expected yield of new planets with characterised radii, masses, and ages is therefore a crucial objective for the mission. 

Up to now, the expected yield of planet detections has been studied in the ESA Definition Study Report (hereafter called "Red Book",~\citealt{PLATORedBook2017}), by \cite{heller2022, matuszewski2023A&A...677A.133M}, and by forthcoming publications like Cabrera et al. (in prep.). We summarise these results here as the status at the current stage of the mission development and for comparison with future PLATO yield estimates by the community. For completion, there are also studies on the fraction of planets that will be impacted by background contaminants which are also valuable for properly determining the scope of follow-up efforts~\citep{2013A&A...557A.139S,2023MNRAS.518.3637B}.

The detection efficiency of a transit survey like PLATO depends on several factors: the stellar target population observed and our good knowledge of it, the observing strategy, the actual planet occurrence rates, the geometrical transit probability, and the detection efficiency of the search methods implemented in the data analysis pipeline. In a second step, ground-based follow-up observations are essential, not only for false positives filtering and confirmation of the detection, but also for the determination of planetary masses. For the purpose of PLATO detection yields presented here, we assume that all planets detected in the prime sample are suitable for radial velocity (RV) follow-up.

\medskip
\noindent{\it The target fields} 

For the Red Book estimates in 2017, simplifying assumptions had to be made regarding the stellar populations in PLATO target fields. At that time, studies to define optimised pointing directions for the mission were still ongoing and Gaia data were not yet available. A preliminary version of the PLATO Input Catalogue (hereafter PIC) was made, providing an initial check whether the required P1 and P5 stellar sample sizes could be met (see Section \ref{sec:StellarSamples} for the requirements on P1-P5 samples, which are defined according to magnitude and signal-to-noise requirements). It was assumed that all stars in these samples were solar-like. \cite{heller2022} focused on PLATO's P1 sample, using as a stellar sample size both the requirement (15000) and the higher goal (20000) of target stars, again assuming all stars are solar-like. Today, we have significantly advanced the PIC based on Gaia data~\citep{montalto2021}. Studies on the best pointing directions have been completed~\citep{nascimbeni2022} and the first pointing field is selected (see Section \ref{sec:TargetFieldsandObsStrat}). \citet{matuszewski2023A&A...677A.133M} and Cabrera et al. (in prep.) use the results of these most recent studies on PLATO's P1 and P5 samples. 

\medskip
\noindent{\it Impact of observing strategy} 

The total planet detection yield depends on the number of target fields observed as well as the duration of continuous observations per field. The baseline for ESA operations of PLATO is an observing period with a total duration of 4 years.  
However, mission extensions are anticipated, subject to ESA review. The satellite is designed with consumables for 8.5 years of operations in total, permitting significant mission extensions. The goal to detect planets in the HZ of solar-like stars drives the observing strategy of PLATO and requires a minimum of two years (taking the Earth-Sun system as baseline) continuous observation per field (with longer periods preferred). However, shorter periods have also been considered to estimate the full planet yield potential to study different kinds of systems. In the Red Book and in~\citet{heller2022}, a baseline observing scenario of two long instrument pointings of two years each ("2+2" scenario) is compared to a "3+1" scenario. The latter allows for one year with shorter observations of, e.g., 60 days duration each. Cabrera et al., in prep., consider also extended observation periods to study the benefits for long-period planet detection (see also below).

\medskip
\noindent{\it Assumed planet occurrence rates} 

This factor presents by far the largest uncertainty in planet yield estimates, in particular for small planets in the HZ. Even today the occurrence rate of terrestrial planets in the HZ of solar-like stars is poorly known, as discussed above. We have to wait for PLATO mission results before this parameter can be better constrained. For the purpose of predicting PLATO planet yields, occurrence rates are input parameters based either on observed data when available (e.g. for close-in planets) or on upper limits/estimates provided in the literature. Results from PLATO will confirm or reject these assumptions.

For the Red Book, planet occurrence rates were taken from~\citet{fressin2013} based on {\it Kepler\/} data for all planets, except for small planets in the HZ which were not well constrained. 
To cover the wide range of uncertainty for HZ-planets (see Table~\ref{tab:etaEarth}), we assumed a planet occurrence rate of 40\% as baseline, but also computed for extreme values such as 2\% and 100\%. \citet{heller2022} assumed planet occurrence rates spanning from 37\% to 88\%  \citep[from][]{bryson2021} for earths in the HZ. Cabrera et al.\ use the same assumptions as in the Red Book for consistency, but also study additional scenarios not shown here. \citet{matuszewski2023A&A...677A.133M} use a very different approach. They address this unknown factor by using planet population models~\citep{emsenhuber2020,emsenhuber2021} to predict the number of planets formed up to the HZ.

\medskip
\noindent{\it Transit detection efficiency }

The efficiency of transit signal detection algorithms depends on several factors~\citep[see, e.g.,][]{christiansen2020}. It is relatively straightforward to quantify how the transit detection efficiency depends on the signal strength. The signal depends on the astrophysical system (planet to star radius ratio, orbital parameters, etc.) and on the level of noise in the light curve (photon noise and systematic -- red -- noise). This noise budget is determined by the instrument design (e.g.\ aperture diameter, number of cameras observing the same target) and the target magnitude (see Section \ref{sec:InstrPerformance}). In addition, temporal variability in the light curves plays a crucial role, whether it is of instrumental origin (e.g.\ effects like random telegraphic pixels, tearing effect, non-stability of temperature or voltages, etc.) or resulting from the stellar properties (e.g. intrinsic variability, binarity). This variability can be partly removed by light curve processing methods (e.g.\ for most systematic trends). Remaining variability is left in the processed light curves  as residual noise.

At the time of writing the Red Book in 2017, detailed instrument performance studies for PLATO as shown in Section \ref{sec:InstrPerformance} were not available. The best existing analogy were {\it Kepler\/}  data. It was therefore decided to account for all the above noise effects by choosing global detection efficiency factors based on the experience from {\it Kepler\/}~\citep{fressin2013} and assuming the PLATO mission noise requirements will be reached. We studied the P1 and P5 samples with the following assumptions: 
P1 sample: 100\% detection efficiency for planets down to Earth size (photon noise dominates over other noise sources for P1). This was justified by i) estimating a signal-to-noise ratio of 13 (respectively 15) observing 2 (respectively 3) consecutive transits with depth 84 ppm
(actually, with Sun-like limb darkening values, the transit depth of a planet with the size of the Earth is closer to 120 ppm, see e.g.~\citealt{heller2019}) 
and ii) considering the detectability fraction originally in \citet{jenkins1996} that is discussed in \citet{fressin2013}.

Residual stellar intrinsic noise could still impact the detection efficiency for the smallest planets, but we anticipated the well-chosen target stars in P1 not to be dominated by this effect (hence choosing quiet stars). 
P5 sample: For large planets (Neptune-sized and larger) 100\% signal detection efficiency was assumed.  Planets with 4 Earth radii produce a signal-to-noise ratio of 20 with just 1 transit around stars observed with 240 ppm, representing the worst of the P5 distribution (magnitude 13 observed at the edge of the field of view). The detection efficiency for P5 targets reduces to 50\% for planets with radii $<$ 2 \rearth\,in light curves with 80 ppm in 1 hour or better. Here the signal-to-noise ratio is 7 for 3 consecutive transits, which corresponds to 50\% fraction as per \citet{jenkins1996}. Again, we took a conservative case. For a P5 star with 50 ppm the signal-to-noise ratio with 2 transits is 9. 
Light curves with higher residual noise levels are not considered for our estimate of small planet yields, although larger planets would still be detectable.

\citet{heller2022} have improved this approach by applying their own transit signal detection algorithm~\citep[Transit Least Squares][]{2019A&A...623A..39H} to transits inserted into simulated PLATO light curves created by the PLATO Solar-like Light-curve Simulator~\citep[PSLS][]{samadi2019}. This approach mimicked the response to the known instrumental noise sources of PLATO in a somewhat more realistic manner. \citet{matuszewski2023A&A...677A.133M} and Cabrera et al. (in prep.) use an updated noise budget based on the currently known instrumental noise sources of PLATO and information from the recent PIC which was not yet available when the \citet{heller2022} study was performed.

\medskip
\noindent{\it Expected PLATO planet yields}

Table~\ref{tab:planet-yield} shows the resulting planetary transit event yields using the various approaches and input catalogues/input population as discussed above. For comparison, we summarise in column 2 the current knowledge of confirmed transiting planets in the literature. These planets result from ground-based detections as well as CoRoT, {\it Kepler\/}, TESS, and CHEOPS observations.  Today, fewer than 1500 confirmed transiting planets around stars brighter than $V=13$\,mag are known, and none of them is a small planet in the HZ of a solar-like star. Fewer than 500 planets orbit stars brighter than $V=11$\,mag and are therefore suitable for effective RV follow-up.


\begin{table}[]
\tabcolsep=1pt
    \begin{tabular}{c|c|c|c|c|c} \hline
                               & known transiting & \multicolumn{4}{c}{2+2 scenario}    \\
 Samples                       & planets   & Red Book        & Heller  & Cabrera       & Matuszewski \\ \hline
all planets orbiting stars     &           &                 &         &               & \\
$<$13 mag in P1+P5 samples     & 1407 & $\approx$4600   & n/a     & 6800-7100     & 4500-46000 \\ \hline
all planets orbiting stars     &           &                 &         &               & \\
V$<$11 mag in P1+P5 samples    &  473 & $\approx$1200   & n/a     & 1200-1350     & 1700-11000 \\ \hline
planets $<$2 \rearth\, in HZ   &           &                 &         &               & \\
orbiting P1+P5 stars $<$11 mag &      0    & 6 - 280         & 11 - 34 & 0 - 95        & $\approx$45 \\ \hline
         \hline
                               & known transiting & \multicolumn{4}{c}{3+1 scenario}  \\
 Samples                       & planets   & Red Book         & Heller  & Cabrera      & Matuszewski \\ \hline
all planets orbiting stars     &           &                  &         &              &  \\
$<$13 mag in P1+P5 samples     & 1407 & $\approx$11000   & n/a     & 10100-10700  & 12000-68000 \\ \hline
all planets orbiting stars     &           &                  &         &              & \\ 
V$<$11 mag in P1+P5 samples    &  473 & $\approx$2700    & n/a     & 2200-2500    & 4000-42000 \\ \hline
planets $<$2 \rearth\, in HZ   &           &                  &         &              & \\
orbiting P1+P5 stars $<$11 mag &        0  & 3-140            & 8-25    & 0 - 60       & $\approx$30 \\ \hline
    \end{tabular}
    \caption{Estimated PLATO planet yields. Red Book: ESA-SCI(2017)1; Heller:~\citet{heller2022}; Cabrera: Cabrera et al. in prep.; Matuszewski: \citet{matuszewski2023A&A...677A.133M}. 2+2 means 2 long pointings of 2 years duration; 3+1 means one 3-year observation followed by one year with six target fields for 60 days each, as in the Red Book. Known (confirmed) transiting planets are taken from the NASA exoplanet archive in Oct.\ 2024 for all planet radii and orbits.}
    \label{tab:planet-yield}
\end{table}


PLATO is predicted to detect about 1200 transiting planets of all sizes in target stars with $V<11$\,mag, which would more than triple our current knowledge of well-characterised planets. 
These planets orbit stars which can be followed by RV ground-based spectroscopy to derive their masses. The number of planets followed will eventually be determined by the availability of telescope resources. For those planets included in the so-called ``prime sample'' (see Section \ref{sec:StellarSamples}), the PMC will provide sufficient ground-based resources by coordinating the follow-up community (Section \ref{sec:GOP}). Planets detected in this sample will be the core of PLATO's planetary systems catalogue. Other planets detected around bright stars will be followed-up by the community around the world and form a long-term legacy of PLATO. 

For host stars of $V=11$ to 13\,mag the resulting planet properties from PLATO will be similar to the main part of the planet populations provided by the {\it Kepler\/} mission, hence masses for small, terrestrial planets will be available only for objects of special interest justifying dedicated follow-up campaigns (i.e. as for CoRoT-7b or Kepler-10b of $V=11$\,mag) or from Transit Time Variations (TTV) analysis. Nevertheless, these planets around fainter stars will be highly useful to study e.g. radius-orbit distributions, etc., just as for similar population analyses carried out with {\it Kepler\/} data. Accounting for an improved age calibration of stars from PLATO also for classical methods, we may be able to go beyond {\it Kepler\/} and study, e.g., the radius gap and orbital distance distributions as a function of stellar age also for this faint population. 
Considering planets of all sizes and all orbital periods, about 4600 planets (see Table \ref{tab:planet-yield}) can be detected by PLATO around stars brighter than $V=13\,$mag by combining the P1 sample with the P5 sample. The PLATO planet yield in this magnitude range is therefore similar to the total planet yield in {\it Kepler}, but most {\it Kepler\/} planets orbit stars of $V>$14\,mag and are consequently difficult to characterise. Thus, at the bright end, where they are more suitable for further characterisation, the number of expected PLATO detections is higher than what has been achieved with
{\it Kepler}. Indeed, although the somewhat smaller effective aperture of PLATO leads to higher noise levels at the same magnitude (see Section \ref{sec:InstrPerformance}), this is compensated by the significantly wider field-of-view of PLATO. This is a consequence of PLATO's modular camera approach and design.

The estimates by~\citet{heller2022}, as well as those by Cabrera et al.\ using more realistic stellar distributions and noise budget levels, are within the uncertainty ranges presented in the Red Book. The high end of numbers from \citet{matuszewski2023A&A...677A.133M} stem from using a planet formation model for the assumed planet occurrence rate. The formation model produces higher planet occurences than assumed by the other authors based on, e.g., {\it Kepler\/} results. The example shows the need for observational constraints to planet formation models.

\begin{figure}[h]
\begin{center}
\includegraphics[width=0.95\textwidth]{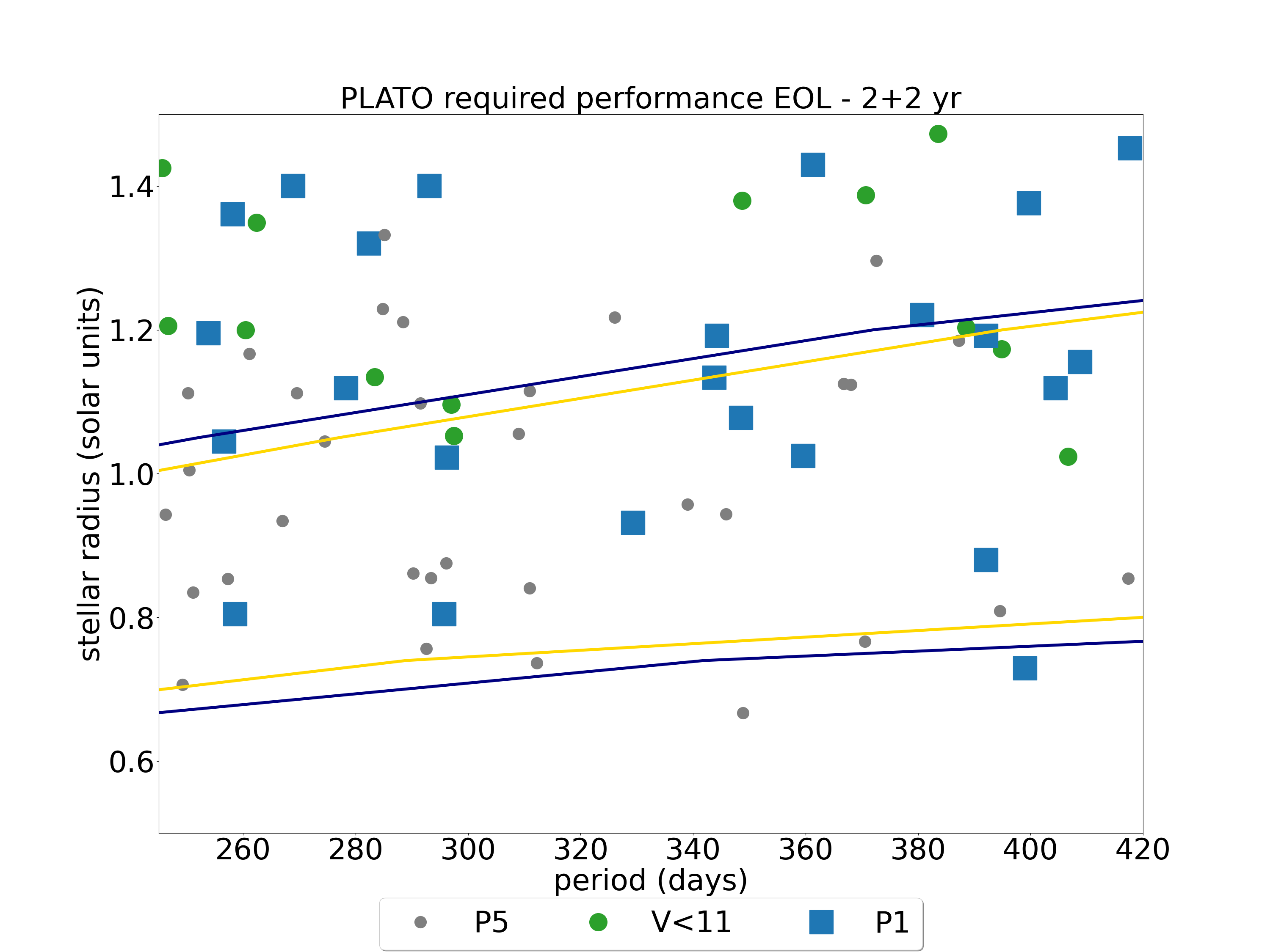}
\end{center}
\caption{Predicted yield of planets with $<$2 \rearth\, in the HZ for the 2+2 scenario assuming 40\% occurrence rate and EOL performance (see Section \ref{sec:InstrPerformance}). We include two definitions for the HZ (the continuous lines): the optimistic~\citep[][blue]{kasting2013N} and the more conservative~\citep[][yellow]{kopparapu2013a}. Symbols indicate the magnitude of host or target samples (P1 and P5) as indicated in the label.}
\label{fig:yield_pl_HZ}
\end{figure}

The key planet sample of PLATO consists of small planets in the HZ of solar-like stars. The Red Book estimates a wide uncertainty range (6--280) for the yield of such planets around bright ($V<$11\,mag) stars, resulting from the large uncertainty in planet occurrence rates.  These planets will mostly be suitable for RV follow-up, eventually depending on telescope time available, and form the main data product from PLATO for exoplanetary science. The study by~\citet{heller2022} predicts a smaller number of planets in the HZ, but 
that is because these authors concentrated exclusively on Earth-sized planets ($<$ 1.5 \rearth) orbiting P1 stars, while the Red Book estimates consider planets of $<$ 2 \rearth\, and stars with $V<11$ mag in the P5 sample too. Cabrera et al.\ (in prep.) agree with previous results with a large uncertainty range, which is again associated with different assumptions for planet occurrence rates. 
\citet{matuszewski2023A&A...677A.133M} use a planet synthesis model to estimate the expected planet occurrences. 
This approach results in much higher numbers of predicted planets, all other assumptions being the same. Figure~\ref{fig:yield_pl_HZ} illustrates the expectations for PLATO for small planets in the HZ of solar-like stars for the 2+2 scenario addressed in the Red Book with current knowledge of the end-of-life (EOL) instrument performance of PLATO (see Section \ref{sec:InstrPerformance}) using PIC version 1.1.0. The figure shows roughly 50 stars in the HZ which is in agreement with the prediction range of 0 to 95 in Table~\ref{tab:planet-yield}, performing simulations as described above. 


In summary, at this point the predictions on the number of planets to be detected by PLATO are highly dependent on the  assumed planet occurrence rates, in particular for temperate small planets. Hence, these predictions merely reflect our poor knowledge on planet formation efficiencies for small planets. Improving this situation with PLATO determined planet occurrence rates from a homogeneously analysed sample is a key science driver for PLATO.

\begin{figure}
\includegraphics[width=1.\textwidth]{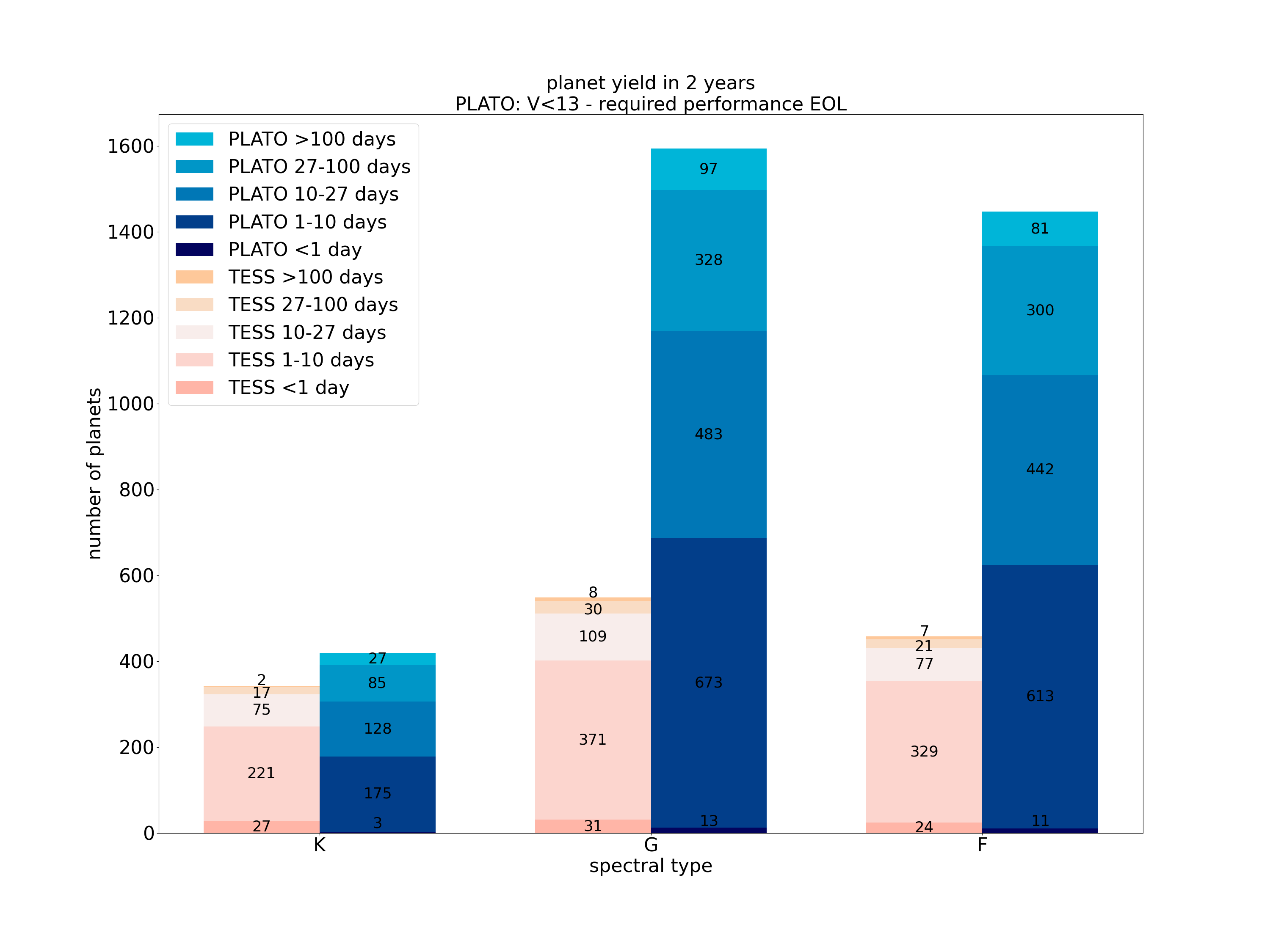}\\
\includegraphics[width=1.\textwidth]{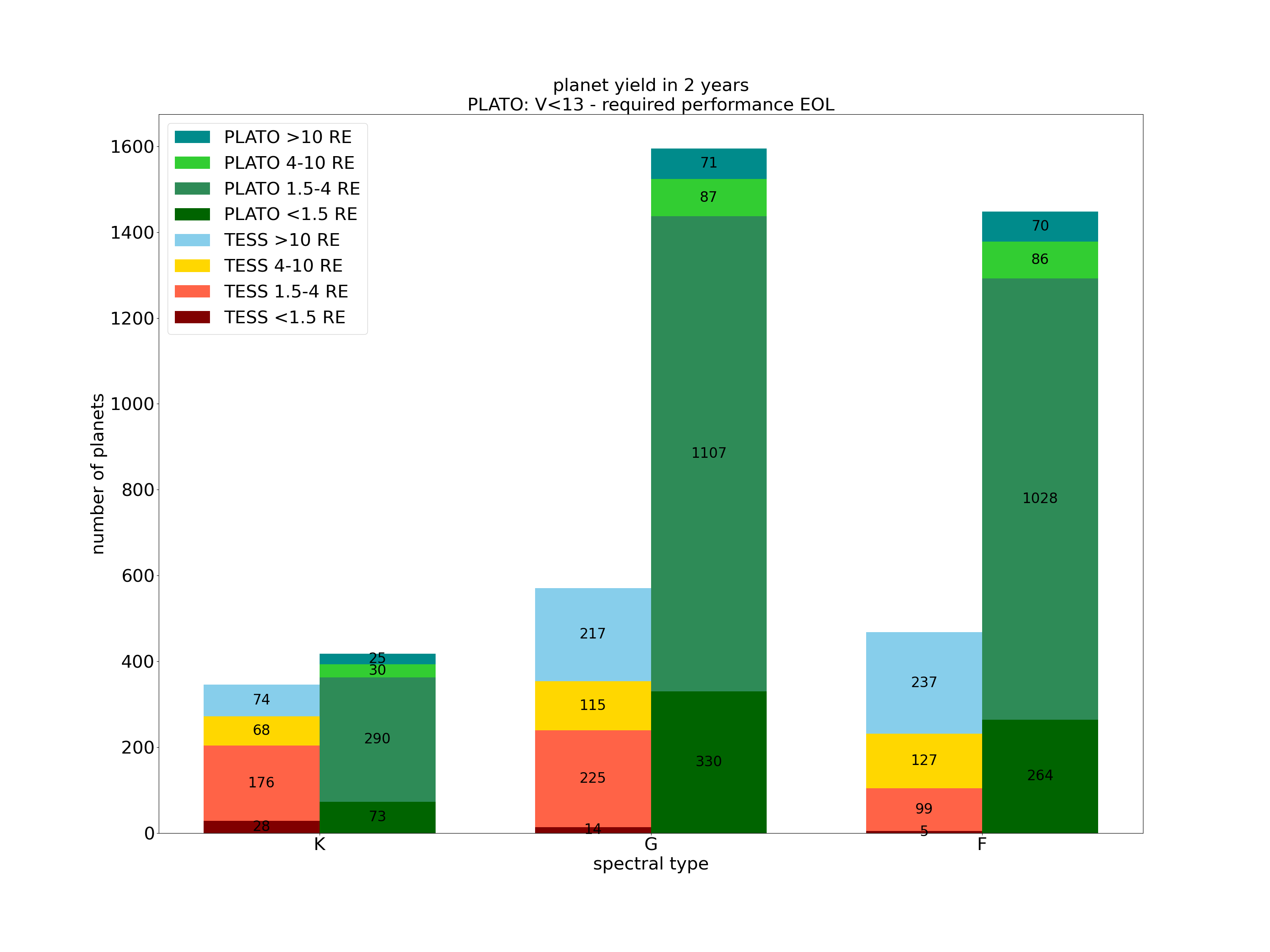}
\caption{Spectral type distribution for TESS Object of Interest (TOI) host stars compared with our expectations for PLATO.
The number of planet detections for each spectral type (K, G, and F) is plotted as histogram.
The values for TESS are taken from~\citet[][their Fig. 7]{Guerrero2021} and correspond to all magnitudes.
False-positive TOIs have been removed.
For PLATO we have taken the stellar population from the PIC~\citep{montalto2021}, which in its current version does not include M or A stars (hence we do not show any counts for these spectral types).
We have considered the end-of-life (EOL) performance as per requirements, which is a conservative approach (see Section \ref{sec:InstrPerformance}).
We compare the nominal mission of TESS (2 years) with half of the nominal mission for PLATO (2 years). 
The PLATO results for the nominal mission (4 years) are about a factor of 2 higher (2+2 scenario in Table~\ref{tab:planet-yield}). Top: breakdown in orbital periods; Bottom: Breakdown in planetary sizes.}
\label{fig:TESS_PLATO}
\end{figure}

\medskip
\noindent{ \it Comparison to TESS}

To illustrate the expected impact from PLATO further, we compare PLATO with TESS choosing the validated TESS Objects of Interest (TOI) of the 2 year nominal mission~\citep{Guerrero2021}.
Figure \ref{fig:TESS_PLATO} shows a comparison of TOIs orbiting FGK stars compared to PLATO expectations for a single field observed for 2 years.
The predicted number of planets orbiting K stars detected by PLATO is only modestly increased compared to that expected from TESS.
For G and F type stars, however, PLATO is expected to outnumber TESS results by factors 2-3 (G stars: 571/TESS and 1595/PLATO; F stars: 468/TESS and 1448/PLATO). It is interesting to compare the performance for temperate small planets of both missions for the same 2-year observing duration. Most TESS TOIs have orbits $<$10 days and only very few have orbits $>$27 days. PLATO planets will show a significantly larger fraction of longer periods from 27 days to $>$100 days. This is also the case for planets orbiting K stars, even though the total number of planets is comparable. 
Figure \ref{fig:TESS_PLATO} shows how the PLATO efficiency to detect small planets around K, G, and F stars largely outnumbers TESS's performance when considering the same observing period. 

In summary, comparing the TOI yield of TESS after 2 years of observations with the expectations for PLATO for the same time span, we clearly see 
the advantage of the PLATO mission design for detecting small and long-period planets.
Note also that only a very small fraction of the TESS long-period planet discoveries will 
occur in the PLATO fields due to the different observing modes.

The expected planet yield for the TESS extended mission has been addressed in~\citet{kunimoto2022}, showing a factor 3 increased number of expected new detections for a total of 7 years of operations.
This includes those to be discovered from analyses of the full-frame images, as well as the higher cadence of 2-minute TESS light curves. However, also for the TESS extended mission most planets are expected to have orbits up to about 30 days and will be larger than 4 \rearth\,(their Fig. 6). 
Regarding small temperate planets in the HZ of solar-like stars (see Fig.~\ref{fig:yield_pl_HZ}) we compare PLATO's capacity with Fig. 5 in~\citet{kunimoto2022}: in 
their bin of $<$2 \rearth\, and orbital periods between 250 and 400 days, no planet detections are foreseen. This illustrates that we have to await the results from PLATO to fill this parameter range, which includes the range Earth would inhabit. It is the larger aperture (see also Section~\ref{sec:InstrPerformance}) and the observing strategy of PLATO that 
will allow breakthroughs in this parameter space.

\medskip
\noindent{\it Impact of extended observations}

Recent studies addressed whether the detection efficiency of PLATO could be improved with adapted observing modes, e.g. by extending the duration of field pointings. As can be seen in Table~\ref{tab:planet-yield}, the total number of detected transits increases with increasing number of target fields observed, e.g. from about 4600 planets for 2+2 years of observations to about 11000 planets when observing one field for 3 years followed by 6 fields of 60 days each, as outlined in the Red Book. Obviously, most of the planets from the step-and-stare year are of short orbital period. 
At the same time, the number of small planets in the HZ decreases in the 3+1 scenario. There is a trade-off 
between observing fewer stars for longer versus more targets but with shorter-duration observations
leading to fewer transits within a given observing time.
Most interesting for PLATO (after the success of TESS on short period planets) are scenarios with long observations of a target field to increase the chances for HZ planets. 
Clearly, the selection of the first target field and options to increase its observing duration will be part of choosing the optimal observing strategy for PLATO. 

\begin{figure}[h]
\includegraphics[width=0.55\textwidth]{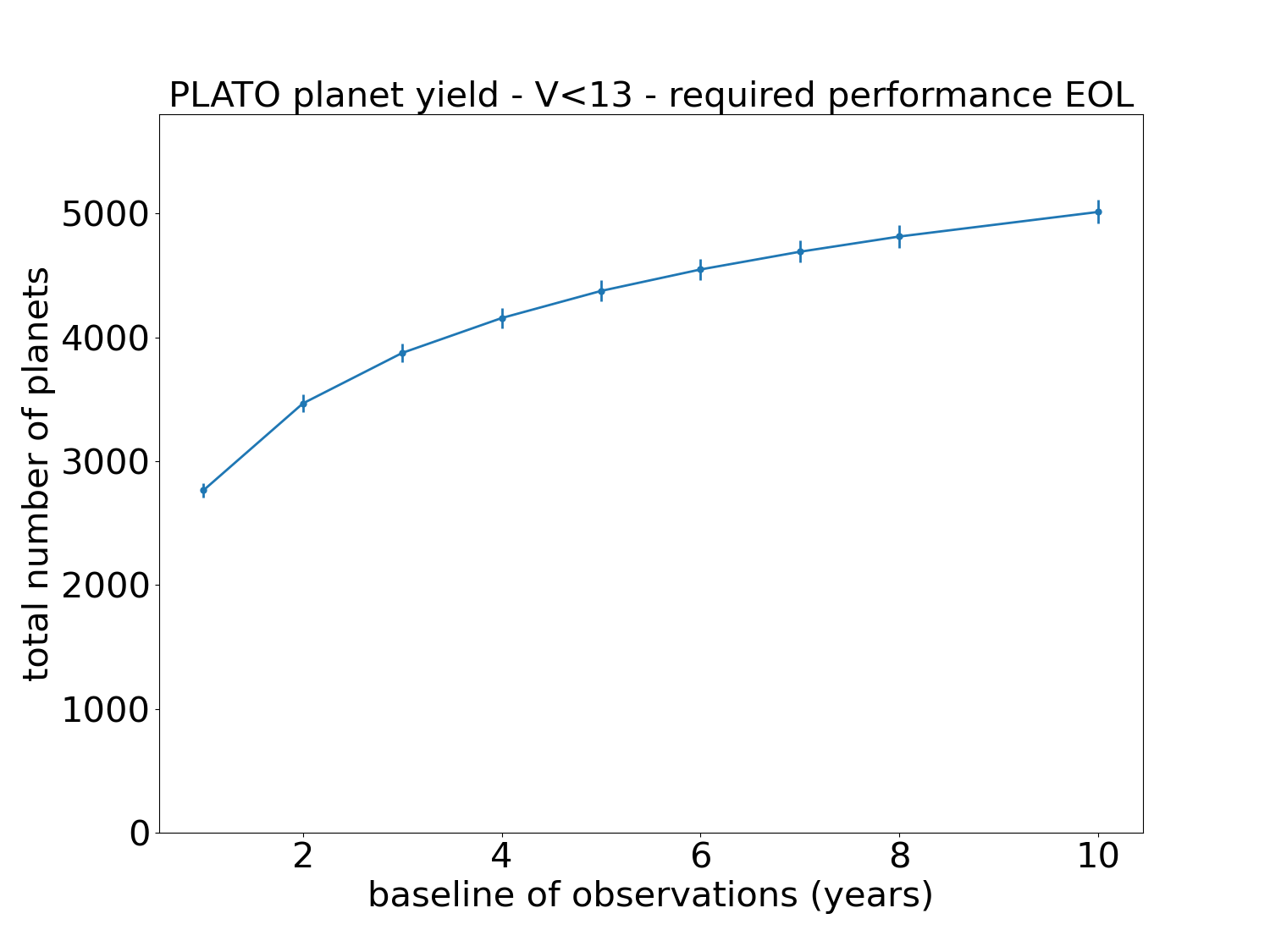}\hspace{-0.67cm}
\includegraphics[width=0.55\textwidth]{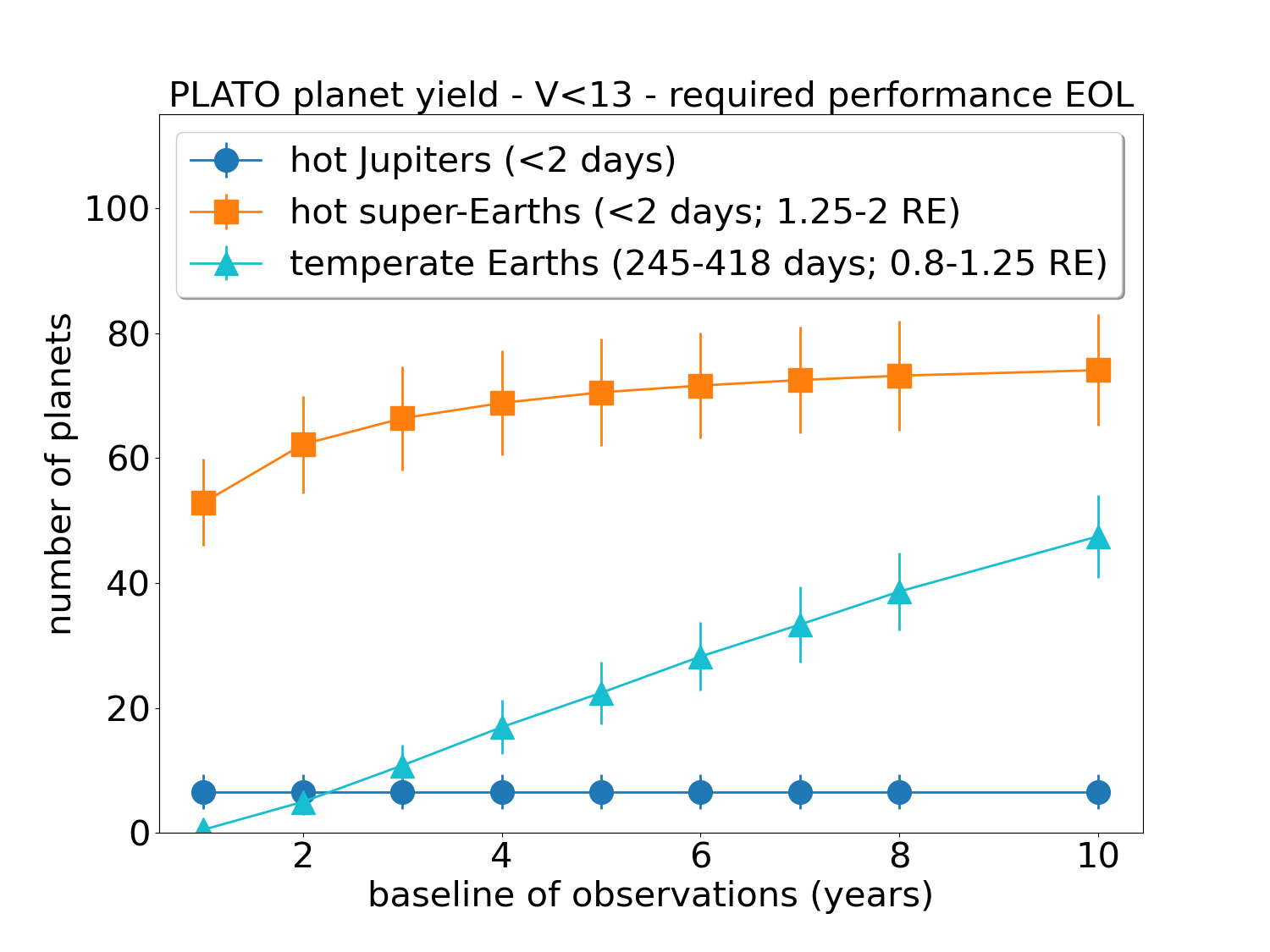}
\caption{The figure shows a prediction of the total number of planets that could be discovered by PLATO as a function of the duration of the observation run. This is done for a single pointing field following Cabrera et al. in prep. The left figure presents together planets of all sizes, all orbital periods, around stars brighter than magnitude 13. The right figure shows the same but for hot Jupiter planets (defined as planets with 6 to 22 \rearth\, and orbital period $<2$ days), hot super-earths (defined as planets with 1.25 to 2 \rearth\, and orbital period $<2$ days), and temperate Earths (defined as planets with 0.8 to 1.25 \rearth\, and orbital period between 245 and 418 days). The vertical lines represent the expected uncertainty in the number of planets. We have considered the end-of-life (EOL) performance as per requirements, which is a conservative approach (see Section \ref{sec:InstrPerformance}).}
\label{fig:detection_yield_with_time}
\end{figure}

Figure~\ref{fig:detection_yield_with_time} shows a prediction of the total number of planets discovered by PLATO as a function of the pointing duration. 
This prediction is done for a single pointing field (Cabrera et al.\ in prep.). Note that only planets with orbital periods up to 418 days were simulated \citep{fressin2013}.
Beyond this value we have no statistics, so longer-period planets will not add to the values in our analysis.
Nevertheless, even if their transit probability is small, theoretical models predict that such long-period planets could be numerous~\citep[e.g.][]{matuszewski2023A&A...677A.133M}.
Figure~\ref{fig:detection_yield_with_time} shows that hot Jupiter detections saturate very quickly, within one year of observation duration. In this saturated regime, observing 2 independent fields for 2 years each would double the amount of planets detected, compared with the yield from one field targeted for 4 years. We note, however, that by the time of PLATO launch, the TESS mission already significantly increased our knowledge on these kind of hot gas giants and they may not be the main driver for selecting PLATO fields and their observing mode. 
The number of detected hot super earths saturates within 2 to 4 years of observations. Hence, once saturation is reached, additional detections can only be gained when changing the target field. In contrast to the hot planets, the number of temperate 
earths grows almost linearly with the pointing duration for up to 10 years. We clearly see the benefit in the detection yield  of temperate planets when increasing the observing duration of a target field. This behaviour points towards increasing the observation of PLATO´s first target field beyond the nominal 2 years. However, before finally deciding on extended observing durations, we also have to consider the availability of resources for follow-up and the real in-flight instrument performance that might affect the expected detection efficiency. Once these factors are better known, a refined assessment on PLATO´s detection efficiencies as a function of target field observing duration can be made.

As outlined in~\citet{rauer2014}, we expect additional planet detections around, e.g., binary (multiple) stars (including circumbinary planets), subgiant and giant stars, as well as the potential for exo-comets, planetary moons, rings or Trojan planets. Estimates for the efficiency of PLATO detections on these targets are subject for future studies and beyond the scope of this review. This applies also for planet detections made via, e.g., Transiting Time Variations (TTV) analyses or reflected light observations. Here, we only recall these extra methods and targets as a reference for the many new discoveries expected from the PLATO mission.

In summary, although recent studies spent some effort to refine the expected transit detection efficiency in PLATO light curves, the largest unknown remains $\eta_\mathrm{Earth}$, the unknown planet occurrence rate of small HZ planets. Current planet yields predict the detection of a few up to a bit more than a hundred small planets in the HZ of solar-like stars, depending
on the value of $\eta_\textrm{Earth}$. These planets are  suitable for radial-velocity follow-up (depending on telescopes availability) and will provide accurate radii, masses, and ages. PLATO data will therefore significantly improve our knowledge of planet statistics of well characterised planets, especially those small in size and with long orbital period - or - in the HZ of their solar-like host.

\subsection{Planet Characterisation}

\label{sec:PlanetCharact}

\subsubsection{Brief summary of state of the art}

PLATO aims to assemble the properties of statistically-significant ensembles of planets, more so than to focus on individual planets. Clear trends and potential clustering in planetary properties will only become apparent once we manage to reduce error bars for the key parameters and can access a large and homogeneously analysed catalogue of planets, such as PLATO will provide.

Derived planetary mass-radius relationships  and mean densities are important indicators of the nature of detected planets. To first order, the mean density-mass relationship can be used as an indicator for planet classification: \cite{hatzes2015ApJ...810L..25H} use it to reveal the turnover from small terrestrial and Neptune-like planets to gas giant planets and further for the transition to brown dwarfs. More frequently mass-radius diagrams are used to obtain a first indication on the nature of planets (starting with, e.g., \cite{seager2007ApJ...669.1279S, adams2008ApJ...673.1160A, 
wagner2011}, and many since). See  \cite{otegi2020A&A...640A.135O} for a recent study of transiting planets up to 120 \mearth. 
Even though a unique one-to-one mapping to particular planetary internal structures and compositions is difficult  based on radius and mass data alone due to the inherent degeneracy of the problem (see, e.g., \citealt{valencia2007ApJ...665.1413V, 
rogers2010ApJ...712..974R, dorn2015}), these data nevertheless provide important constraints on formation models when ensemble properties become apparent. 

The situation on deriving internal structure and composition of planets improves when simplifying assumptions can be made like elemental abundances taken from host stars (e.g. \citealt{dorn2017, brugger2017}) or information from our knowledge of planets in the Solar System. 
Although this greatly reduces the various degeneracies, it is clear that many assumptions currently made are over-simplified. For example, the host star photospheric composition today may not reflect that of its planets  \citep{plotnykov2020} as additional processes including devolatilization trends \citep{wang2019MNRAS.482.2222W} or diverse impact scenarios can alter the planet’s composition, whereas material falling into the star might affect the measured stellar composition. Therefore, planet characterisation models should not rely only on stellar constraints, in view of large error bars that prevent a definite conclusion. While degeneracies can likely never be completely resolved, they can be reduced  when the primary parameters are precisely measured (as done with PLATO) and when combined with additional data, in particular providing information on the planetary atmospheres
 (e.g. with JWST, ARIEL, ELT, LUVOIR/HABEX (now the Habitable World Observatory, HWO), LIFE). A key factor will be to increase the available statistical sample of well-characterised small planets.

A recent example of planet characterisation by combining density measurements with further atmospheric data are TRAPPIST-1 b and c. Emission photometry of planet b suggests a bare rock surface with apparently no atmosphere \citep{greene2023Natur.618...39G}, while planet c is unlikely to have a thick CO$_2$ atmosphere \citep{zieba2023Natur.620..746Z}. \cite{acuna2023A&A...677A..14A} use these measurements to constrain the Fe-to-Si ratio of TRAPPIST-1 b based only on the density of the planet, without making assumptions on the composition of the star, by performing a retrieval on the mass and radius using a two-layer interior structure model. PLATO will provide a wealth of data to further follow such approaches to planet characterisation.

The case is particularly challenging when considering planets in-between a predominantly gaseous or rocky nature, the so-called mini-neptunes (or ´mini gas planets´, hence small gaseous planets) and also when considering masses from earths to super-earths. If one assumes for simplicity that such planets would consist of four distinct layers (i.e. an iron core, a silicate layer, a water/ice layer and an atmosphere), several combinations of these layers of different inherent densities or mass fractions can result in the same mean planet density. Early studies of super-earths with little or no hydrogen atmosphere showed that planet radius measurements to better than 5$\%$ together with masses determined to better than 10$\%$, such as PLATO will provide, would allow us to distinguish between an icy or rocky composition (e.g. \citealt{valencia2007ApJ...665.1413V}). Recent studies with more advanced interior models, allowing for additional phases and internal layers in the interior,  confirmed these requirements on radius and mass precision when constraining the planets interior structure \citep{otegi2020A&A...640A.135O, dorn2021ApJ...922L...4D, baumeister2023A&A...676A.106B}. 

Furthermore, depending on the internal energy and in turn on the age of a planet, large portions of the silicate layer may be molten, leading to water being dissolved in the magma ocean rather than a separation of the two layers, which would directly influence the planet’s observed density \citep{dorn2021ApJ...922L...4D}. The internal energy further determines if the metallic core can efficiently separate from the mantle, which also influences the measured planet radius and hence density \citep{elkins-tanton2008ApJ...688..628E, lichtenberg2021ApJ...914L...4L}. The precise measurement of planetary age will help to assess the likelihood of different endmember scenarios, since cooling and differentiation are strongly time-dependent processes.

The planetary bulk composition and evolving interior structure are also important for assessing the possible evolutionary pathways of such planets (especially for low-mass, rocky planets) and their surface processes. Studies focusing on Earth-like compositions have already shown that key processes such as plate tectonics, volcanic activity or magnetic field generation (all directly linked also to the atmospheric evolution) strongly depend on planetary mass, metallic core size and surface temperature \citep{wagner2011, wagner2012, valencia2007ApJ...665.1413V, kite2009ApJ...700.1732K, ortenzi2020, dorn2018A&A...614A..18D, bonati2021JGRE..12606724B, kislyakova2020A&A...636L..10K, baumeister2023A&A...676A.106B}. The atmospheric evolution is further influenced by the orbital distance of the planets, another observable of PLATO, leading to a bifurcation in atmospheric evolution \citep{hamano2013Natur.497..607H} and erosion efficiency \citep{godolt2019A&A...625A..12G, moore2020MNRAS.496.3786M}, which can furthermore influence the measured planetary radius.

Apart from precise mass and radius measurements alone, constraints on planet composition and internal structure will be improved  significantly when combined with additional observables, e.g. atmospheres as discussed already above.  However, there are also additional observables from PLATO data alone, such as the fluid Love number for giant planets near the Roche limit  
\citep{correia2014A&A...570L...5C, kellermann2018A&A...615A..39K, 
padovan2018A&A...620A.178P, akinsanmi2019A&A...621A.117A,
csizmadia2019A&A...623A..45C, hellard2019ApJ...878..119H, hellard2020ApJ...889...66H, baumeister2023A&A...676A.106B},  stellar parameters (e.g. heavy element content, activity), and especially age (e.g. to compare with timescales of atmospheric loss processes). In particular the correlation of planetary parameters with ages has the potential to provide significant new insights into the development of planets. 
For example, for warm Jupiters PLATO ages will help to break the degeneracy with respect to their heavy element component \citep{mueller2023arXiv230412782M}. How planetary properties correlate with age (e.g. due to contraction, atmosphere loss, tidal interaction, etc.) can be studied once ages are available in sufficient numbers and accuracy from PLATO. Correlating terrestrial planet properties with age is entering a new area of understanding planets similar to our own. PLATO ages will be an important element in the full characterisation of these planets with JWST as well as future HWO and LIFE type missions. 

Unfortunately, our current knowledge on mass-radius and planetary densities is still significantly observationally biased. Concerning close-in planets, we expect many new ultra-short-period planets and hot gas giants to be detected using NASA´s TESS mission (see Section \ref{sec:PlanetYield}) and characterised with follow-up, including atmosphere studies by JWST and ARIEL. For those planets in the  PLATO fields, additional data such as TTVs, albedos, phase curves, and high accuracy radii and ages from PLATO will complement TESS, JWST and ARIEL data.

\begin{figure}[h]
\includegraphics[width=.6\textwidth]{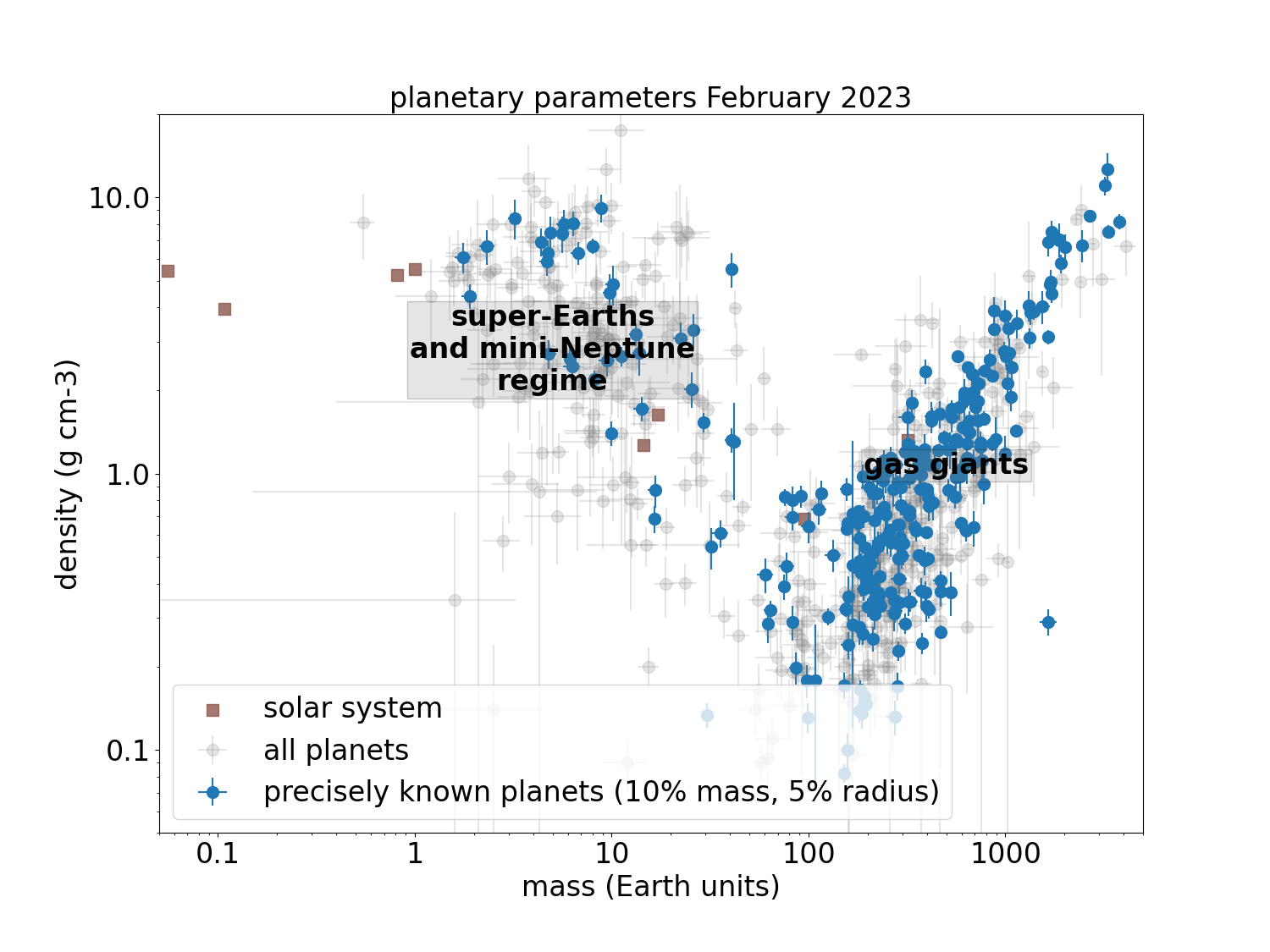}
\includegraphics[width=.6\textwidth]{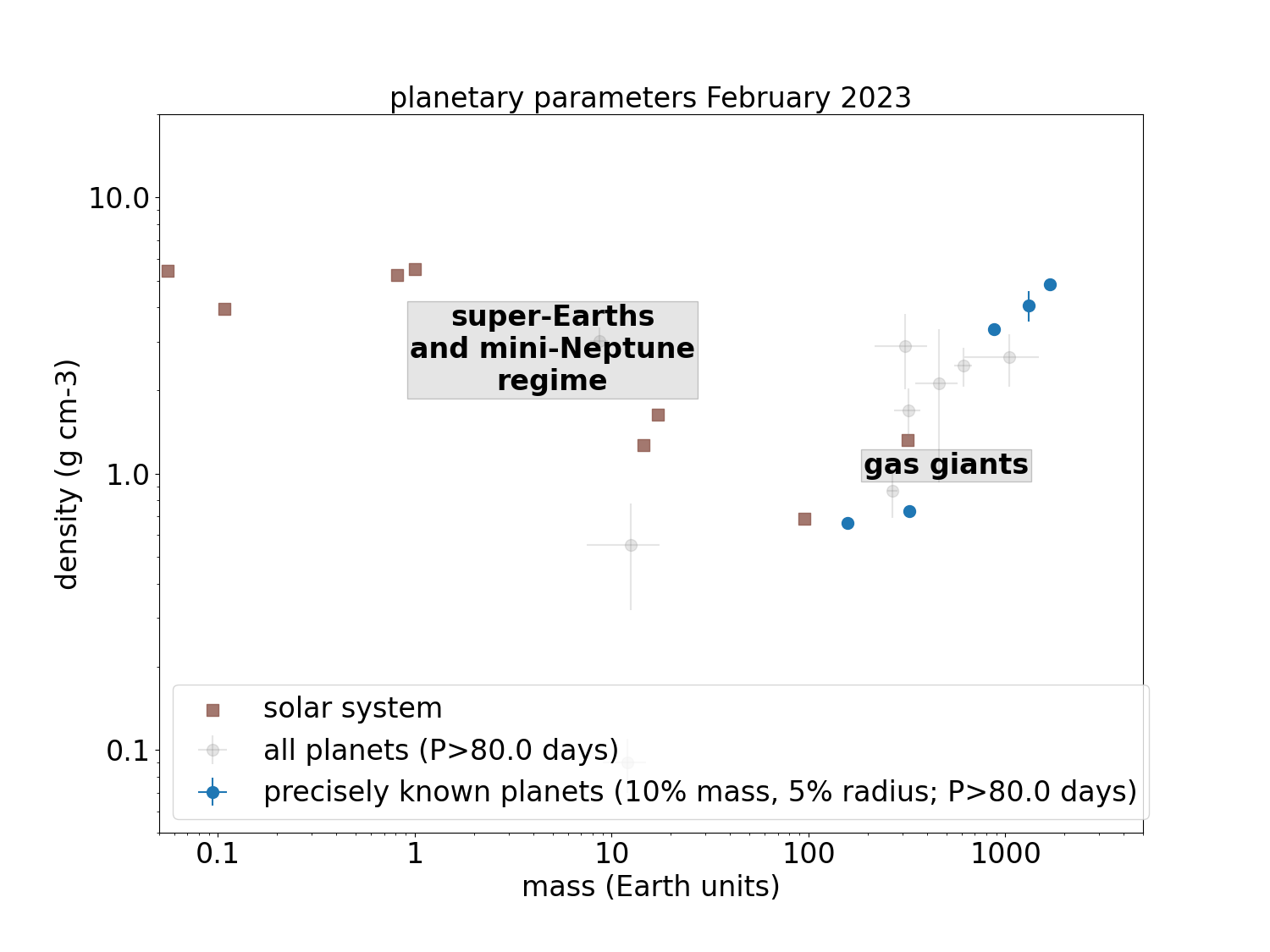}
\caption{Known planetary mean densities versus mass (status February 2023). Grey symbols indicate confirmed planets, blue symbols show planets with precisely known parameters. Brown squares indicate Solar System planets. Left: all planets; Right: only planets with orbital period $>$80 days. Each value in the plot is taken from the most recent publication on the target that provides a consistent analysis of the planetary system (stellar parameters and planetary parameters) and includes measurements from radius and mass.}
\label{fig:densities}
\end{figure}

How planet densities (hence planet nature) correlate with orbital distance (orbital period) will be among the key findings of PLATO.  
Fig. \ref{fig:densities} shows our current knowledge of planets with known mean densities. Most of the planets with precise mass and radius, hence density, are in the gas giant regime (Fig. \ref{fig:densities}, left). When considering planets with orbital periods beyond 80 days, however, only few gas giants with precise densities are known to date (Fig. \ref{fig:densities}, right). The left branch of this diagram, where terrestrial planets and mini-Neptunes are located, is empty to date for periods beyond 80 days, except for the planets in our Solar System.  How small planets are distributed at intermediate orbital distances around solar like stars will be revealed by PLATO. These results will form a major legacy of the mission. Such observations will also show whether terrestrial planets are present beyond the ice line in M dwarf planetary systems, which forms another 'unknown' to be revealed by PLATO.

Understanding the nature of exoplanets is a significant puzzle to be resolved, requiring the combination of all available observational information. The role of PLATO in solving this puzzle is in the provision of exploring accurate planet parameters (including the ages) in a homogeneously processed sample of objects. 

\subsubsection{Expected planetary radius accuracy}
\label{subsection-planet-radius}

PLATO requirements for planetary radius and mass are defined for a reference Earth-Sun scenario: an Earth-sized planet orbiting a G0V star as bright as $V=11\,$mag at 1 AU (see Section 
\ref{sec:SGS}). The required planet radius accuracy is 5\% for this reference case. For the brightest targets ($\le$10 mag) a radius with 3\% accuracy should be achieved. This corresponds to an accuracy for R$_\mathrm{planet}$/R$_\mathrm{star}$ of 2\%. We note that these requirements are in many ways the most difficult case and larger planets and/or smaller host stars will provide even better precisions, depending on respective signal-to-noise levels and stellar models. 

For now, we assume that the required stellar radius accuracy of 2\% is reached (see Section \ref{seismicperformance} for a discussion on PLATO's capacity to deduce stellar parameters). Hence the precisions on planetary radii discussed here provide \textit{accurate} radii in terms of our terminology. 
Planetary radii are then obtained by fitting the shape of observed transit events in the processed light curves. Residual noise sources (instrumental or stellar) therefore could affect the achievable precision. In Section\,\ref{sec:InstrPerformance} we show that PLATO light curves will be dominated by white noise for stars between $V=8$ and 12\,mag. Hence, PLATO instrumental noise effects are minimal for the P1 sample and even for the bright end of the P5 sample. What remains is a potential impact by residual stellar variability in the light curves. In a first step we neglect such an impact (see the literature overview of detailed studies made at the end of this section, showing that such an assumption is justified for the bright samples).

\begin{figure}
\includegraphics[width=0.5\textwidth]{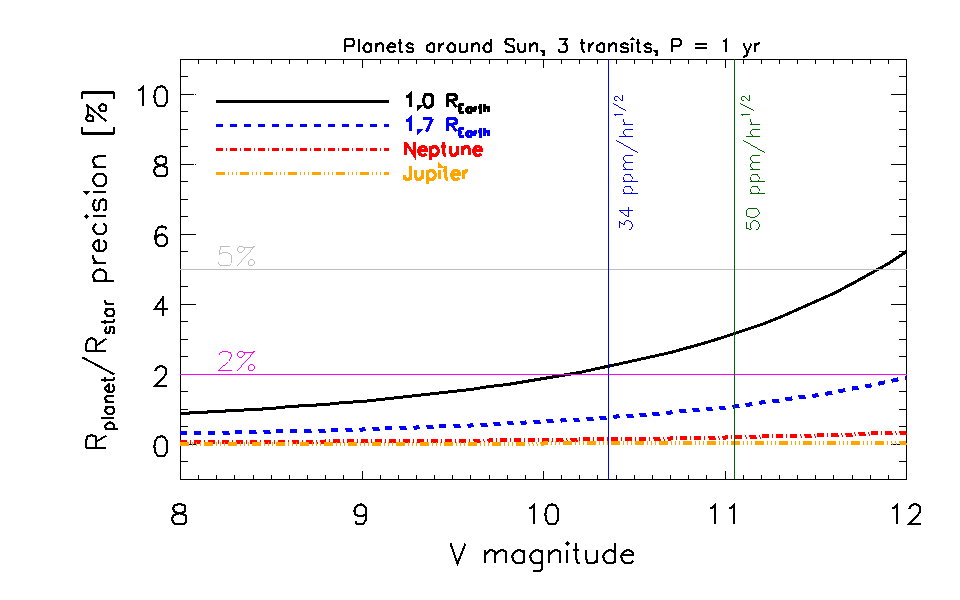}
\includegraphics[width=0.5\textwidth]{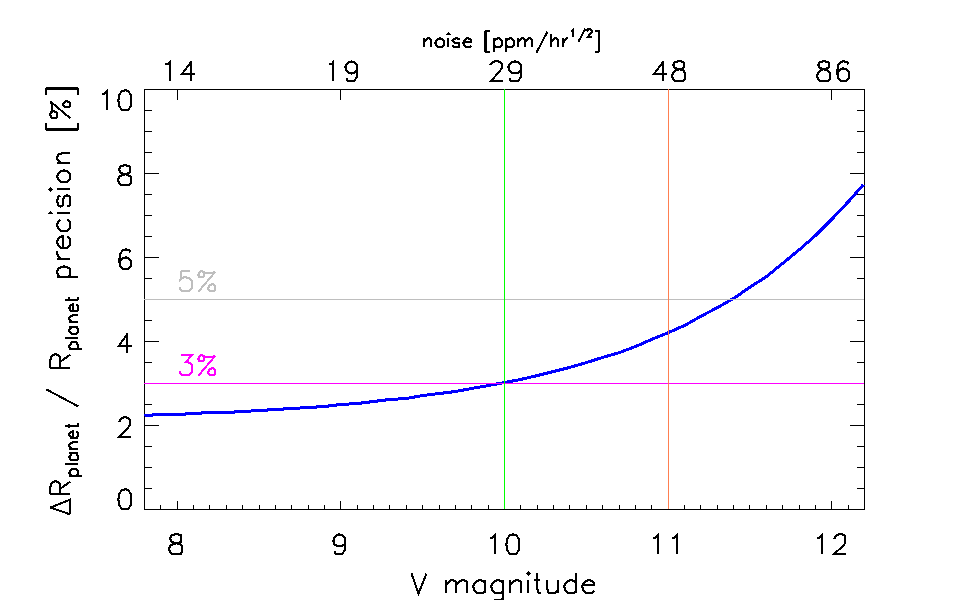}
\includegraphics[width=0.5\textwidth]{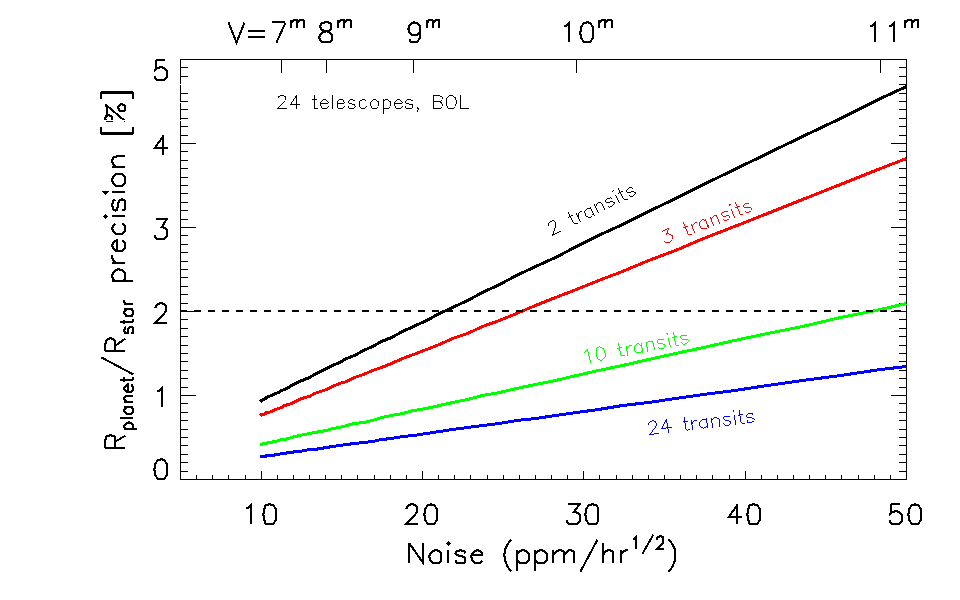}
\includegraphics[width=0.5\textwidth]{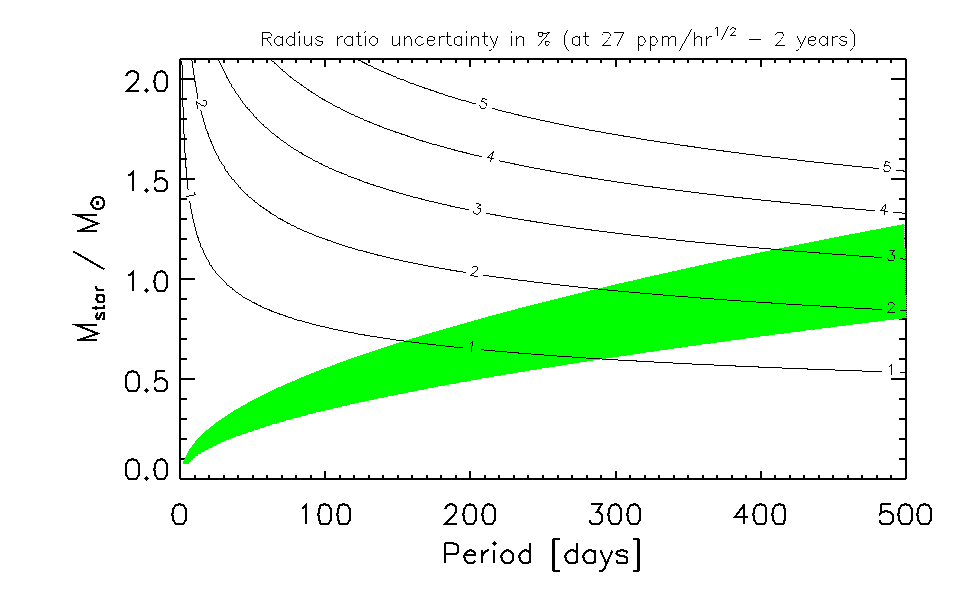}
\caption{Simulated radius precision for planets orbiting a Sun-like star. Top left: radius ratio precision for 3 observed transits and different planet types. Top right: The ``Earth-around-a-Sun'' case for 3 transits. The desired radius accuracies are noted, where we assumed that the stellar radius - independently of the magnitude of the host star - is known to 2\%. Bottom left: radius ratio precision for the ``Earth-Sun'' case and different number of transits, where a larger number of transits can be reached at shorter orbital periods when the duration of the pointing is unchanged. Bottom right: Earth-sized planets transiting different host stars. The approximate position of the HZ is indicated in green. The stellar radius was estimated via $R_\mathrm{star} = R_\odot (M_\mathrm{star}/M_\odot)^{0.95}$. The number of transits depends on the orbital period, where we assumed that all transits are detected during a 2 year long pointing. We have considered the end-of-life (EOL) performance as per requirements, which is a conservative approach (see Section \ref{sec:InstrPerformance}).}
\label{fig:radiusprecision}       
\end{figure}

Figure \ref{fig:radiusprecision} (top left) shows that for a $V=10\,$mag Sun-like host star, a radius ratio precision of 2\% can be reached for an Earth-sized planet with 3 transits. Better precision is reached for larger planets and brighter host stars, as expected. The top right shows that a planet radius precision of 3\% can be obtained down to $V=10.5$\,mag and increasing to a radius error of 5\% at $V=11.6$\,mag. These results were obtained analytically (\ref{app:planetary radius estimation}) and agree well with the detailed numerical models \citep{morris2020,2023A&A...675A.106C}. The results confirm that the PLATO design is in agreement with its respective science requirements.  
To give an estimate accounting for possible catastrophic events during the mission, the impact of the loss of two cameras has been studied. In such a case the radius precision would decrease by about 10\% to 5.5\% precision.
In the bottom left the dependence on observing duration, hence the number of detected transits is shown. As expected the radius error decreases with more transits observed, e.g. due to longer target field observation but also for planets with short orbital periods.
Hence, very precise radii are expected for hot terrestrial exoplanets around solar-like stars. The lower right of Figure \ref{fig:radiusprecision} shows the expected precision for an Earth-sized planet orbiting different stellar types. 

For a more detailed analysis of the expected resulting planet parameter precision additional effects need to be accounted for. These are, however, rather independent of the instrument design but instead depend on the quality of data reduction and analysis methods. At least three factors need to be considered: residual red noise effects, stellar limb darkening, and the knowledge of relevant orbital parameters.

The most challenging problem is caused by the detected baseline stellar flux which is rarely constant in time. It is changing because of instrumental effects, cosmic ray impacts, straylight changes, variable contaminating sources in the aperture (e.g. a variable star) and because of astrophysical reasons intrinsic to the star: stellar variability and activity, etc. All of these effects cause correlated,  or so-called red, noise effects in the light curves. 
The issue of residual red noise in the stellar baseline and in-transit data points was investigated by, e.g. \cite{barros2020}, \cite{morris2020}, and \cite{2023A&A...675A.106C}. \cite{morris2020} utilised SOHO solar images to simulate transits to be modelled. \cite{barros2020} tested the applicability of Gaussian Processes. \cite{2023A&A...675A.106C} assessed the 
performance of wavelets to model the stellar and instrumental noise. 
These authors used {\it Kepler\/} Q1 short cadence light curves with injected transits for different planetary and stellar parameters to mimic the red noise effects. They modelled the light curves with wavelets and analysed the results with retrieval methods. Their findings show  that above a certain signal-to-noise ratio (SNR), preferentially $\text{SNR}>45$\footnote{In their work the signal-to-noise ratio was defined as: SNR$\equiv (R_{\rm planet}/R_{\rm star})^2 / \sqrt{\sigma_w^2 + \sigma_r^2} \times \sqrt{N_{\rm transit} D/t_{\rm exp}}$, where $N_{\rm transits}$ is the number of observed transits, $D$ the transit duration, $t_{\rm exp}$the exposure time, and $\sigma_w$ and $\sigma_r$ the white-noise level and the red noise per data point.}, the planet radius as well as other parameters can be retrieved with highest accuracy, even for the difficult Earth-Sun reference case, as long as the stellar radius is known to better than 3\% (we recall that detection is possible already at lower SNR -- e.g. 7 to 10 --, but provides larger uncertainties in the planetary parameters).
Thus, combined Gaussian processes and a wavelet technique to model the noise provide appropriate tools to obtain the radius ratio with the desired precision and remove the red noise effects while solving the stitching of PLATO light curves. Further analyses of respective data processing methods are ongoing in the PMC to support the present conclusion that PLATO will deliver the planetary radii with an accuracy better than 5\% for the aforementioned baseline scenario.

\subsection{Constraints on planetary atmospheres}

\label{sec_atmospheres}

Although the PLATO mission is not primarily designed to study exoplanetary atmospheres, its light curves can nevertheless provide relevant information on atmospheric properties. The amplitude and shape of white-light orbital phase curves, for example, can provide a first constraint on atmospheric meridional transport, mass and albedo. High precision white light photometry has already been performed to measure geometric albedos e.g. for the Hot Jupiter, HD209458b by \citet{2022A&A...659L...4B} and by \citet{2022A&A...659A..74D} for the Ultra Hot Jupiter, WASP-189b. Basic colour information (``red'' and ``blue'') from the broadband filters on the PLATO fast cameras (with fast read-out cadence) can constrain the bulk atmospheric composition from estimating the Rayleigh spectral absorption feature. Also, PLATO's prime data products, namely planetary radius, mass, and age, are crucial for retrieval and interpretation of atmospheric properties from spectroscopic data.

\begin{figure}
\includegraphics[width=0.9\textwidth]{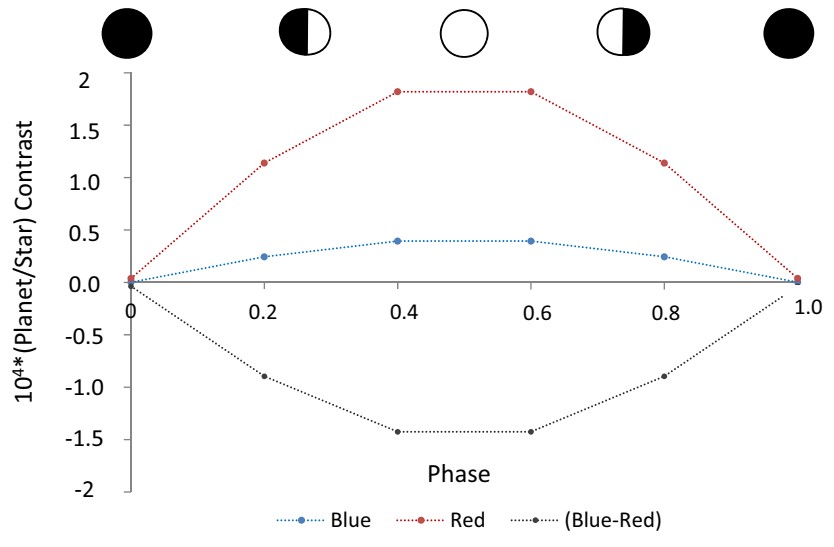}
\caption{The phase curve of the contrast (Planet/Star) enhanced by a factor $10^4$ as would be observed by the PLATO fast cameras with their ``red'' and ``blue'' filters for a hypothetical, nearby (10\,pc) Ultra-Hot-Jupiter, assuming the planetary properties of WASP-103b. Data begins at the nightside during conjunction and shows 6 equidistant points.  Figure adapted from~\citet{2020ExA....50....1G}.  }
\label{fig:phasecurve}       
\end{figure}

\citet{2020ExA....50....1G} found that planetary geometric albedos, as well as moderate-to-strong Rayleigh extinction can be detected for targets up to 25-100\,pc distance based on simulated SNR for 
observations of nearby hot and ultra-hot Jupiters (UHJs) with the PLATO fast cameras. Their work suggests that initial constraints can be deduced from UHJ phase curve amplitudes for nearby targets closer than 25\,pc, although this will be challenging. To illustrate this, Figure~\ref{fig:phasecurve} shows a simulated phase curve for a hypothetical close-by (10\,pc away) UHJ as would be observed by the PLATO fast cameras. From phase curve data on UJHs as shown in Figure~\ref{fig:phasecurve} one can infer the dayside and nightside temperatures and thus the energy balance and intrinsic heat flow as well as planetary albedos, which are very important planet properties. Given that at present there are about only two dozen planets with observed phase curves, any additions that PLATO can make will be valuable. 

Regarding warm sub-Neptunes and Super-Earths, the~\citet{2020ExA....50....1G} study suggested that basic atmospheric bulk compositions and haze properties in the upper atmosphere can be distinguished for some nearby ($<$10\,pc) favoured targets, although this will likely be a challenging task. \citet{carrion2020, carrion2021}  point out that having prior information from PLATO on the planetary radius is an important input for model studies constraining atmospheric clouds and composition from direct imaging, since being able to fix the radius helps to disentangle degeneracies between planet radius, cloud layers and major atmospheric absorbers when studying planets in reflected light. \cite{turbet2019, 2024PSJ.....5....3S} suggest that PLATO measurements of planetary radii for hot Super-Earths with giant steam atmospheres will allow us to pinpoint and characterise runaway greenhouse regimes with strongly inflated atmospheres, yielding an empirical test for the habitable zone hypothesis \citep[see also][]{boukrouche2021}. \cite{ortenzi2020} offer a way to distinguish lighter, more extended Super-Earth atmospheres (outgassed by more reducing mantles \citep{lichtenberg2021ApJ...914L...4L}) from more compact atmospheres outgassed by more oxidised mantles from the high-precision PLATO measurements of the planetary radius.

Clearly, PLATO data alone are not sufficient to resolve the complex optical properties of planetary atmospheres. They can, however, serve as a guide to trigger follow-up observations with spectroscopic instruments. The full benefit of a large sample of exoplanets with well-known PLATO ages will become apparent when spectra of their atmospheres become available. As such, PLATO will provide constraints on the evolution of gaseous planets in combination with missions featuring spectroscopic capabilities such as JWST and ARIEL. In case of lucky detections of transiting planets which are sufficiently close to us and on sufficiently wide orbits can be made, they would even provide the opportunity to compare their transmission spectra to direct imaging spectroscopy.
In the somewhat more distant future, well characterised terrestrial exoplanets with accurate age determinations will allow 
to obtain observable constraints on the typical evolution of such planets \citep{lichtenberg2019NatAs...3..307L, lichtenberg2022ApJ...938L...3L}. We will then be able to compare the terrestrial Solar System planets and their evolutions with a major sample of terrestrial exoplanets. While this goal can only be fully achieved by combining results from various future space missions, PLATO will provide a significant piece to this puzzle.

\subsection{Constraints on planet formation and evolution}

\label{sec_planet-formation}

A key science objective of PLATO is to further our understanding of how planets and planetary systems form and evolve. It was recognised at the inception of the mission concept that complementary theoretical modelling efforts in planet formation and evolution will be essential to achieving this goal. The large and well-characterised population of planets in diverse Galactic environments to be obtained by PLATO, with precise measurements of radius, mass, density, age and host star properties such as metallicity and stellar type, extending out to orbital periods $>80-100$~days, will provide the most comprehensive testing ground for the modelling of formation and evolution processes to date. In particular, the discovery of Earth-like planets would help our understanding of how terrestrial planets form.

State-of-the-art global models of planet formation are in continuous development and already account for many of the important processes involved in the building and shaping of planetary systems during their first 10-100~Myr, leading to synthetic planet populations that can be compared with planet population data \citep{Benz2014,Draz2023}.

These processes include protoplanetary disc evolution \citep{Lesur2023}, the growth of planetary embryos through pebbles and/or planetesimals \citep{Morbidelli2012,Johansen2017}, the accretion of gas onto forming planets \citep[e.g.][]{Cimerman2017,Lambrechts2019b,Nelson2023}, and disc-driven migration \citep{Kley2012}. Although much progress has been made in the past decade on these topics, some key physical mechanisms and their relative importance remain poorly understood. For example, a variable mass budget available in pebbles could regulate the formation and migration of either Earth-like or super-Earth-like planets in the terrestrial zone \citep{Lambrechts2019a}. Disc evolution in turn regulates migration rates and resonant-trapping of planet systems \citep{Kajtazi2023,Batygin2023,Huang2023}. Finally, the composition of planets is a complex product of the full planetary growth process, starting from the growth and drift of primordial dust grains \citep[e.g.][]{Johansen2021,Schneider2021}.
 Models also exist of processes that act on longer timescales, such as tidal evolution, core-powered and photoevaporative gas envelope loss, and the role of stellar cluster membership in shaping planetary systems \citep[e.g.][]{Papaloizou2011,WuMurray2007,Rogers2021,Li2023}.

Progress will be made through comparison between these models and the PLATO-observed planet population which will then shed light on the relative importance of these processes, and will point to areas where model improvements are required, either through the inclusion of hitherto neglected physical effects or through improved understanding and modelling of fundamental planet formation processes \citep[e.g.][]{ColemanNelson2016, Bitsch2019, emsenhuber2021, Izidoro2021, Qiao2023}. Through the diverse composition of PLATO's stellar sample, predicted and newly emerging demographic trends can be studied.
In particular, the occurrence rate of small planets around cool M stars will be determined and put into context with models constraining their formation \citep{Liu2019, miguel2020, burn2021, schlecker2022}. PLATO will further shed light on the existence and occurrence rate of terrestrial planets with significant water content (e.g., \cite{kimuraikoma2022}), thereby yielding important constraints on water delivery on Earth and other terrestrial planets. Studies of architectural trends within multiplanet systems will be facilitated by PLATO's large expected planet yield, shedding additional light on the factors that control planet formation (e.g., \cite{gilbert2020, mishra2023a, mishra2023b}.)

PLATO will not only find planets around single main-sequence stars, but is expected to find substantial numbers of more exotic planets, such circumbinary planets, that provide more extreme environments that can act as stringent tests of planet formation theories. For example, allowing discrimination between in situ scenarios and those that require migration \citep[e.g.][]{Pierens2021}. Development of population synthesis models of these types of systems \citep{Coleman2023} will allow comparison with PLATO data, constraining theories of planet formation in diverse environments.

\section{Stellar Science }
\label{sec:StellarScience}

\medskip
\subsection{State of the art of seismic stellar characterisation}
The characterisation of exoplanets highly depends  on the characterisation of their host stars. 
 The more precise and unbiased the characterisation of the latter is, the more accurate  will be the characterisation of the planets they host. 
Stellar properties such as  mass or age are most often derived from stellar modelling constrained by classical 
data. Yet seismic information, whenever available, helps tremendously to improve such modelling because it provides direct information from the deep stellar interior.

 Starting with the Sun, two  decades of observations  have   demonstrated that solar-like  seismic constraints are the most powerful tool to derive precise  stellar masses, radii, densities,  and ages,  provided that  high-quality  seismic parameters  are available \citep[for reviews see][]{chaplin2013, jcd2018b,
 Garcia2019, serenelli2021}.  However, this requires short-cadence (less than 1 min for dwarfs and subgiants),  ultra-high photometric precision (at the level of parts-per-million or ppm)  and  nearly-uninterrupted   long-duration (from months to years)  monitoring as best provided by space observations.

Beyond the precise characterisation of specific stars, seismology allows us to probe the physical interiors. In this way, it allows the identification of shortcomings in  stellar modelling, to guide the theoretical developments for improvements in the physical description  and then validate those improvements. This in turn leads to a more accurate characterisation of {\it all\/} stars -- in particular their age-dating, not only for those with seismic 
features. Compared to classical modelling without asteroseismology, higher accuracy is achieved for the properties and parameters measured for a given star from the use of seismic constraints and more realistic stellar modelling 
 \citep{silvaAguirre2017, nsamba2018b} or through the use of seismically calibrated relationships such as gyrochronology \citep[][and references therein]{angus2015, vansaders2016,hall2021} and abundance ratios -- age relations 
\citep{nissen2017, morel2021}.

\subsubsection{Ultra-precise photometric seismology: the space era }
 After the proof of concept space missions -- either by accident \citep[WIRE: ][]{buzasi2005} or dedicated  (MOST: \citep{walker2003MOST} -- the French-led CoRoT satellite 
\citep[2006 -- 2009;][]{baglin2006, auvergne2009}
 really opened the space-seismic era with the detection of solar-like oscillations in about twelve bright dwarfs and subgiants 
 \citep{michel2008a} and thousands of red-giant stars \citep{deridder2009, Miglio2013}. 
 This has led to the first detailed asteroseismic studies of solar-like oscillations in main-sequence stars,  as well as to the new field of red giant seismology \citep[e.g.][]{baglin2016legacy}.

The NASA space mission {\it  Kepler\/}  \citep[2009 -- 2018;][]{borucki2010a} monitored  more than 150000 main sequence stars, of which about 2600 dwarfs and subgiants were observed for one month in short-cadence mode (58.85 sec). 
These measurements turned detailed
seismic investigations into a reality for large samples of dwarfs \citep{gilliland2010b,chaplin2014}. Overall, the nominal  {\it Kepler\/} mission  led to the detection of solar-like oscillations for more than 600  dwarfs and subgiants. Most of them were observed for only one month while about  150 stars were observed for longer, up to 4 years. 
The so-called {\it Kepler\/} Legacy sample of 66 dwarfs revealed the major power of asteroseismology of Sun-like stars \citep{lund2017, silvaAguirre2017}. 
A few dozen of them host planets and these are now among the most precisely characterised host stars
\citep{huber2013, SilvaAguirre2015, betrisey2022, hatt2023}.

The NASA TESS space mission launched in 2018 
\citep{ricker2015} is currently in operation. It  provides lower-quality photometry than the {\it Kepler\/} mission but focuses on bright stars ($5 < V < 11$) and  carries out a nearly-all-sky survey.  The observing time is short, 27 days for most stars, but it reaches up to 352 days
in the continuous viewing zone. During its 2 year nominal mission, TESS had a short-cadence mode of 2 min. During the first extended mission an additional 20 second mode was operational, which is better suited for seismic analyses as advocated by \cite{huber2022}. As demonstrated by the first results,  TESS  is capable of detecting solar-like oscillations with high amplitudes and therefore suitable to  study seismology of  subgiants and red giants \citep[e.g.,][]{campante2019,Huber2020, 
chontos2021}.

\begin{figure}[t]
 \begin{center}
   \includegraphics[width=11.7cm]{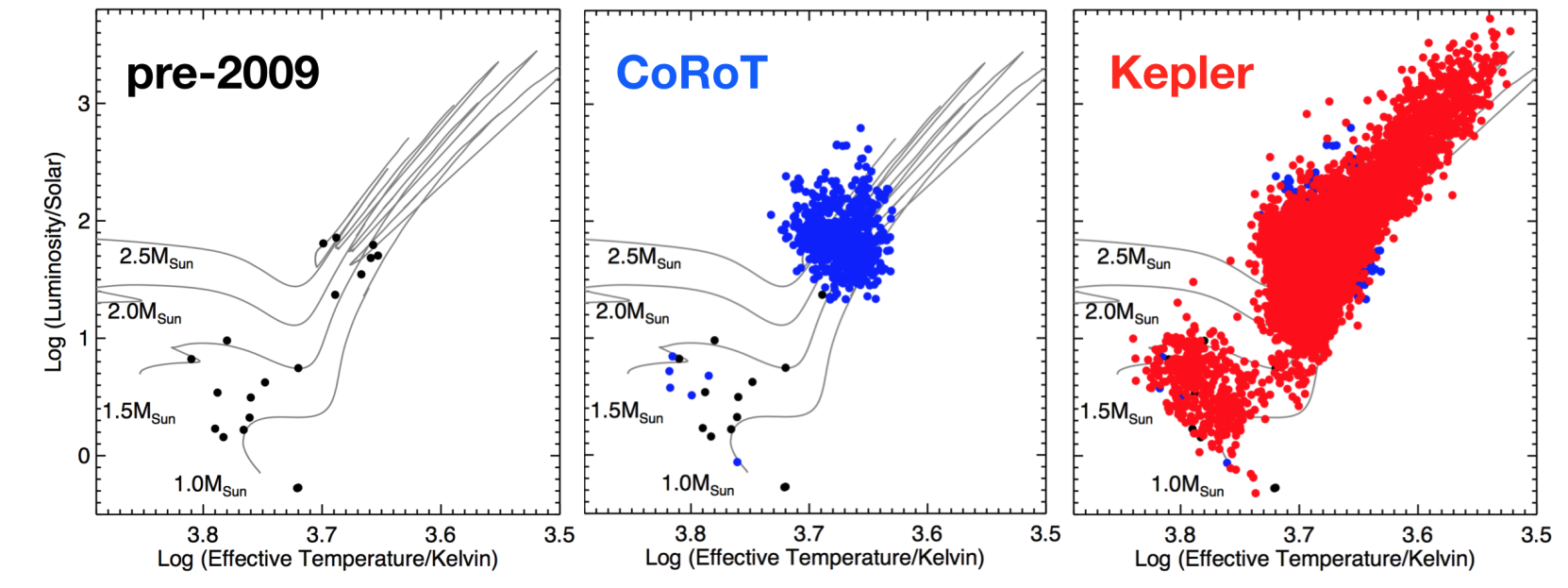}
   \caption{ \label{HRdiag1} 
   Stars with detected solar-like oscillation from 
   ground-based instruments prior to the seismic space era (left) and from the space missions CoRoT (middle) and {\it Kepler\/} (right). Figure reproduced from ~\citep{Huber2020}}.
 \end{center}
\end{figure}

Figure \ref{HRdiag1} shows 
samples of stars covering the main sequence to the red-giant branch for which solar-like oscillations have been detected at three epochs marking the major progress of the space revolution in asteroseismology. Progress went from a handful of stars with solar-like oscillations detected with ground-based instruments to hundreds of detections from CoRoT focused on evolved stars and thousands of them resulting from the {\it Kepler\/} (and now TESS) space missions. 
The PLATO P1 and P5 samples will further increase the samples appreciably compared to those shown in Fig. \ref{HRdiag1}, particularly for dwarfs and subgiants. With its 25\,sec cadence during the entire long pointings, PLATO will not only deliver new detections of solar-like oscillations in thousands of stars, but the precision of the oscillation frequencies will also be unprecedented for such a large homogeneously assembled and analysed ensemble of dwarfs and subgiants, among which will be thousands of exoplanet host candidates.  

\subsubsection{Ultra-precise seismic  stellar characterisation }

The {\it Kepler\/} seismic observations have motivated a large number of investigations over the past years to make the best use of seismic data to characterise stars. 
Here, we present some lessons of interest for the PLATO mission that one can draw from those studies.

The detection of solar-like oscillations, the quality of the measured seismic parameters, and the accuracy with which seismic stellar characterisation is obtained all depend on the SNR, the cadence, and the observation baseline for each star. 
The SNR depends not only on the instrumental observational noise and the apparent magnitude of the star but also on the type of star. 
Hence, depending on the quality of the observations and the type of star, we have different levels of stellar characterisation. 
PLATO is designed and optimally suited to perform asteroseismology of sun-like stars.
However, we learned from the {\it Kepler\/} data that, for  K dwarfs and M-type stars, either the oscillation amplitudes are too low for detection or the stars do not oscillate at all \citep{rodriguez-lopez2015MNRAS.446.2613R, rodriguez-lopez2019FrASS...6...76R}. 
Even for F to G-type dwarfs, it is thought that the impact of strong surface magnetic activity on envelope turbulent convection can cause a significant decrease of the  oscillation amplitudes, which can hamper their detection~\citep{garcia2010, chaplin2011, kiefer2017, santos2021}.
Some other causes of non-detection are also possible, such as metallicity effects and binarity~\citep{samadi2010, karoff2018, mathur2019}.
On the hot side, late F-type stars have larger mode linewidths than G-type stars, causing  larger uncertainties in their seismic measurements. 
These translate  into larger statistical uncertainties for their stellar parameters~\citep{appourchaux2012, compton2019}. 
However, those stars with their large radii (larger than  1.2 solar radius) and high temperatures are not the main priority for PLATO's exoplanet search and characterisation. 
Actually the P1 sample is truncated at temperatures on the hot side, such that one still encounters solar-like oscillators.  
We therefore focus hereafter on the case of late F, G, and early K-type dwarfs, and subgiants.  

\medskip
\noindent{\it Averaged seismic indices and scaling relations}
  
When the SNR is low and/or the observation time duration is short (one month or so), the solar-like oscillations are detected mainly as a power excess in a power spectrum. 
One then takes  the regularity of the frequency pattern to determine a first average seismic quantity, namely the mean large frequency separation, $\Delta \nu$, that scales with the square  root of the mean stellar density. 
Several techniques also allow the determination of the frequency at maximum oscillation power, $\nu_{\rm max}$, in a power spectrum.
This frequency is related to the surface gravity of the star and to some lesser extent to the effective temperature. 
Given the effective temperature, these two quantities give direct access to the mass and radius of the star~\citep[see][and references therein]{hekker2020}.
Hence, uncertainties of the resulting mass and radius determination directly depend on the uncertainties of $\Delta \nu$, $\nu_{\rm max}$, and $T_{\rm eff}$. 
For a sample of about 500 {\it Kepler\/}  stars observed in short cadence mode over a  period of nearly a month, the quoted average median statistical uncertainties are $\sim$ 1.5 -- 2\% for $\Delta \nu$ and $ \sim$  4\% -- 5.8\% for $\nu_{\rm max}$. 
For rather high SNR~\citep{verner2011} these measured seismic observables and their uncertainty lead to a median relative uncertainty of the order of 5.5\% and 10\% in  radius and mass, respectively~\citep{chaplin2011}. 
As shown by~\cite{goupil2024}, most of the stars within the PLATO P1 sample and a large number of stars in the P5 sample will benefit from such uncertainties.

\medskip
\noindent{\it Seismic stellar modelling with scaling relations}

A more precise characterisation, including  age determination, is obtained when one infers the properties of a star by means of a fitting technique.
Usually a grid-based approach is adopted, where one selects the best stellar models from a large grid in such a way that the oscillations predicted for these models satisfy the observational constraints to within a defined matching criterion.
In that spirit, several  pipelines have been developed for asteroseismic modelling, each with the same goal but differing in various aspects.
The inter-comparison of their results shows good agreement if all other aspects, such as input data and stellar models, are kept the same. 
One of the most popular techniques searches the optimal stellar models in a pre-computed grid and  is therefore  referred to as  grid-based modelling (GBM)
\citep[e.g.][and several others]{SilvaAguirre2015, rodrigues2017, rendle2019, lebretonReese2020, farnir2021, aguireeBorsen2022}. 

Because it is easy to use and relatively fast, the GBM approach is favoured for the PLATO  baseline stellar modelling pipeline. This is motivated by the \emph{Kepler} experience, which shows that for long duration observations (up to 1050 days), we obtain average median precisions of $\sim $ 1.7\% and $\sim$  4\%,  respectively, for $\Delta \nu$ and $\nu_{\rm max}$ and subsequently median statistical errors of about   2.3\% for radius, 4.2\% for mass, and 15\% for age. This result has been obtained using a GBM stellar characterisation \citep{serenelli2017} and implies that the faintest stars in PLATO's P1 sample as well as  a number of the brightest stars in the P5 sample are likely to be in this situation, 
meaning that for their mass, we are already within the PLATO requirements as far as precision is concerned. 
The PLATO age-dating requirement, however, still needs to be scrutinised with better models and modelling methodology, which is under current development.

\medskip
\noindent{\it Ultra-precise  stellar characterisation based on individual mode frequencies}

While acceptable results are obtained by using the average seismic quantities, 
the most precise stellar characterisation is obtained when a detailed measurement of the frequencies of individual oscillation modes is possible.  This occurs for the brightest stars  observed over a long temporal baseline.
In such cases,  individual  mode frequencies can be fitted with methods such as  the popular  ``peak-bagging'' technique 
\citep[e.g.][]{Appourchaux2003, Handberg2011,Corsaro2014}. 
Combined with optimisation procedures such as a GBM approach,  this allowed for a very precise determination  of the properties of  these stars. The precision obtained with detailed modelling  are somewhat dependent on the chosen quantity to be reproduced (frequencies or frequency combinations), as well as the optimisation algorithm and the 
density of the stellar evolution model grids in terms of the free stellar parameters involved in the fitting procedure.

At present, two  major  catalogues of stars exist 
from the viewpoint of containing the best homogeneously derived oscillation frequencies,  stellar masses, radii, and ages: the so-called Kages sample and the Legacy sample. 
The Kages sample includes 35 exoplanet host stars, which were  observed over the whole nominal {\it Kepler\/} mission, resulting in oscillation frequency uncertainties  at the level of  0.3 $\mu$Hz (best cases 0.1 $\mu$Hz) in the vicinity of  $\nu_{\rm max}$ for dipolar  modes \citep{davies2016}. \cite{SilvaAguirre2015} used combinations of the individual frequencies and a GBM approach to derive stellar properties with  median statistical uncertainties of  1.7\% (density), 1.2\% (radius), 3.3\% (mass), and 14\% (age). 
 The Legacy sample includes 66 dwarfs of low-mass with  short-cadence {\it Kepler\/} observations spanning at least 1 year and up to 4 years. With a white noise level ranging between  0.1  and 8 ${\rm ppm}^2/\mu$Hz  (5.3 -- 47 ${\rm ppm\,h}^{1/2}$),  frequency uncertainties at $\nu_{\rm max}$ for
 dipolar modes are found in the range $0.03 -0.35 \mu$Hz depending on the star  \citep{lund2017}. 
Quoted average uncertainties range from 0.5\% to 2.6\% in density, 1.3\% to 4.2\% in radius, 2.3\% to 4.5\% in mass, and 6.7\% to 20\% in age \citep{silvaAguirre2017}.
   
These results are in agreement with the level of mass and age uncertainties  determined from hare-and-hound exercises \citep{reese2016}  which give 1.5\% (radius), 3.9\% (mass), 23\% (age), 1.5\% (surface gravity), and 1.8\% (mean density). For two 1 $M_\odot$ stellar targets, the precision on
the age is better than 10\%. Moreover, for the best (mostly brightest) stars in those samples, a seismic determination of  the stellar rotation rate and inclination was also  possible and valuable for studies on spin-orbit alignment and  orbital eccentricities of exoplanetary systems. Those {\it Kepler\/} stellar sets are, however, biased towards main-sequence stars hotter and 
more evolved than the Sun because the oscillation amplitudes are higher in such cases. It is a major goal of the PLATO mission  to extend and  better  populate  this seismic dwarf sample
with thousands of dwarfs instead of tens.

Summarizing at this stage,  the Kages and  Legacy stars together with a few other individual seismically studied stars from CoRoT, {\it Kepler}/K2 and TESS represent  today's set of most precisely characterised field  stars.  
The addition of asteroseismology to stellar modelling permits to reach levels of  $\sim$ 10\% precision in stellar ages at least for stars similar to the Sun~\citep[see for the case of Kepler 93][]{betrisey2022}.  
This complies with the requirements for the PLATO mission as far as precision is concerned. This was further confirmed by a PLATO dedicated hare-and-hound exercises  \citep[][see below]{cunha2021}. Together with the Sun, those stars are now being used to calibrate 
gyrochronology relations, which can then be used to age-dating other stars without detected oscillation modes. Those stellar samples also serve as reference stars to validate other techniques capable of constraining stellar properties, such as the derivation of $\log g$  from photometric variability due to surface granulation (\cite{bugnet2018} and references therein).

\subsection{Precision versus accuracy} 
 Precision   on stellar  radius, mass,  and age determinations as mentioned above is  associated  with  statistical  uncertainties due to the propagation of observational errors. 
In addition to those uncertainties, one must also account for systematic errors or biases for assessing the accuracy of the seismic results. 
Such  systematic  errors contribute to  the final  error budget but do not directly depend on the observations themselves. Rather, they depend on our ability to improve  the data-analysis procedures, adopted choice of constraints, optimisation procedures and strategies, and most importantly approximations adopted for the physical description of the stellar models used in seismic inferences.
When highly  precise and accurate  seismic observations are available together with    precise and  accurate   classical stellar parameters (effective temperature, chemical composition), the statistical errors on the derived stellar properties decrease to a level where the systematic errors due to stellar modelling dominate the uncertainties.
 
A number of studies have used several fitting methods to derive the stellar mass, radius and age of oscillating solar-type stars 
with an internal structure similar to that of the Sun (e.g. \cite{lebreton2014}, \cite{SilvaAguirre2015}, 
\cite{reese2016}, \cite{cunha2021}). They found that, given the same set of input information (same grid of stellar models, same number of free parameters to adjust, same choice of seismic diagnostics, \ldots), the different data analysis and optimisation methods give similar results  (with uncertainties well below PLATO  requirements) and introduce much smaller biases than those induced by systematic errors
  due to the choice of seismic diagnostics, to the adopted number of free parameters or, more importantly, to the physical description of the stars.
  
  In contrast,  even restricting the case to low-mass main-sequence and subgiant stars as we do here, important sources of  systematic errors result from our poor knowledge of various physical processes that affect the stellar structure and evolution, as well as the oscillation frequencies. This is particularly true for the age,  which of all stellar properties, is by far the most challenging to determine accurately. 
  In most cases, stellar ages can  only be  determined through stellar modelling and  their accuracy therefore strongly depends on the degree of reliability  of the available  stellar models \citep[e.g.,][]{christensen-dalsgaard2018}.   Particularly critical (and uncertain) are physical processes that mix the chemical elements in the stellar cores, affecting the relation between the age and the composition profile and, thus, the seismically inferred age.
 
The stellar properties of the targets in the Legacy sample modelled by \cite{silvaAguirre2017} were actually determined with varying degrees of uncertainties 
 depending on the adopted pipeline (including different input grids of stellar models) while using the same observational data. When considering the results of the  three  pipelines -- out of the  seven considered --    which used the same seismic diagnostics (individual frequencies),
    the median age uncertainties range from 0.5 -- 0.8\%, 1.3 \%, and 6.7 -- 10\%  for the three respective pipelines. The  spread of  these uncertainties can be mostly attributed to the use of different input physics assumed by the three pipelines and  not to the pipeline operational strategies themselves. For instance, the lowest values of these uncertainties can be partly attributed to a different adopted GBM procedure. Yet the systematic uncertainties are mostly due to the use of a grid of stellar models where the
values of two free parameters are  held fixed, while for the other
two pipelines they are left either variable or free to take more than one value. Fixing the values of the parameters (usually taken as those of the Sun) decreases the uncertainties but at the cost of being less accurate since there is usually no reason for these parameters to take exactly the solar
values. This is confirmed with PLATO hare-and-hound exercises \citep{cunha2021}.

Therefore, besides the precision, one must also be concerned about the accuracy of  the central or median values themselves: how can we assess the accuracy of these results? Improving stellar modelling not only requires theoretical developments but also a set of very well and accurately characterised stars. The latter serve to diagnose the dominant shortcomings  in stellar modelling on one hand and to validate the theoretical developments designed to correct
   for those shortcomings  on the other hand. Characterisation of the properties  of such stars must  of course be as independent from stellar models as possible (e.g. benchmark stars such as eclipsing binaries, stars with interferometric radii) or  enable  ensemble studies to  constrain some  physical ingredient or to calibrate a free parameter involved in an empirical physical  formulation (stars in clusters, unevolved massive stars, red giants). 

\medskip
\noindent{\it Accuracy tests: the Sun. } A   routinely  used  test  of the accuracy of  seismic modelling is to look at the results for 
 the so-called ``degraded'' Sun for which seismic and non-seismic data were built to match the typical quality  of the {\it Kepler\/} Legacy sample ($\sim $ 0.15 $\mu$Hz  for a $l=1$ mode at $\nu_{\rm max}$). 
The derived  values  depart from the independently  known solar values by $\sim $ 3\%, 0.8\%, 1\%, and 9.8\%  for the  solar luminosity, radius,  mass,  and age, respectively,   the differences being  mainly due to the adopted chemical composition. This must be  compared to 1$\sigma$ uncertainty given by the pipelines, which are in the typical range  0.5 -- 4\% and  3 -- 8\% for the mass and  age. While the  net error budget for the ``degraded'' Sun  is  at the level of  $10\%$  accuracy  for the age, one must keep in mind that the seismic Sun is still  discrepant  in several aspects compared to our real Sun \citep{Christensen-Dalsgaard2021}.
 The same discrepancies, still not fully identified, will have an impact of unknown magnitude for other (solar-like) stars for which no independent age is available and differing in mass, chemical composition, evolution, and environment. 
 
\medskip
\noindent{\it Accuracy tests: independent additional observational  information. }
  For {\it Kepler} stars using interferometric or astrometric observations as external constraints,  scaling relations have  been reported  to be accurate at the  level of $\sim 2\%$ in density, 2 -- 5$\%$ in radius,  and about $5\%$ in  masses for dwarfs and subgiants \citep[e.g.][] {coelho2015, huber2017, zinn2019}.
  Taking into account several different sources of systematic error in a GBM approach,  \cite{serenelli2017}  quoted a systematic error  of the order of 1.2$\%$ in radius,  3$\%$ in mass and 12$\%$ in age   with a total combined error of  approximately   2.6$\%$ in radius, 5.1$\%$ in mass, and $\sim$ 19$\%$ in age~\citep[see Tab.2 in][]{serenelli2017}.
Those studies also emphasised  the importance of having accurate and precise classical parameters  to exploit  the  full potential of seismology.
Getting such classical parameters requires the development of specific and sophisticated pipelines based on as accurate model atmospheres as possible \citep{morel2021}. This by itself represents a whole branch of stellar astrophysics, which contributes to properly characterise stars.
 We note that the determination of the classical parameters is an integral part of the mission, as developing a specific pipeline that makes use of various types of observations to tightly constrain these quantities is one of the ground-segment activities (\cite{gent2022}, Orlander et al., 2024 submitted). In parallel, there are efforts to set up collaborations with large-scale surveys (e.g. 4MOST; \citet{guiglion19}) in order to optimise the homogeneity and completeness of the preparatory spectra available for the PLATO targets. It will allow to meaningfully examine the correlations of planet properties and their occurrence rates with, for instance, the metallicity of the stellar host.
 
 \medskip
\noindent{\it Accuracy tests: ``differential'' studies. }
 Another commonly  used way of at least partially assessing the accuracy  is rather differential than absolute: it consists of  considering the  dispersion of the results provided by different pipelines (which differ in many aspects of input physics) or the results  coming from a single pipeline but  varying the  input physics of the stellar models.
 In this context, several studies have revisited the characterisation of the  Kages and {\it Kepler\/} Legacy stars.
 It was found that significant systematic  differences for  the mean values of the mass and the age occur.  
 The  dispersion in medium age can range between  $\sim$  15\% and  $\sim $ 33\%.
 This is attributed to different options about the optimisation strategy  but mostly to  different options of the stellar modelling  \citep{bellinger2016, bellinger2017, creevey2017, farnir2020, nsamba2018a, nsamba2018b, nsamba2021}. 
The most studied Legacy stars are the brightest (G1.5V + G3V) solar analogue components A and B of the  multiple  system \mbox{16 Cyg}  (with apparent {\it Kepler\/}  magnitudes $K_p \sim 5.86$ and $6.09$, respectively). 
Their masses are precisely determined at the level of 3\% or below.
Most studies passed the accuracy test of  both stars  having the same age -- as expected for a binary system --  within the quoted  uncertainties (reported to be  in the range 2 -- 6\% depending on the study) \citep[e.g.][]{silvaAguirre2017, creevey2017,bellinger2016, bellinger2017, buldgen2016, verma2016, bazot2020}.
However, examining the mean values derived by those various studies tells us that the seismically derived age  of \mbox{16 Cyg A} covers a range of  $\sim 6.4 -8.3$\,Gyr depending on the  assumed input physics. \cite{farnir2020} modelled both stars varying one input physical ingredient at a time and found   ages   in the range  6.4 ($\pm 0.08$) to 7.5 ($\pm 0.1$) Gyr.      This represents  an age  dispersion up to  about 16\%  when one takes 6.95\,Gyr as a reference mean value for \mbox{16 Cyg A}.
All these above dispersion estimates are larger than the uncertainties provided by each individual pipeline and currently remain larger than the 10 \% accuracy  for such a type of star as required for the PLATO mission. 

\begin{figure}[t]
 \begin{center}
  \includegraphics[width=12cm]{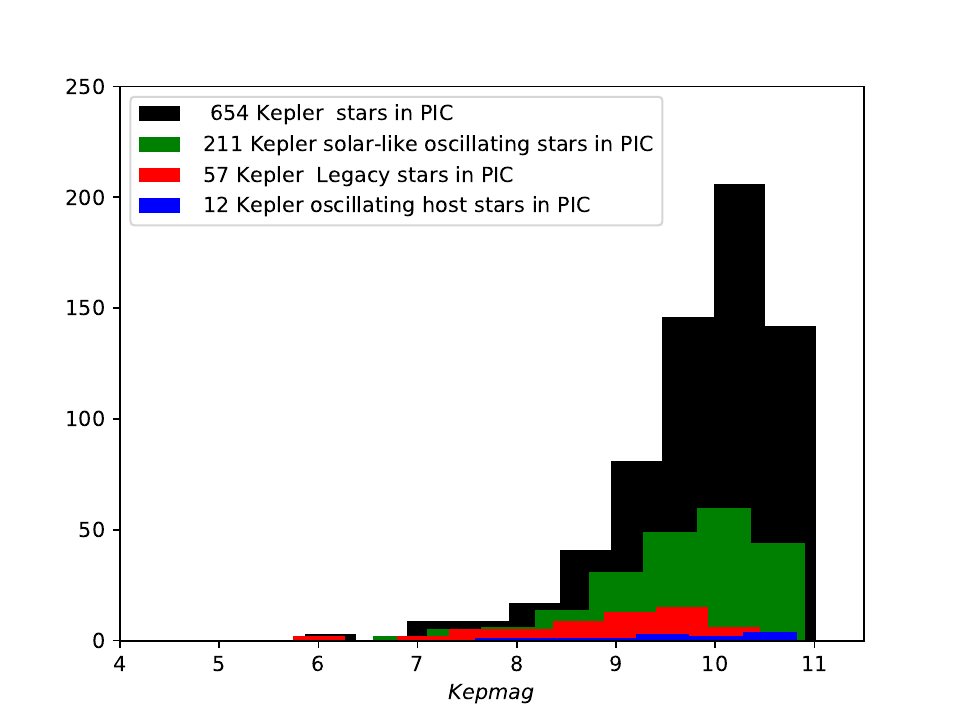}   
 \caption{  \label{PICKepler} 
 Histogram of various types of {\it Kepler} stars in the PIC~\citep{nascimbeni2022} over their {\it Kepler}-magnitude. according to their brightness, as labelled according to the legend.  }
 \end{center}
\end{figure}

    \medskip
\noindent{{\it Accuracy tests: seismic inversions.} They represent  the  most efficient way  to  diagnose shortcomings in stellar modelling by providing  model-independent constraints \citep{buldgen2017, buldgen2018, buldgen2019, bellinger2020}.
We must however stress that the inversion techniques  need to be performed from a reference model that is assumed to be already fairly close to the actual stellar structure.
For  dwarfs  other than the Sun, their use remains difficult due to the small number of significant modes.
Inversions are therefore restricted to the brightest stars with the highest SNR such as \mbox{16 Cyg AB} \citep{buldgen2016, bellinger2017}.
 The seismic inversions for  the \mbox{16 Cyg AB} system reveal discrepancies in the sound-speed profiles  when using the currently   adopted physics of stellar models.  The stellar modelling of the most constrained system \mbox{16 Cyg AB}  is therefore still not satisfactory as it cannot reproduce simultaneously  all the seismic and non-seismic constraints.
For the main-sequence star  KIC\,622571 \citep{bellinger2019b} and the subgiant HR\,7322 \citep{bellinger2021} discrepancies were also found in the sound-speed profile at the border between respectively the convective core or the helium core and the layers above.
In all these cases, the origins of the discrepancies are not yet identified, although it is likely due to some missing or poorly modelled  transport  processes  with a so far unknown impact on the age accuracy.  

Although the various sources of systematic errors due to stellar modelling cannot be detailed here,  it is  clear that  improvements in stellar modelling  are still necessary in the coming years in order to reach the PLATO age-dating requirement. This is within reach from intense on-going theoretical work within the PMC addressing the main inaccuracy issues.
Real advances in this activity will come from the confrontation of the updated modelling by the time of the PLATO commissioning with observations during the first long pointing of a sample of  well-characterised stars (i.e., with the highest possible precision and accuracy, mainly via seismology).

We point out that the above discussion on systematic uncertainties only concern solar-like stars in the core science case of PLATO. The seismology of stars born with a convective core involve other uncertainties, particularly for stars that are not subject to magnetic braking due to lack of a convective envelope. We refer the reader to the reviews by \citet{Aerts2021} and \citet{Aerts2024} for modelling procedures of such objects and the accompanying uncertainties. These stars are part of PLATO's Complementary Science program.

\subsection{Detecting solar-like oscillations with PLATO}

PLATO will  complement the above  samples with a much larger number of  bright 
main-sequence stars (hereafter MS-stars) and subgiants, 
increasing significantly the number of  low-mass seismically-characterised stars. 
A total of 630 {\it Kepler\/} stars   (see Fig.~\ref{PICKepler})  belong to the PLATO observable fields of view }
mentioned in \cite{nascimbeni2022}. Of this set, 296 {\it Kepler\/} stars belong to  the P1 sample (14886 stars) and 331 {\it Kepler\/} stars are in the P5 sample (266673 stars). Among the 57 Legacy stars in the PLATO Input Catalogue (PIC), 49 are in the P1 sample.

The requirements of the PLATO mission are extremely challenging, especially for the stellar age (e.g. 10\% accuracy for a  reference star with $V=10\,$mag with $1 M_\odot, 1 R_\odot, T_{\rm eff}= 6000\,\text{K}$ --  that is a solar analogue slightly hotter than our Sun).
 In the following, the assessment of the seismic performance of PLATO is discussed    at two levels:
the detection of oscillations as an excess of power in a power spectrum that comes with the measurement of the two averaged seismic parameters
and the detection of oscillation modes for which individual frequencies will be  measured. 
   We will consider the performance for the sample P1--P2 on the one hand and the sample P5 on the other hand. The estimates given below are 
   based on what {\it Kepler\/} data taught us (for details see, \cite{goupil2024} ).

 The P1--P2 sample is deliberately designed to be composed of stars 
 with a noise level (random and systematic residual, non-stellar) 
 below $50\,{\rm ppm\,h^{1/2}}$ at magnitude $V = 11$.
  It is therefore made up of a large number of targets for which one expects the
 detection of solar-like  oscillations in the majority of cases  (K-stars remain an open issue)  as well as the measurement of individual mode  frequencies with a precision high enough  to provide high quality seismic masses, radii, and ages. We expect a considerable improvement in the quality of stellar modelling in terms of accuracy thanks
to the P1--P2 seismic sample.

\begin{figure}[ht!]
 \begin{center}
\includegraphics[width=8.5cm]{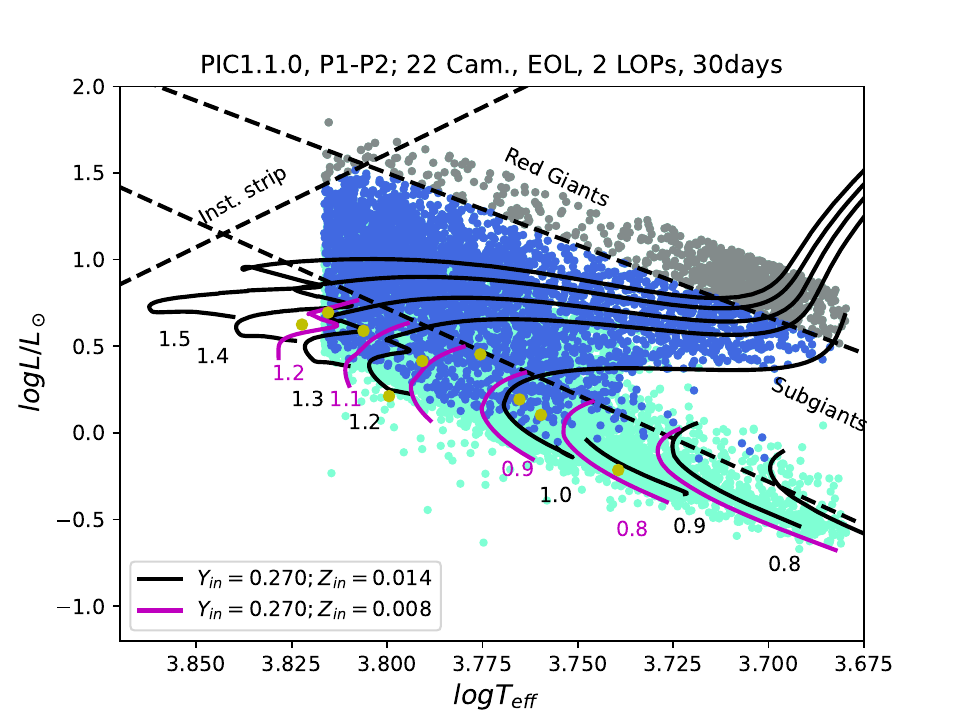}
\includegraphics[width=8.5cm]{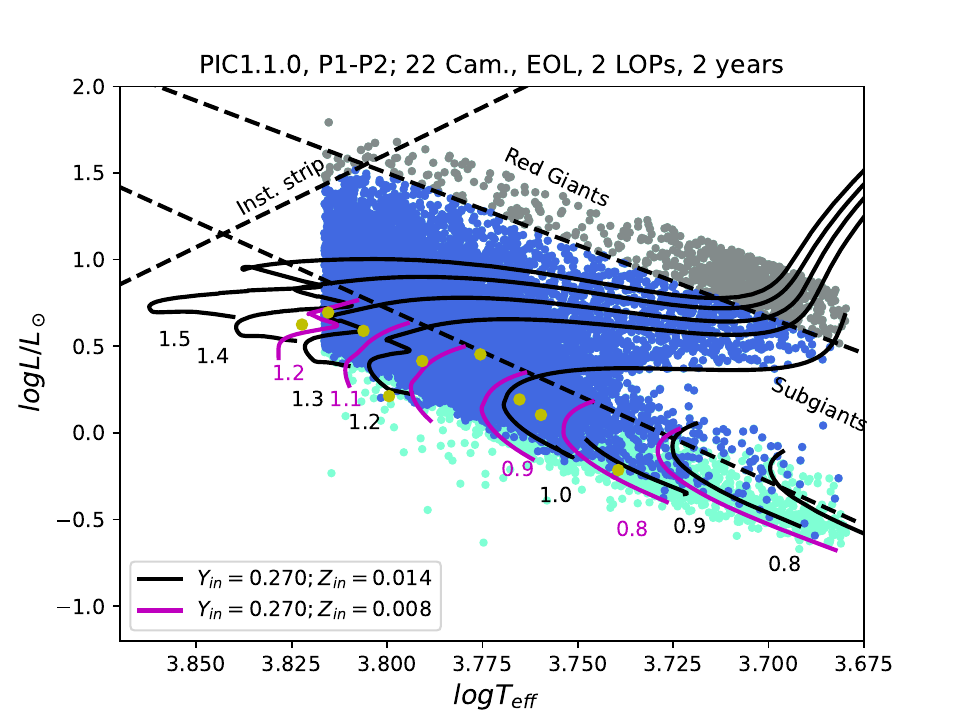}\\
\caption{  \label{HRdiag2} 
HR diagram showing the  subsample of  P1--P2 dwarfs and subgiants of the PIC catalogue 
 for which    a theoretical  calculation led to a probability of detecting solar-like oscillations  equal or above 99\% 
 (blue dots). Stars for which  the  detection level 99\% was not reached are represented with  cyan dots.
 Early red giants and hot stars beyond the instability strip  - included in the PIC catalogue but not in P1--P2-  are  shown in grey dots. 
 The detection probability assumed  an observing run of 30 (top) and 730 (bottom) days.
   Yellow dots represent selected stars from the {\it Kepler\/} Legacy sample.
  We classify stars as main sequence stars when they satisfy  $\log T_{\rm eff} \geq 3.7282+0.10 \log L/L_\odot$ which corresponds to a 
central hydrogen relative abundance in mass greater than $10^{-6}$. Subgiants are then located above that threshold in a HR diagram as represented by a dashed line.
 Dashed lines also delineate the instability strip   
  and the arbitrary separation between subgiants and early red giants.  
  The coloured solid curves represent evolutionary models for masses
  ranging from 0.8 to 1.5\,$M_\odot$ and for two initial metallicities. 
 } 
 \end{center}
\end{figure}

The P5 sample, on the other hand, consists of stars to be observed with a lower SNR. 
Its targets will mostly be observed with a cadence of 600\,s, which is too long to properly detect solar-type oscillations of dwarfs and subgiants. However, at least 10 \% of these stars will be observed with a cadence of 50\,s (see Section \ref{sec:StellarSamples}), suitable for the detection of stellar oscillations and  for a measurement of the two global seismic parameters that can provide seismic  masses, radii, and  ages. The precision will be degraded  compared to the results for the P1--P2 sample but will nevertheless greatly improve their modelling compared to the  stars without seismic constraints.

\subsubsection{Expected solar-like oscillations within the PLATO P1-P2 sample }
The first question is to what extent solar-like oscillations can be detected for PLATO targets. 
Figure \ref{HRdiag2} shows the stars in the P1--P2 sample for two long-pointing  fields in an HR diagram. 
They are taken from the PIC version PICv1.1.0 \citep{nascimbeni2022}, 
which provides effective temperatures and stellar radii, hence luminosities.
In Fig. \ref{HRdiag2}, we distinguish between stars for which a theoretical calculation has led to a probability of detection of 
solar-type oscillations and those for which the  threshold for a positive seismic detection has not been reached. 
The probability of detection were not calculated for hot stars in the instability band defined according to the criterion adopted by 
\citep{chaplin2011}.   
 This concerns only a few stars because the restriction on the temperature of the hot side adopted to construct the PIC is more severe. 
 We also did not consider the early red giants --also contained  in the PIC-- which will be included in 
a specific scientific calibration sample catalogue as part of the overall PIC \citep{Aerts2021}. 
Their performance will be addressed elsewhere. 
Once early red giants and hot stars have been removed, the total subsample  (dwarfs and subgiants)  contains 14083 stars.
We considered a  positive seismic detection  when the probability  of the signal being due to noise is below
0.1\% and the probability of detecting  solar-like oscillations is equal to 99\% or above. 
Details of the computation can be found in \cite{goupil2024}.

The seismic detection probability depends on the observing duration. We therefore show two cases in Fig.\ref{HRdiag2} :
  observing runs of 30 days and 2 years long.
The probability also involves the signal-to-noise ratio in a power spectrum. 
We then  used a combination of the formulations by \cite{chaplin2011}  and by \cite{samadi2019} for the oscillation amplitudes calibrated with {\it Kepler\/} data (a compilation of oscillating stars from the catalogues of \cite{serenelli2017} and \cite{mathur2022}
  and  short-cadence stars found with no detection from the \cite{mathur2019} catalogue. For the noise, we used  
the PLATO (random and systematic residuals) noise level included in the PIC,
to which we added the stellar granulation background noise. 
For comparison, we included in the plot the positions of some {\it Kepler\/} stars from the Legacy sample \citep{lund2017,silvaAguirre2017} as characterised by \cite{creevey2017}.

 The calculation of the probability to detect oscillations depends on the width ($\delta \nu_{\rm env}$) of the assumed Gaussian-shape envelope due to oscillations   in a power spectrum. 
From  Kepler observations, $\delta \nu_{\rm env}$ ranges from $\nu_{\rm max}$ to $\nu_{\rm max}/2$. 
The first option ($\delta \nu_{\rm env}=\nu_{\rm max}$  together with a probability threshold set at 0.90)  (later option 1) provides a number of positive detections in agreement with  the number of {\it Kepler\/} stars with detected oscillations, but a too large number of false-positive detections among the {\it Kepler\/} stars with no detected oscillations. For  the other option ($\delta \nu_{\rm env}=\nu_{\rm max}/2$ together with a conservative probability threshold of 0.99)  (later on option 2), it is the reverse, i.e., the number of predicted non-detections is in agreement with the number of {\it Kepler\/} stars without oscillations, but too few detections are found compared to  the  {\it Kepler\/} stars with detected oscillations.  Unless indicated otherwise,  we remain conservative and use the second option to derive the number of expected detections for PLATO.

Tab. \ref{table:MSstars} 
shows the significant impact on the number of oscillation detection  of   the observing time (30 days and 2 years) assuming option 2 for different subsamples of the P1P2 sample.  The figures are limited to the case of MS-stars with masses  $M \leq 1.2 M_\odot$.

Had  we assumed option 1 for the width of the oscillation envelope,  the number of expected detections for stars with mass $\leq 1.2 M_\odot$  for instance  would 
  increase  from about 55\% of stars (option 2) up  76\% of stars (option 1)  for a 2 years run.
\mj{
\begin{table} [h]
\centering
\caption{\label{table:MSstars} Impact  of the observation duration  on the number of expected detections   of solar-like oscillations  for  P1P2 stars in 2 LOPs.  
}
\begin{tabular}{ ||l | c | c| c||}
\hline
\hline
cases             &    730 days    &      30 days      \\
\hline
Detections         & &\\
\hline
MS ($M<1.6$) +    &                 &   \\
subgiants         &  11109          &  3188 \\
\hline
MS ($M<1.6$)      &   4972          &  477           \\
\hline
subgiants         &   6137          &   2711           \\
\hline
 MS + subgiants   &                 & \\       
 ($M\leq 1.2$)        &    2344     &   302     \\  
\hline
 MS   ($M\leq 1.2$)   &    1844      &    166        \\ 
\hline
\hline
Perfomances ( MS  with $M\leq 1.2$)  &         &    \\
$\delta M/M <15\%,\delta R/R <2\%$ & 1844      &  166\\
\hline
$\delta M/M <15\%,\delta R/R <2\%$  & 1844 &      153   \\
and $\sigma<0.5 \mu$Hz & &\\
\hline
$\delta M/M <15\%,\delta R/R <2\%$  & 1600 &     0   \\
and $\sigma<0.2 \mu$Hz              &         &\\
\hline
\hline
\end{tabular}
\end{table}  
}

Not surprisingly, stars for which one might not detect solar-like oscillations for  too short an observing time are MS-stars of low mass because
 their oscillation amplitudes are too small. 
  In contrast,   detection of solar-like oscillations   - when applied to the {\it Kepler\/} sample- 
   is not  obtained for some stars for which they are theoretically expected \citep{chaplin2011, chaplin2011b,mathur2019}. 
Several reasons may occur, one being the strong magnetic activity. As an 
example, we found $\approx$8 \%  false-positive detections for the  \cite{mathur2019} {\it Kepler\/} sample. Based on this,
the number of  PLATO detections for main-sequence stars with  mass $\leq 1.2M_\odot$ decreases to $\sim$ 51\%.  

  Uncertainties in the probability calculation and the number of stars with expected detections of solar-like oscillation may 
  originate from the use of the PIC1.1.0 radius and effective temperature adopted to compute the global seismic parameters  and to derive the seismic mass.    
  The scaling relations used to derive the seismic masses remain themselves approximate
 and as a result the number of stars in the different mass regimes remains also approximated, albeit with the same order of magnitude.   
       
\subsubsection{Expected solar-like oscillations within PLATO P5 sample }
 
\begin{figure}[t]
 \begin{center}
   \includegraphics[width=12cm]{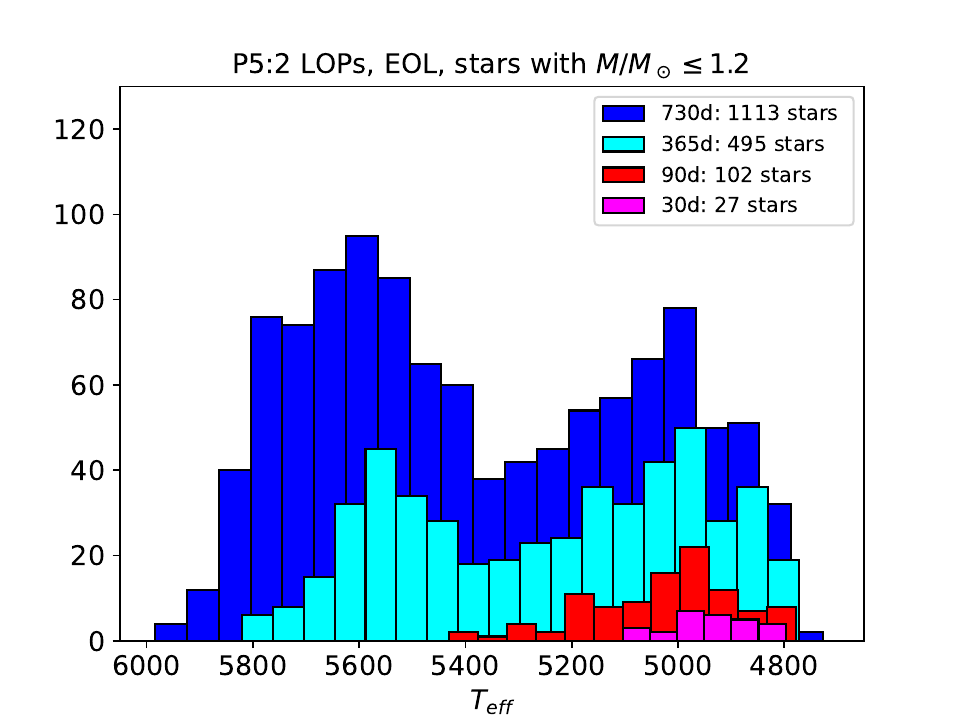}
   \caption{ \label{histoprobaP5} Histogram of the number of stars  with masses $M\leq 1.2 M_\odot$ 
    from the P5 sample with an expected detection of solar-like
    oscillations with a at least 99\%  probability for uninterrupted observations lasting 730, 365, 90  and 30 days.}
 \end{center}
\end{figure}

Let us note that   10\% of the P5 sample will be observed with a 50\,s cadence. This  will make it possible to detect  solar-like oscillations
 and    to measure   at least the global seismic parameters $\Delta \nu$ and $\nu_{\rm max}$. 
 We  performed the same probability calculation as  for the P1--P2 sample 
  after  eliminating the same types of stars and assuming a detection probability equal or greater than 99\%  (and $\delta \nu_{\rm env}=\nu_{\rm max}/2$).                                                          
    The number of seismic positive detections is estimated to be 9941 stars for an observation period of 2 years.
    This number falls   to   5637  stars after only 1 year of observation and only 401  stars after 30 days of observations. In the last case, 
  none are expected to be in the main sequence: the sample is dominated by subgiants because their amplitudes (roughly $\propto L /M $)  are higher than for
   main-sequence stars. 
 The drastic increase of detections of oscillating stars with the observing time in the P5 sample is illustrated in Fig.~\ref{histoprobaP5}.

\subsection{PLATO seismic performance for stellar  mass, radius, and age characterisation  for the  P1--P2 sample}
\label{seismicperformance}

For the subset of  P1--P2 stars with expected solar-like  oscillations,  the detection and highly precise  measurement of   individual  frequencies for  a significant number of modes 
 is ensured by the selection of a high signal-to-noise ratio by construction. As an illustration of the expected PLATO performance, \cite{samadi2019}, to which we refer for details, illustrated 
the excellent PLATO performance for the  {\it Kepler\/} legacy star, \mbox{16 Cyg\,B}, a 6th magnitude dwarf observed over  815 days. 
The simulated PLATO power spectrum built assuming the expected noise  for a $V=10$ star observed with 22 cameras at the end-of-life (EOL, see Section \ref{sec:InstrPerformance}) conditions and for 2 years of observations with the PSLS simulator reveals the strong capacity of the mission to detect the oscillations (note that the noise level was for 28 telescopes (34 ppm h$^{1/2}$) at the time).
 
 This will allow the   determination of the stellar age at the level of 10\%  for F and G stars.
To give a general  estimate of the PLATO stellar  performances in terms of stellar ages is difficult, 
here we use a proxy  based on the expected  frequency uncertainties.
 
  \medskip
\noindent{\it Oscillation frequency uncertainties} 

Dipole ($l=1$) modes produce the highest amplitude and smallest uncertainty, so we focus on these here.
Denoting the uncertainty of the $l=1$ mode closest to $\nu_{\rm max}$ as  $\sigma_1$, {\it Kepler\/} data  taught us that a precision on frequencies of the order of   $\sigma_1$ = 0.2$\mu$Hz for a few modes around the  frequency at maximum power can provide an age precision at the level of 10\% for a Sun-like star. Several hare-and-hound exercises using artificial data constructed for the PLATO noise characteristics were conducted by PMC members (e.g. \citealt{goupil2017},
\citealt{cunha2021}) and confirmed this result. 
 
 Another practical yet  more conservative estimation results from using $\Delta age/age = {\sigma_1}/2.2$ \citep{appourchaux2016}.
  
\begin{figure}[t]
 \begin{center}
 \includegraphics[width=9.cm]{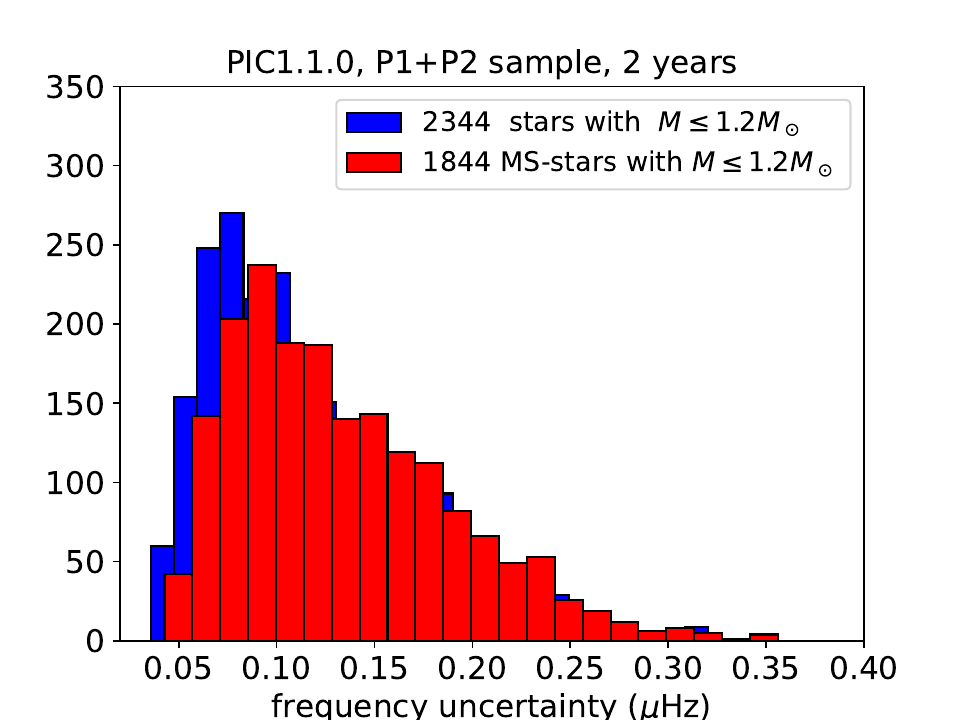}
   \caption{   \label{signoise}  
  Histogram of the frequency uncertainties  for an $l=1$ mode at $\nu_{\rm max}$ for the sample of P1--P2 low mass stars
  with detection probability equal or larger than 99\%.
 }
 \end{center}
\end{figure}


\cite{lund2017} computed how the frequency uncertainties for individual modes decrease with increasing observation duration. A light curve was simulated for a star  of $V=10.5\,$mag  with $M = 1.12 M_\odot, R = 1.20 R_\odot, T_{\rm eff} = 6129\,\text{K}$.
The foreseen PLATO reference noise level of $34\,{\rm ppm\,h^{1/2}}$ at magnitude 11 for 28 telescopes at the time of this study was used. With an updated current reference noise level of $50\,{\rm ppm\,h^{1/2}}$ at magnitude 11~\citep{boerner2024}, the frequency uncertainties correspond to a simulated star of $V\sim 9.7\,$mag.  The
frequency uncertainty increases with increasing magnitude but decreases with decreasing  effective temperature. In this way, it is found that a level of  $0.2\,\mu$Hz can be reached after more than roughly 1.5 years of  observations.  We will then take   that criterion  to estimate the number of stars for which one can reach a statistical error of 10\% for the stellar age of a reference star.

 Assuming individual frequencies are available, we   determine the uncertainty on the frequencies using 
 the \cite{libbrecht1992} formula, which depends on the total SNR and the duration of the observation and has been  proven to yield
the right order  of magnitude
\citep[we refer to the monograph by][for thorough discussions on the basic proportionality with 
$\sim 1/\sqrt{({\rm total\ time\ base})}$ for the frequency uncertainty occurring in this formula]{BasuChaplin2018}.
   
Figure.~\ref{signoise} shows the histogram of the  frequency uncertainty   $\sigma_1$ 
computed from the formulation by \cite{libbrecht1992} for each P1--P2 star for which 
the detection probability is equal to or larger than 99\%  and with  masses less or equal to 1.2$M_\odot$.
The bulk of stars have frequency uncertainties below about  0.12 $\mu$Hz. Those estimates are purely theoretical but give 
the same order of magnitude than for the {\it Kepler} Legacy sample.   

\begin{figure}[t]
 \begin{center}
   \includegraphics[width=6.2cm]{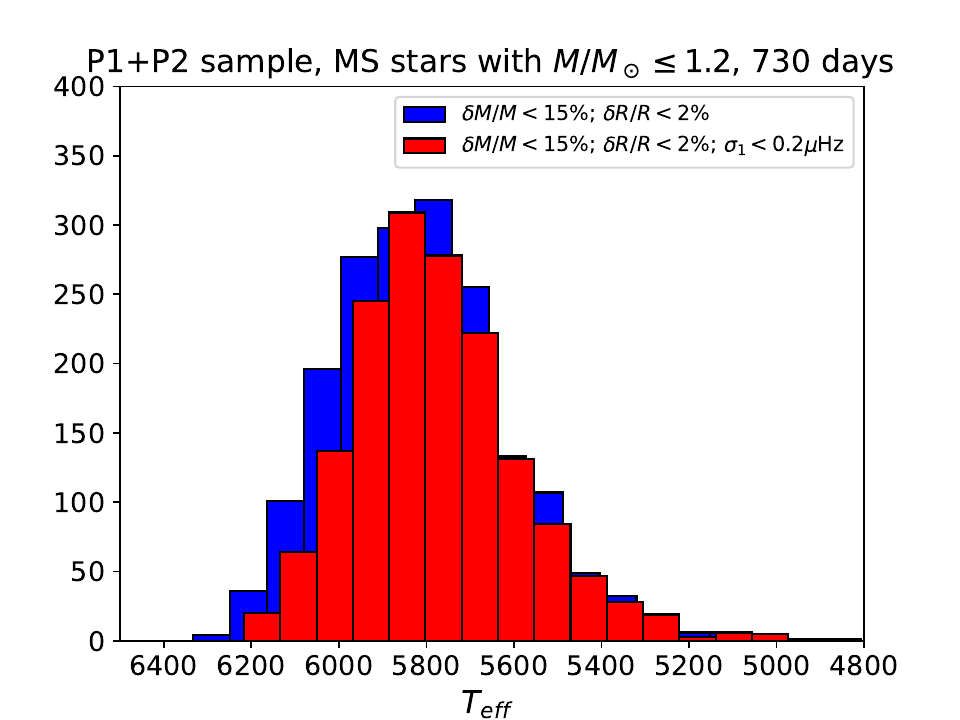}
   \includegraphics[width=6.5cm]{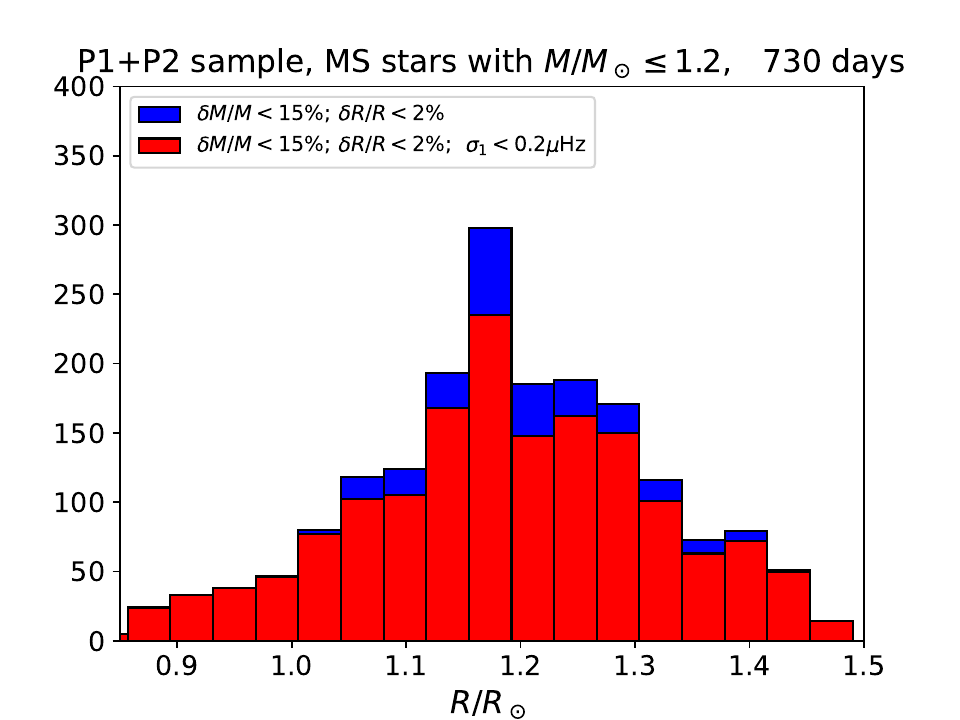}
   \caption{ \label{histoP1Teff} 
   Histograms of $T_{\rm eff}$ (left) and stellar radius (right) for MS-stars with masses $M\leq 1.2 M_\odot$ 
   in   the P1 -P2 sample  with expected detection of solar-like  oscillation after 730 days 
   of observation and satisfying  $\delta M/M   < 15 \% $   and $\delta R/R\leq 2$\% (blue, 1844 stars);
     $\delta M/M   < 15 \% $, $\delta R/R\leq 2$\% and 
  \cite{libbrecht1992}  frequency uncertainty of the dipole mode with frequency closest  $\nu_{max}$, 
  $\sigma_1   < 0.2 \mu {\rm Hz} $     
   (red, 1600  stars) and.
   }
 \end{center}
\end{figure}
   \begin{figure}[ht!]
 \begin{center}
   \includegraphics[width=6.cm]{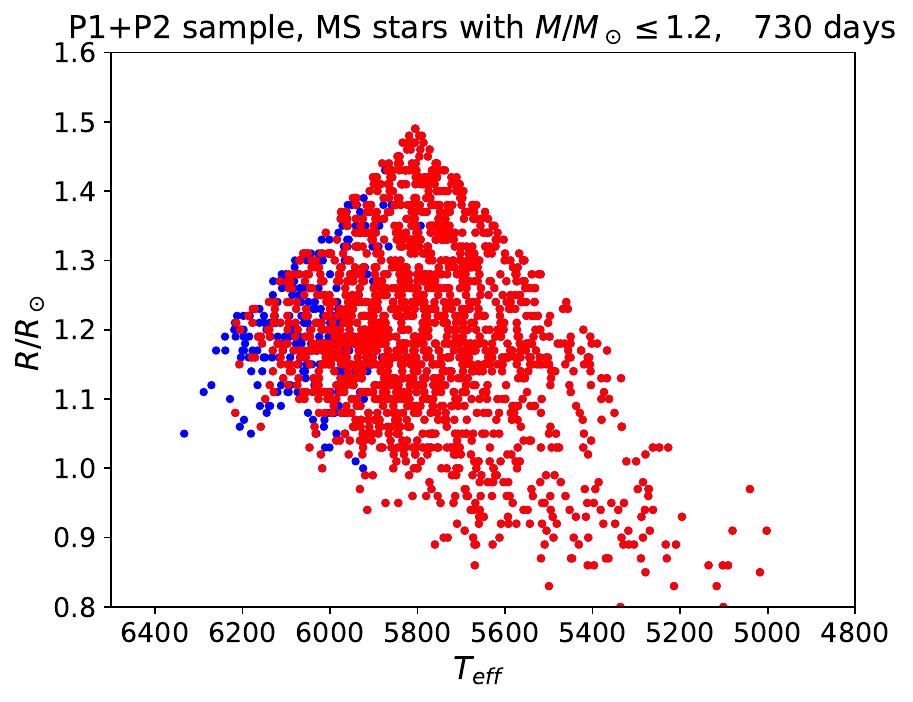} 
   \caption{ \label{histo2D} 
    Same as   Fig.\,\ref{histoP1Teff} but represented as  2D histograms ($T_{\rm eff}$  on the x-axis and radius on the y-axis). Blue points   represent stars satisfying on the left: $\delta M/M < 15 \% $ and $\delta R/R\leq 2$\% and red dots $\delta M/M < 15 \% $ and $\delta R/R\leq 2$\% and $\sigma_1 < 0.2 \mu {\rm Hz} $ . }
\end{center}
\end{figure}

\medskip
\noindent{\it Mass, radius and age seismic uncertainties}   

We now turn to the expected accuracy of the seismic determinations of  masses, radii, and ages. 
For the mass and radius relative  uncertainties, we used empirical formulations obtained as fits  of the results of 
 seismic inferences of mass, radius and age of synthetic stellar models (see \cite{goupil2024} for detail).
 For the age uncertainty, as mentioned above, we consider the  criterion $\sigma_1 \leq 0.2 \mu$Hz  as a proxy for 
 the requirement on the age uncertainty (i.e. $\leq$ than 10\%). Those estimates are given in Tab. \ref{table:MSstars} above.
   Figures~\ref{histoP1Teff} and \ref{histo2D} focus on  MS-stars   with masses $M/M_\odot \leq 1.2$
and show the $T_{\rm eff}$, and radius histograms for  
stars  satisfying the above constraints : mass  and radius relative  uncertainties  better than 15 \%  and 2\%  respectively and $\sigma_1 \leq 0.2 \mu$Hz.    Of interest for the  exoplanet yields, 1640  stars with masses $M\leq 1.2M_\odot$ and  937 stars with  radius $\leq 1.2 R_\odot$  have  $\sigma_1 \leq 0.2 ~\mu$Hz. 

 We recall that our estimates  are based on a PLATO noise level corresponding to observations with 22 cameras at EOL. 
Note also that the above uncertainties  refer to precision and  do not account for systematic errors due to limitations in 
 the models of stellar interiors and atmospheres
discussed above (such as surface effects, \cite{jorgensen2020}, or stellar activity).  
  Currently the total uncertainties  increase roughly by $\sim$ 3 to 5\% due to lack of accuracy for a solar-like star and to $\sim $ 15\%-- 50\%   for a more massive star with a convective core. Theoretical work is ongoing  to decrease the impact of lack of accuracy. 
The impact of the main systematic effects were 
 investigated    by \citep{cunha2021} in view of PLATO applications
 using also  seismic inferences of mass, radius and age of synthetic stellar models.

\subsection{Expectations for M dwarfs (P4 Sample)}

One does not expect  to detect solar-like oscillations for  M-dwarfs,   mainly because of their high level of activity. 
Without seismic data from the PLATO mission, one will have to resort  to classical methods and stellar models 
to characterise such stars (see for instance the input Carmenes catalogue, \cite{cifuentes2020} and references therein).

 In the framework of the PLATO project, an updated and accurate preliminary library of stellar models 
for M-dwarfs - stars with $M < 0.5 M_\odot$ - 
has been produced. A sub-set of this library has been
already published (e.g. \cite{hidalgo2018, pietrinferni2021}, but for the aims of the
PLATO project, the set has been hugely extended by adopting a very fine grid as far as it  concerns the total stellar mass and metallicity (e.g. Orlander et al. 2024, in preparation). All those models are available to the scientific community. 
   
  This model set is based on the best available input physics as far as opacity and thermodynamical tabulations, nuclear cross sections, and outer boundary conditions are concerned.   
  
Comparison with suitable observational benchmarks has shown the existence of a good agreement  between model predictions and observations, although some discrepancies occur. More in detail,  in the regime of fully  convective stars ($M < 0.35 M_\odot$), stellar models over-predict the effective temperature of suitable benchmark stars by on average 3—4\%, and underestimate the stellar radius by about 5\%. Current theoretical mass – IR-band magnitude luminosity relations to derive the mass of M dwarfs deliver an accuracy of the order of 4\%  (e.g., \citealt{passeger2019}). The presence of these discrepancies is a clear proof that there is still a significant uncertainty in some input physics adopted in the stellar modelling, and/or some physical process such as magnetic field/rotation is not properly accounted in the computation of models for such stars. In order to solve this problem and crucially improve our capability to characterise host-planet M dwarfs, a detailed search for well studied single M dwarfs and, in particular, suitable binary systems formed by an M dwarf and a more massive star is ongoing within the PMC. The selection of these crucial benchmark M dwarf stars will allow us in the context of the PLATO stellar science, to achieve a better knowledge of the atmospheric layers of these stars, and hence of the outer boundary conditions needed in their modelling, as well as of the link between convection and magnetic fields. In this specific context, the availability of a detailed characterization of the magnetic field activity in these stars - obtained by combining the PLATO data analysis with the results collected from related follow-up surveys - is mandatory to shed light on the origin and physical properties of the magnetic fields in M dwarfs and their impact on their structural properties. On a different ground is the issue of a realistic estimation of the uncertainty on the age estimate for these stars; this is made very difficult by the intrinsic extremely long evolutionary lifetime of these peculiar stars.

\subsection{Measurement of surface rotation period, stellar activity indices and flares}

Stellar rotation, whether uniform or differential,  is an important input to both the
Exoplanet and Stellar Analysis Pipelines of PLATO. Indeed, in order to characterise and model  the host star and its cohort of planets, information about the occurrence of stellar activity and its magnitude is required for both applications.
 Measurements of various quantities related to stellar rotation and activity are therefore among the top objectives of the PLATO mission. 
  The measurement procedures rely on our experience obtained from the {\it Kepler data}
\citep{Garcia2011, handberg2014, mcquillan2013a, garcia2014, mcquillan2014ApJS..211...24M, aigrain2015MNRAS.450.3211A, breton2021, santos2021}.

 All the PLATO photometric timeseries will be analysed to search for surface rotation periods, surface differential rotation, and  magnetic indicators and activity cycles. The analyses will be carried out on the data collected during each quarter, and on consecutive stitched quarters as  the mission progresses.  Such techniques are very sensitive to  contamination from instrumental modulations. The long-term stability of the instrument at low-frequency is therefore crucial as discussed  by \cite{santos2019} for instance.
 The time series will be analysed using a combination of different algorithms, such as
 the Generalised Lomb-Scargle (GLS; \citealt{handberg2014} and references therein) and the AutoCorrelation Function  (ACF; \citealt{mcquillan2013a, garcia2014}), to produce a composite spectrum (hereafter CS)
 \citep{ceillier2016,ceillier2017}.
Only highly-significant periodicities in the GLS power spectrum will be considered.
Once the rotation period is identified in the CS, a machine learning (ML) random-forest approach is used to decide if the final rotation  period will be the one computed from the GLS, the ACF or the CS (see, e.g., the Random fOrest Over STEllar Rotation (ROOSTER) methodology; \citep{breton2021, santos2021}). The light curves included in the training set will be carefully selected, with the possibility 
to incorporate relevant light curves from previous space missions (Kepler/K2, TESS) together with PLATO simulations. 
It is planned that the module performance will be evaluated within the PLATO consortium after launch, 
once the first observations are available, and the possibility to include a subset of real PLATO targets with rotation measurements validated independently from ROOSTER 
is under consideration \citep{breton2024A&A...689A.229B}. Some additional stellar parameters such as luminosity or effective temperature can also  be used to help in   the selection   of the best rotation period.

An important point of such a methodology is that a random forest allows us to use more parameters in the decision than a manually-set threshold when assessing the robustness of a rotation detection and selecting the corresponding rotation period. This methodology has the advantage of preserving the uncertainty and the posterior distribution of the method selected by the random forest. This type of machine learning algorithm can be combined with any inference techniques applied on the light curve or the periodogram.
In the baseline version of the PLATO pipeline, the uncertainty associated with the rotation period and the activity cycle length will be inferred from the width of the corresponding power spectrum peak, and it is expected to improve as the timeseries get longer.
Candidate values for starspots and levels of differential rotation will be inferred from the relative difference between the rotation period and other, if any, significant periodicities. Such candidate values will be validated using priors from theoretical models, as well as information on either the single or binary nature of the star, since unresolved secondary components can also produce detectable periodicities in the flux timeseries. On the other hand, starspot evolution and differential rotation may affect the determination of the mean stellar rotation period for some targets. In most solar-like stars, we expect an increase in the broadening of the periodogram peaks well within our adopted 10\% uncertainty. 
 However, a systematic investigations of these effects for a proper consideration in PLATO data analysis is under way
  (see \cite{breton2024A&A...689A.229B}).  The presence and length of long-term periodicities, likely arising from either activity cycles or  from beating of close frequencies, will be inferred as done for the rotation period, but focusing on the lower-frequency region of the power spectrum computed from the stitched timeseries.

In the future, some complementary methods will be evaluated in order to be included along the current baseline. 
  Among possible additions, we mention the wavelet decomposition (\citealt{torrence1998, liu2007, mathur2010}),
   a time-frequency analysis that eases discriminating stellar rotation signatures with instrumental modulations or neighbours
    contamination. A new development around this wavelet approach, the Gradient Power Spectrum (GPS) was recently proposed in 
    order to measure rotation in stars with low levels of activity (e.g. \citealt{amazogomez2020, reinhold2022}).

The Stellar Analysis pipeline will also detect and remove stellar flares from PLATO
light curves. Flares are both a nuisance for the detection of exoplanet
transits or any other astrophysical signal in PLATO light curves and an important input parameter
for the modelling of the evolution of planetary atmospheres \citep{tilley2019, gronoff2020}.
The duration, peak flux, and total energy of these events in the PLATO passband will be measured.
Typical durations of optical flares are on the order of few minutes up to about one hour, and the flare
duration and energy are correlated \citep{namekata2017}. This makes PLATO with its superb data
cadence combined with unprecedented photometric precision ideally suited to study these fast transients.

\section{The PLATO Complementary Science} 
\label{sec:ComplScience}

Previous photometric space missions built for exoplanet detection by the transit
method, such as CoRoT \citep{auvergne2009}, {\it Kepler\/} \citep{koch2010a}, and TESS~\citep{ricker2015} have already shown that the uninterrupted high-precision photometric light curves they assembled triggered studies far beyond the nominal mission goals. 
 Following this, the PLATO Complementary Science programme (PLATO-CS hereafter) 
stands for a broad activity initiated by ESA's Science Working Team and supported by the PMC. It has the global goal to get the maximum scientific return from the PLATO mission. In order to achieve this, PLATO-CS has defined several operational tasks. 

First of all, PLATO-CS will 
offer the community an electronic database containing a variability classification and characterisation of all the stars brighter than about 13.5\,mag in the TESS band and situated in PLATO's fields. This database for the first long pointing (see Section\,\ref{sec:TargetFieldsandObsStrat})
will be made available about a year prior to the launch. A meta-classifier is currently being trained for it, by means of supervised and
unsupervised classification. It relies on
machine-learning tools and is trained on {\it Kepler\/} and TESS 
light curves for the class
definitions following \citet{Audenaert2021}. The PLATO-CS variability catalogue will be made publicly available and will be continuously updated as newer results from the classifiers become available 
\citep{Audenaert2022}. 
This variability catalogue will be a major
source of information for GO applicants  (see Section\,\ref{sec:GOProgram}).

Secondly, PLATO-CS actively extends 
the PLATO image and light curve simulator. This software tool takes into account all known instrumental noise sources to date to simulate realistic PLATO data \citep[{\tt PlatoSim},][]{jannsen2024A&A...681A..18J}.
{\tt PlatoSim} is a versatile and publicly available suite of modules for the community to help assess PLATO's capacity for a large variety of scientific goals. PLATO-CS also offers mock data simulated with {\tt PlatoSim}. Along with   {\it Kepler\/} and TESS light curves, the PLATO mock simulations serve as 
preparatory tools and testbeds to understand PLATO's performance outside the core programme (Jannsen et al., in prep.).  

Finally, PLATO-CS triggers the worldwide
community to prepare competitive GO programs (see
Section \ref{sec:GOProgram}) on any topic not covered by the core science.
PLATO-CS has predefined a series of such topics to ensure that these will in any case be the subject of GO applications \citep[see][for an overview of these]{Tkachenko2024}, including asteroseismology of massive and compact stars (Aerts et al., submitted), binarity and multiplicity, galactic structure of the Milky Way and Large Magellanic Cloud, transients and more broadly extragalactic science. However, many more topics are anticipated following ESA's GO call. By means of illustrations, we now briefly discuss three examples defined by PLATO-CS as topics suitable for GO applications from the community.

\begin{figure} \begin{center}
\rotatebox{270}{\resizebox{8.cm}{!}{\includegraphics{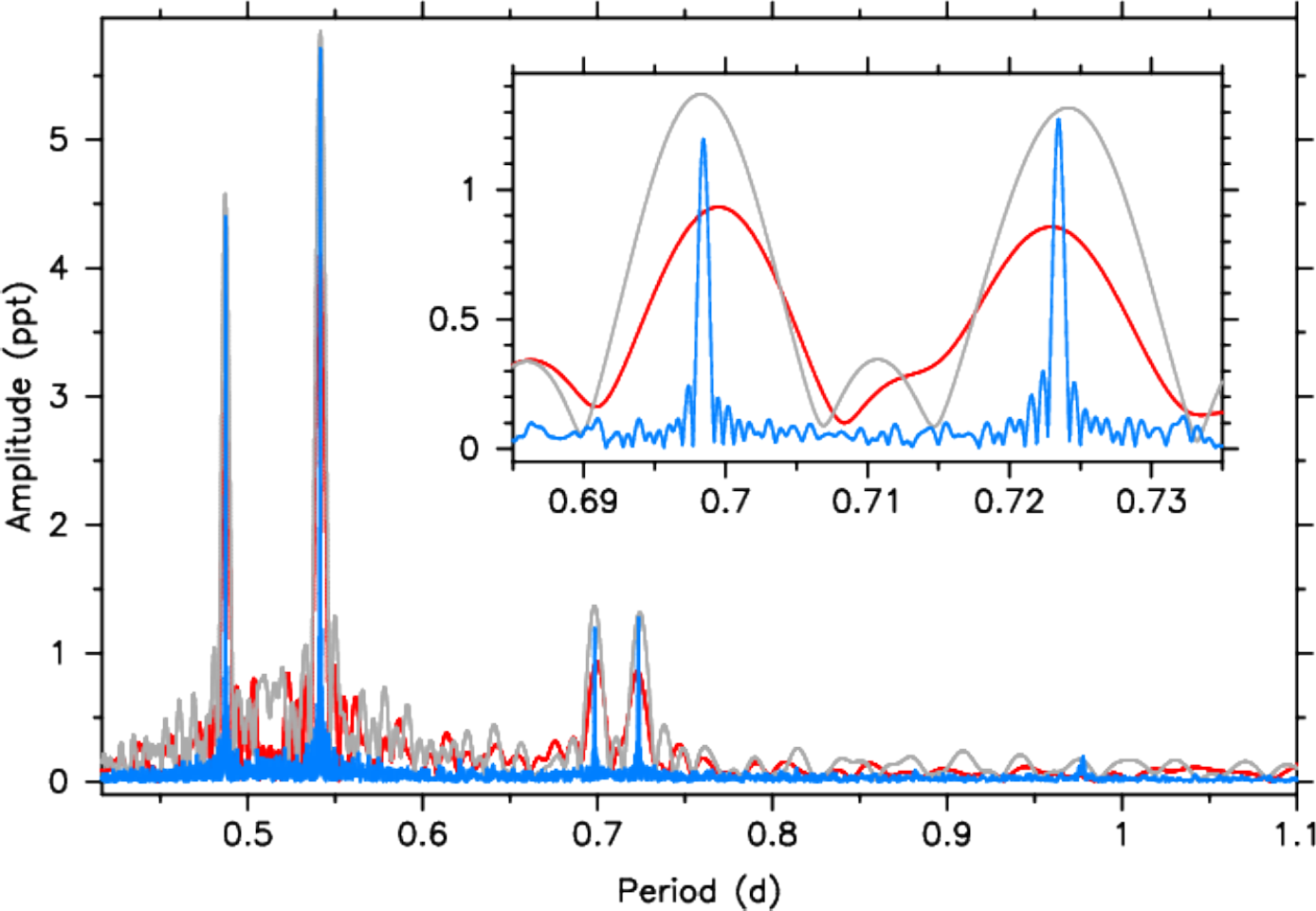}}}\end{center} \caption{\label{fig:gdor} Amplitude spectrum of the $\gamma\,$Dor
g-mode pulsator KIC\,8645874 (V mag of 9.9)  plotted as a function of period. Blue: periodogram deduced from the {\it Kepler\/} light curve covering 684\,d, red: periodogram from TESS data covering 54\,d; grey: periodogram for an excerpt of the {\it Kepler\/} data with a time base of 54\,d. The dipole prograde modes deduced from the blue periodogram reveal a period spacing value of 2322\,s (or 0.026876\,d) as can be seen in the inset. 
This diagnostic gives direct information on the internal rotation and buoyancy frequencies of the star \citep{VanReeth2016}. PLATO will deliver a g-mode periodogram similar to the blue one should the star be observed by all 24 normal cameras.}
\end{figure}

\subsection{PLATO-CS Example 1: gravity-mode asteroseismology}

Gravity-mode (g-mode) asteroseismology got kick-started by the CoRoT mission with the discovery of period spacing patterns due to high-order g modes in the B-type dwarf HD\,50230 \citep{Degroote2010}.   
The 5-month CoRoT light curve of this star led to the first detection of a g-mode period spacing pattern of this slowly rotating pulsator. However, the real breakthrough 
in the derivation of the internal rotation and mixing inside rotating
dwarfs came from the 4-years long {\it Kepler\/} light curves. These led
to the near-core rotation rates of $\sim 700$ dwarfs covering the mass range from 1.3 to 9\,M$_\odot$  \citep[see][for a summary]{Aerts2021}. 
With PLATO we have the opportunity to embark upon g-mode asteroseismology on a much larger scale, for a wide variety of masses, metallicities, and evolutionary stages, opening up
asteroseismic probing of stars that will eventually explode as supernova as well
as the most massive exoplanet host stars outside of PLATO's core science. 

PLATO's potential for g-mode asteroseismology 
is illustrated in
in Fig.\,\ref{fig:gdor}, which shows the Fourier transform in the form of a
periodogram for the 684-d {\it Kepler\/} light curve of the $\gamma\,$Doradus
pulsator KIC\,8645874. This young early-F type star 
has an effective temperature of 7240$\pm$90\,K, $\log\,g=3.88\pm 0.27$,
$v\sin\,i=21.4\pm 0.9$\,km\,s$^{-1}$ and solar metallicity \citep{Tkachenko2013}. 
It is a dipole prograde g-mode pulsator with an asteroseismic mass determination of $1.52\pm
0.02$\,M$_\odot$. The star is found to rotate quasi-rigidly with a rotation period of $\sim2.7\,$d, which is about 5 times slower than the period of its dominant dipole g mode shown in Fig.\,\ref{fig:gdor} \citep{VanReeth2015,Mombarg2021}. 

Figure\,\ref{fig:gdor} also shows the
periodogram derived from this star's 54-day TESS light curve in red
and a 54-day string from the {\it Kepler\/} light curve in grey. This illustrates that the increased pixel size
of TESS versus {\it Kepler\/} does not bring any contamination issues for this
particular star, which has a visual magnitude of 9.92. The inset in
Fig.\,\ref{fig:gdor} zooms in on two of the star's g modes and reveals that the mode
period precision downgrades tremendously by reducing the light curve from 684\,d
to 54\,d (grey versus blue). 
It is also seen from the different shapes of the red and grey periodograms
that the multiperiodic g-mode beating remains unresolved in a 54-day light curve.

The {\it Kepler\/} data of this star led to the size and mass in its
convective core 
and allowed to deduce its evolutionary
stage in terms of its central hydrogen mass fraction with a relative precision of $\sim\,10\%$ \citep{Mombarg2021}.  
PLATO's 2-year light curves can deliver similar powerful asteroseismic probing to the blue
curve in Fig.\,\ref{fig:gdor}, but now for tens of thousands of targets in the Milky Way.

\subsection{PLATO-CS Example 2: galactic archaeology}

Step-and-stare phases of duration shorter than two years are currently not foreseen in the nominal 4-year mission, but the instrument design makes it possible to take up such an observing strategy during the extended mission (see Section\,\ref{sec:TargetFieldsandObsStrat}). Such step-and-stare phases 
would be suitable 
for galactic archaeology. This science case is extensively discussed by
\citet{Miglio2017}, who showed that asteroseismic age-dating of red giants at
$\sim\,10\%$ level requires light curves with a duration of at
least 150\,d. 
 
The internal rotation of red giants, on the other hand, can only be deduced
from rotational splitting of dipole mixed modes. This requires a minimal
duration of 2 years for the light curves, as revealed from the detections
of core rotation from {\it Kepler\/} data \citep{Beck2012,Mosser2012}. 
PLATO offers this capacity from its onboard 
light curves, again for tens of thousands of red giants.

\subsection{PLATO-CS Example 3: transient studies}

Outbursts in accreting compact binaries with white dwarfs, neutron stars or black holes hold the potential to  gain crucial information on disc instabilities and accretion processes. In particular low-mass X-ray binaries display outburst precursors in optical wavebands prior to the X-ray detections and subsequent 
delays in optical to X-ray emission during the onset of outbursts 
\citep{Russell2019,Goodwin2020}. 


Accreting white dwarf binaries of Cataclysmic Variable (CV) type have orbital periods ranging
from $\sim$5 min to half a day and contain a white dwarf accreting
material from a late type main sequence star or another white
dwarf. Their outbursts arise through instabilities in the accretion disc and
can have timescales from a few weeks to decades.
The dozen CVs in the {\it Kepler\/} field
allowed the accretion process to be studied in an
unprecedented way. 
Both normal and superoutbursts were observed
from V344\,Lyr. These revealed positive and negative superhumps at
different stages of the outburst cycle. These superhumps have periods
slightly longer (positive) and shorter (negative) than the binary
period and are due to a precessing and/or tilted accretion disc and
accretion streams
\citep{Still2010,Wood2011,OsakiKato2013}. {\it Kepler\/} observations of 
V1504\,Cyg and
V344\,Lyr also revealed that superoutbursts are preceded by a normal
outburst giving insight to the physics of superoutbursts in general
\citep{Cannizzo2012}.

\begin{figure}
\includegraphics[width=\linewidth]{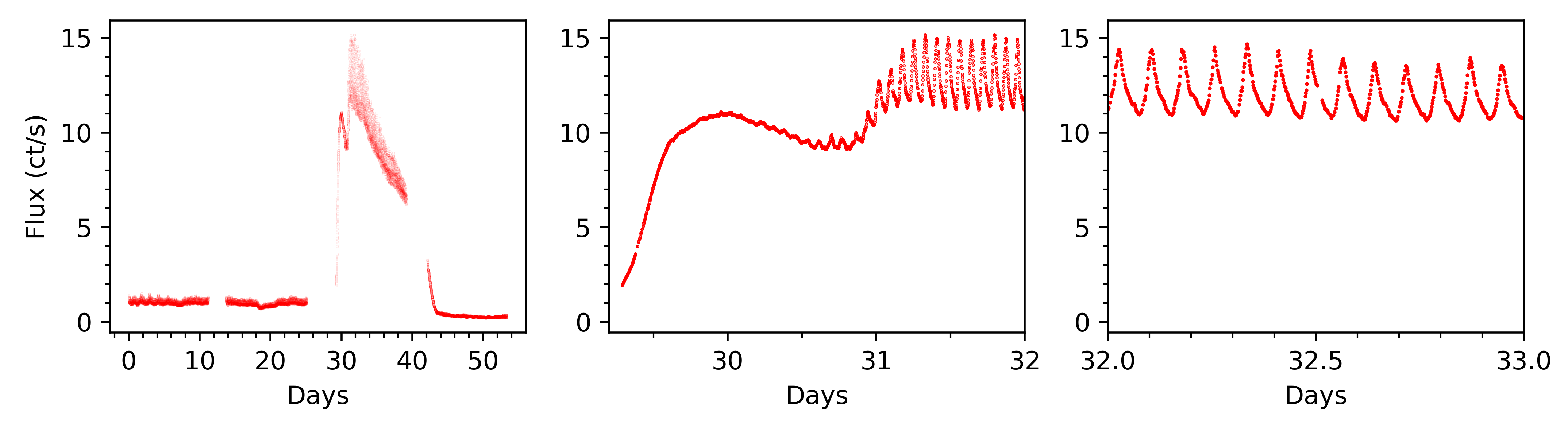}
\caption{A snapshot of TESS 2-min 
cadence observations of the CV\,VW Hyi made in
Cycle\,1 between 2019-03-28 and 2019-05-20 and showing a
superoutburst. Positive superhumps are seen at the point of maximum
light (middle panel); their period starts to shorten as the outburst
progresses (right panel).}
\label{fig:VWHyi}       
\end{figure}

Since then, many more CVs and accreting binaries have been studied using K2 and TESS - see the 
example of VW\,Hyi in Fig.\,\ref{fig:VWHyi}.
It shows a normal outburst preceding a superoutburst, with
superhumps appearing at maximum brightness. The period of these positive superhumps decrease as the outburst progresses. Observations
such as these can be compared directly with models of superoutbursts \citep{Thomas2015,Kimura2020}.
Other types of CVs have also been discovered to display unusual behaviour that went unnoticed using ground-based telescopes, such as the switch on and switch off of accretion in TW Pic \citep{Scaringi2022a} and fast optical enhancements in a number of magnetic CVs interpreted as micronova events \citep{Scaringi2022b}.

PLATO will allow long term observations of relatively bright CVs, or
systems which are usually faint but experience rare outbursts. More
importantly, the potential exists to obtain high cadence, {\sl
multi-colour} photometry over the outburst cycle from PLATO´s fast cameras. This will allow for
a search for variations in colour over orbital cycles and
super-humps. Observations of VW\,Hyi made several decades ago
\citep{vanAmerongen1987} showed such colour variations can be
present. PLATO-CS will also allow for the detection of longer
period dwarf nova oscillations which can be used to probe the accretion
process. We expect that outbursts from accreting binaries will be prime targets for ToO observations.

\section{PLATO Stellar Target Samples} 
\label{sec:StellarSamples}

\subsection{Stellar Samples Requirements}
\label{sec:samplesDefinition}

The PLATO mission has defined seven overall science objectives and four types of stellar samples (P1, P2, P4 and P5, P3 was dropped during mission development). Here, we summarise the key parameters of the stellar samples requirements set for the 4 year mission baseline (see Table \ref{tab:StellarSamp}, and Fig. \ref{fig:mission-comp} for a comparison to other missions). The actual target field is discussed in Section \ref{sec:TargetFieldsandObsStrat}.\\

\subsubsection{The bright samples (P1 and P2)}

These samples are at the core of the PLATO mission and consist of the brightest targets that PLATO will observe. Most of them will be part of the prime sample forming the main body of the final ``PLATO Catalogue''. The sample P1 includes at least 15,000 dwarf and sub-giant stars (types F5 to K7), cumulative over the nominal mission, with $V\leq11$ mag and a noise level of $<$50 ppm in 1 h (see Table \ref{tab:StellarSamp}). We note that for the brighter stars in the sample ($V<10$ mag) a noise level as low as 34 ppm in 1 hour can be reached (see Section \ref{sec:InstrPerformance}). Sample P2 includes at least 1000 targets of the same type and noise performance with brightness $V \le 8.5$. The P2 targets are therefore a sub-set of the P1 sample, ensuring a certain number of very bright stars to be observed. Targets in these bright stellar samples observed by the normal cameras (N-CAM) are obtained with 25 sec sampling cadence. 
 
A sample of 300 stars will be observed with the two "fast" cameras (F-CAM) with 2.5 sec cadence, providing 'red' and 'blue' colour information. We note, however, that the F-CAMs cover only part of the central FoV of the normal cameras. 

For all targets in the P1 and P2 samples, imagettes consisting of a small cut-out image of the respective targets of configurable size (typically 6$\times$6 pixels)  will be obtained and down-linked to ground without on-board pre-processing (see Section \ref{sec:lightcurvesImagettes} for a discussion of the share of data on light curves versus imagettes). 
 
The P1 and P2 samples are bright enough to determine the prime planetary parameters mass, radius and mean density in a wide range of systems, including terrestrial planets in the habitable zone of solar-like stars. Host stars are bright enough for asteroseismology. Table \ref{tab:NoiseLevels} illustrates the expected performance for planet characterisation for different noise levels and samples (see also Section \ref{sec:InstrPerformance}). Known eclipsing binaries will be part of the sample (also of P5), after a revision of their suitability for the potential detection of circumbinary planets.

We expect a sample of $>$100 (goal: 400) exoplanets characterised for their radii with better than 3\% accuracy  when their host stars are brighter than $V=10$ mag, and better than 5\% radius accuracy for host stars brighter than $V=11$ mag. Their masses are expected to be determined with an accuracy of $\sim$10\%. This planet sample will span over a wide range of physical sizes and mean densities, including $>$5 (goal: 30) (super-)Earths in the habitable zone of solar-like stars. See Section \ref{sec:PlanetYield} for an analysis of the expected PLATO planet yield and a comparison to other missions.

\begin{table}[]
\tabcolsep=4pt
    \centering
    \begin{tabular}{c|c|c|c|c|c|c} 
Sample  & target   & noise       & type  & \# targets  & sampling & data        \\
        & $V$ mag      & ppm in 1 h  &       &             & sec      &             \\ \hline
P1      & $\le$11  & $\le$50     & F5-K7 & $\ge$15000 & 25       & imagettes   \\ \hline
P2      & $\le$8.5 & $\le$50     & F5-K7 & $\ge$1000  & 25       & imagettes   \\ \hline
P4      & $\le$16  &             & M     & $\ge$5000  & 25       & imagettes   \\ \hline
P5      & $\le$13 &              & F5-K7 & $\ge$245000 & 600     & light curve    \\ 
including: &      &              &       & $\ge$10 \% 
of the targets        & 50     &  light curve    \\
        &         &              &       & $\ge$5 \%  
of the targets       & 50     & centroids      \\
        &         &              &       & $\ge$9000  
        targets  & 25     & imagettes      \\ 
   \end{tabular}
\caption{Summary of PLATO stellar sample requirements on dwarf and sub-giant targets.}
\label{tab:StellarSamp}
\end{table}

Asteroseismic measurements will be performed for $>$5,000 stars in the bright samples to obtain precise ages of planetary systems. Asteroseismic modes can be analysed with high precision to improve stellar models. This is expected to result in a sample of $>$100 (goal: 400) bright planetary host stars with accurate ages ($\sim$10\%). This data set is therefore of fundamental importance for the mission and will also be used to e.g. calibrate classical age determination methods applicable to hosts which do not allow for asteroseismic investigation.\\

\subsubsection{The "M dwarf sample" (P4)}

This sample is dedicated to survey cool late-type dwarfs (i.e., late K to M dwarfs) in the solar vicinity and shall include at least 5000 objects with $V \leq 16$ mag monitored during long pointings. \\
   
    \begin{table}[]
        \centering
        \begin{tabular}{c|c|c} \hline
           noise    & sample          & performance  \\ 
       ppm/sqrt(1h) &                 & for solar-like hosts  \\ \hline
            34      & P1 and P2, $V \le 10$ mag & 3\% r$_p$, 10\% age  \\
            50      & P1, $V \le 11$ mag & 5\% r$_p$, 20\% age  \\
            80      & P5, $V \le 12$ mag & detection of 1 R$_\textrm{Earth}$ \\ \hline
        \end{tabular}
        \caption{Overview of expected performance for different samples and noise levels. }
           \label{tab:NoiseLevels} 
    \end{table}
       
\subsubsection{The "statistical sample" (Sample P5)}

At least 245000 dwarf and subgiant stars (F5-K7) with $V \leq$ 13 mag are observed in this sample, cumulative over at least two target fields. For most of these targets, stellar light curves are computed on-board the satellite. The final sampling of these light curves will then be 600 sec. However, the data volume allows for some of the P5 targets, e.g. the brightest ones in this sample or targets of special interest, to be transmitted to ground as imagettes, see discussion below.
For stars with light curve noise levels better than 80 ppm in 1 hour we expect that detection of Earth-sized planets around solar-like stars will still be possible in the P5 sample, even if their bulk parameters cannot be characterised with the same accuracy as for the P1/P2 samples.
    
 The "statistical sample" will be used for, e.g., planet frequency determinations, studies of planet parameter correlations with stellar parameters or with the environment of planetary systems. We expect $>$4000 (goal: 7000) planets with well determined orbital parameters. For $>$100 (goal: 400) of them, orbiting bright stars, accurate masses will be determined via RV. In addition, mass determination from TTVs will provide upper mass limits for suitable systems.  We note, however, that the majority of targets in the statistical sample will be too faint for large-scale asteroseismic analysis and RV follow-up activities with highest precision.\\


\subsubsection{The "prime sample"}

A "prime sample" of up to 20000 targets will be selected out of the above stellar samples (P1, P2 and P5), supervised by the ESA PLATO Science Working Team (SWT), for observations with highest accuracy and will be treated with highest priority throughout the mission. The PMC will organise ground-based follow-up observations to confirm the planetary candidates for this sample and measure the mass of the planets via its Ground-based Observing Programme (GOP). Level-2 and Level-3 data products for the "prime sample" will form the core of the final "PLATO catalogue" consisting the final planet and stellar parameters derived by the mission.

 The prime sample with targets in the first long pointing sky field is to be defined at least nine months before launch and updated six months before every satellite sky field pointing. The SWT decided that 15000 prime sample targets should be selected in the first pointing field of PLATO (LOPS2, see Section \ref{sec:TargetFieldsandObsStrat}), putting emphasize on the first field.  The metrics to select the "prime sample" targets in the first field is currently under definition in the SWT. The criteria will be chosen such as to optimize the detectability of small planets with long orbital periods via transits in PLATO lightcurves as well as in the ground-based follow-up.

\subsubsection{The propriety sample}
  
The ESA Science Management Plan grants the PMC proprietary rights over a small, pre-defined set of a maximum of 2000 targets in total over the 4 years of nominal mission duration. These targets will be selected using the first three months of PLATO observations of each field. From the "prime sample" stars with brightness $V \leq$ 11 for each sky field, the stars belonging to the lowest quartile (25\%) of the noise distribution will be identified. The PMC proprietary targets will be 25\% of these, with the condition that they will have a noise distribution similar to that of the original sample.

\subsection{Light curves versus Imagettes}
\label{sec:lightcurvesImagettes}

The processing and telemetry resources of PLATO easily cover the needs of the science requirements concerning the number of imagettes and light curves as outlined in Section \ref{sec:samplesDefinition}. It is, however, not possible to transfer imagettes for all PLATO targets to ground. Therefore a choice has to be made on how to distribute the resources between further imagettes and on-board processed light curves. Table \ref{tab:use-cases} shows two examples of such possible "use cases (UC)". The final choice will be made once the target fields and guest observer programs are selected.

\begin{table}[]
    \centering
    \begin{tabular}{l|c|c|c|c}
     Data product  &  UC1 & UC1 & UC2 & UC2\\ 
                                  & \#/pointing & \#/mission & \#/pointing & \#/mission\\ \hline
     24 x N-cameras/12 N-DPU             &        &        &       &        \\ \hline
     imagettes (36 pixels, 25s sampling) &  22000 &  44000 & 41300 &  82600 \\
     light curves (600s sampling)        & 147000 & 294000 & 97210 & 194420 \\
     light curves (50s sampling) & 62700 & 125400 & 62700 & 125400 \\
     Centroids (50s sampling) & 7400 & 14800 & 7400 & 14800 \\
     background (25s sampling) & 6000 & 12000 & 6000 & 12000 \\ \hline
          2 x F-cameras/2x F-DPU    & & & & \\ \hline
     imagettes (36 pixels, 2.5s sampling) & 325 & 650 & 325 & 650\\
     background (2.5s sampling) & 100 & 200 & 100 & 200 \\
     fine guidance data (2.5s sampling) & 40 & 80 & 40 & 80 \\     \hline
    \end{tabular}
    \caption{Examples (use cases, UC) of possible shares between imagettes and light curves. Numbers are given per pointing for the whole FoV and over the mission assuming 2 pointings in total (current baseline).}
    \label{tab:use-cases}
\end{table}

Let us have a look at the two examples to illustrate the capabilities of PLATO. In Table~\ref{tab:use-cases} we assume a homogeneous distribution of targets over the field of view and the maximum capabilities of the data processing system at camera level.
In UC1  22,000 imagettes with 25 s sampling rate have been allocated per pointing. With two target fields, therefore, 44000 imagettes can be obtained. 
This number has to be compared to the requirements for the mission, which include at least 16000 imagettes for P1+P2, at least 9,000 imagettes in P5 and optionally additional imagettes for the P4 sample, adding to 30,000 required imagettes. The 44000 imagettes allocated here therefore include a substantial number of additional imagettes. This use case also foresees resources for 147,000 light curves per pointing with 600 s sampling. Assuming again 2 pointings over the course of the mission, this is much more than the required P5 sample size. In addition, resources have been devoted to light curves with higher sampling rates, centroids and housekeeping data, and calibration data. 
It is therefore an example where on-board light curve processing is maximised. 

UC2 illustrates another option, aiming at maximizing the number of imagettes. In this case 41300 imagettes/pointing have been allocated. In addition, more than 97000 light curves with 600s/pointing are possible, plus the additional sampling rates and housekeeping budgets. These values are indicative and can vary by about 10\% depending, e.g., on the targets chosen. 

Table \ref{tab:use-cases} also includes a budget for the two fast cameras. We will obtain 325 imagettes with 2.5 s sampling rate per pointing from the F-CAMs, resulting in 650 imagettes of the brightest stars over the mission (assuming 2 pointings). In addition, a budget of 40 targets is maintained for fine guidance.

The final distribution and absolute numbers of imagettes and light curves observed by PLATO will depend on the distribution of stars in the field of view, the data processing prioritization chosen by the ground segment, the in-flight performance of the instrument, and the system-level margins released after final development. The numbers in Table~\ref{tab:use-cases} serve as illustration for the sizing of the design.

We note that it is in principle possible to change during science operations from imagette to light curve mode, and vice versa, for a given target. This is however subject to the operational planning cycle and will be addressed on a case by case basis depending on the resulting impacts on mission performance due to observational gaps. The same applies for changing/adding targets during science operation phases.

In conclusion, the requirements for target samples sizes recalled in the previous sub-section from PLATO´s Science Requirement document are well within the instrument telemetry and computing resources. The share between transmission of additional light curves and/or imagettes has to defined once the target stars are selected.

\section{Target Fields and Observing Strategy}
\label{sec:TargetFieldsandObsStrat}

\begin{figure}
\includegraphics[width=\linewidth]{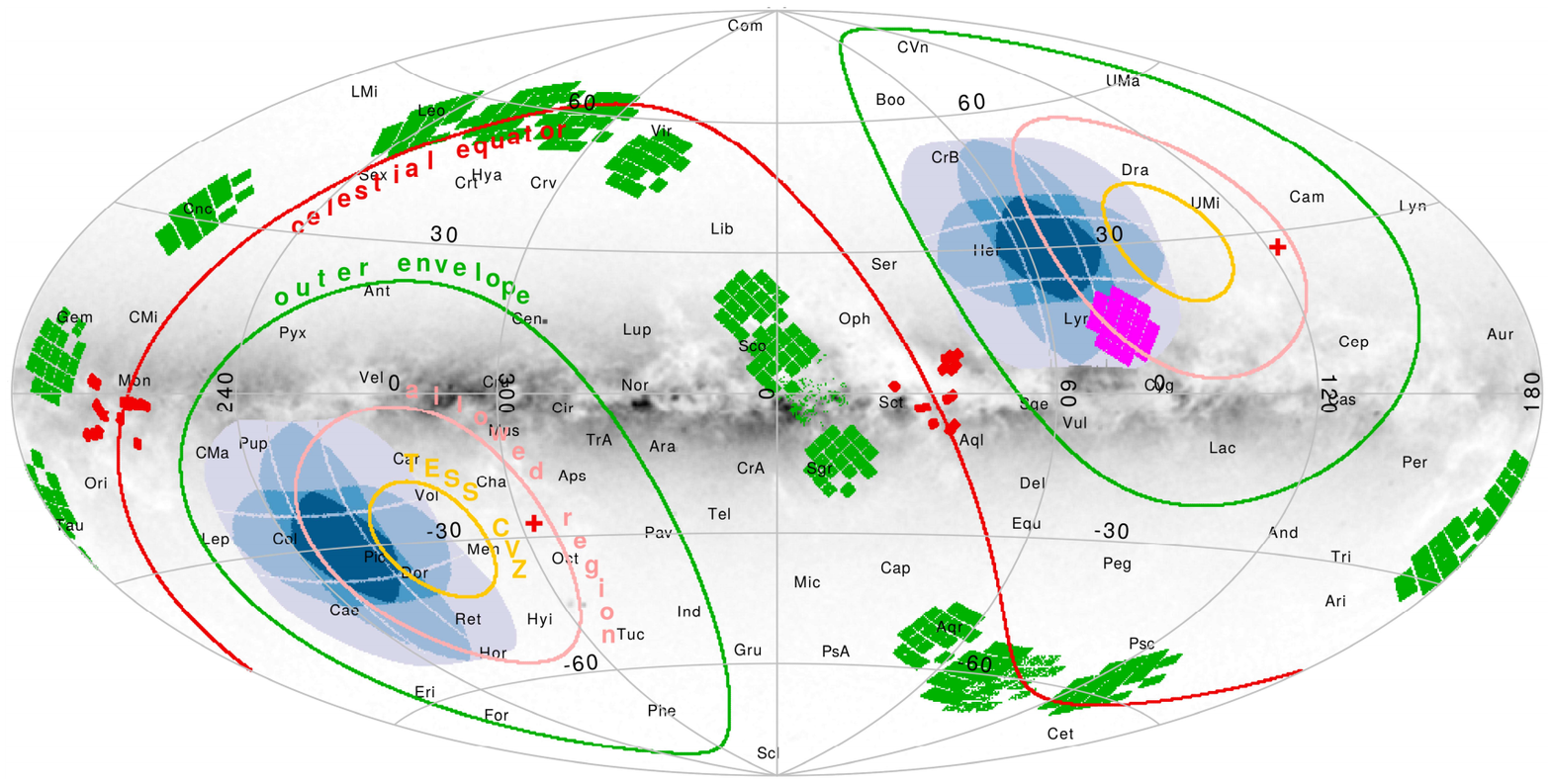}
\caption{Illustration of PLATO LOPN1 and LOPS2 pointings (blue) in comparison to {\it Kepler\/} (pink), K2 fields (green) and CoRoT mission fields (red). Lines indicate the TESS continuous viewing zone (yellow) and the technically allowed region (pink) for the PLATO field centers, respectively. See \citealt{nascimbeni2022} and Nascimbeni et al. 2024 (submitted) for details on the PLATO field selection and target populations.}
\label{fig:sky}       
\end{figure}

PLATO is a wide-field instrument whose field-of-view (FoV) covers 2132 deg$^2$ on the sky (see Section \ref{sec:payload}). For comparison, the {\it Kepler\/} mission FoV was 115 deg$^2$ \citep{koch2010a}, and the TESS satellite FoV covers 2300 deg$^2$~\citep{ricker2015}, while CoRoT covered 8 deg$^2$~\citep{auvergne2009}. For illustration purposes, Fig.~\ref{fig:sky} shows approximate PLATO field positions on the sky in comparison to {\it Kepler\/}/K2  (K2 being the extended use of the Kepler satellite \citep{howell2014PASP..126..398H}) and CoRoT mission fields.

PLATO will be launched into an orbit around the L2 Lagrangian point. The cruise and commissioning phases after launch can take up to 90 days. The duration of the nominal science operation phase of PLATO is then planned for 4 years. Extensions of the mission science operations are possible as the satellite has been designed with consumables for up to 8.5 years (see Section \ref{sec:mission}). During long observing periods, the spacecraft has to be turned by 90 degrees every 3 months for sun protection reasons (see Section \ref{sec:mission}). For the design of the spacecraft, on-ground operations and mission performance studies, a set of baseline assumptions concerning the science observing strategy have to be made. The assumed baseline observing scenario therefore splits the 4 years science observation phase into 2 observing blocks of 2 years each, so-called long-duration Observing Phases (LOP). Alternative scenarios are possible, e.g. a LOP of 3 years duration followed by shorter Step-and-stare Observations Phases (SOP) with a minimum of 2 months each. Another possible scenario would be, for example, to stare at one field only for the whole science operation phase. The spacecraft provides the technical flexibility to choose from such scenarios, and even adapt the strategy during the science operation phase if needed. 

The PLATO Input Catalogue (PIC) for the first pointing of the satellite must be defined by ESA´s PLATO SWT at the latest 2 years before launch. An update is planned for 9 months before launch, including the definition of the so-called "prime sample" (see Section \ref{sec:StellarSamples}). The subsequent target fields and their "prime sample" members are defined 6 months before the start of each field. Once this mission is completed and in case an extended mission is granted, another evaluation for key regions which would deserve dedicated PLATO pointings can be made, unless extended observations of already covered fields are given priority.

PLATO target sky regions are constrained by the mission science requirements defining the stellar samples to be observed (see Section \ref{sec:StellarSamples}) as well as technical constraints (e.g. Sun avoidance angles). In fact it turned out that due to the large field size and the requirement for long phase pointings, the possible sky target regions are constrained rather well. In a first step, stellar properties meeting the requirements of PLATO samples were investigated to produce an all-sky version of the PLATO Input Catalogue~\citep[asPIC,][]{montalto2021}. In a next step, \cite{nascimbeni2022} presented the identification of possible sky regions for PLATO´s LOP pointings, taking into account technical boundary conditions, Gaia data and simulated  signal-to-noise ratios for targets in these regions. First two provisional LOPs (LOPN1 (north) and LOPS1 (south)) were presented, one for each hemisphere  (Fig. \ref{fig:sky}).  

\begin{table}[]
    \centering
    \begin{tabular}{c|c|c} \hline
                 & LOPS2         & LOPN1 \\ \hline
$\alpha$ [deg]   & 95.31043      & 277.18023 \\
$\alpha$ [hms]   & 06:21:14.5    & 18:28:43.2 \\ \hline
$\delta$ [deg]   & -47.88693     & 52.85952 \\
$\delta$ [hms]   & -47:53:13     & 52:51:34 \\ \hline
l [deg]          & 255.9375      & 81.56250 \\
b [deg]          & -24.62432     & 24.62432 \\ \hline
$\lambda$        & 101.05940     & 287.98162 \\
$\beta$          & -71.12242     & 75.85041 \\ \hline
\end{tabular}
    \caption{
    \label{tab:TargetField}
Coordinates of two long pointing target fields of PLATO (Nascimbeni et al., in prep.). LOPS2 has been selected by the SWT as the first target field of the mission. }
\end{table}

Meanwhile the final choice of two long pointing target fields has been made (Table \ref{tab:TargetField}), see Nascimbeni et al. 2024 (submitted) for details on the selection procedure and for the target field properties in terms of stellar populations and known transiting planets covered. A slight offset of 5 degrees of the southern field in comparison to the first choice was introduced to maximize pointing efficiency, leading to an updated southern field called LOPS2.  It turns out that both fields (LOPN1 and LOPS2) meet the science requirements for the P1-P5 sample equally well. The choice of the first field was therefore made based on the availability of ground-based facilities for radial-velocity follow-up, in particular to provide masses for small planets. As a result of this assessment, the ESA PLATO SWT finally decided in June 2024 to start observations with the southern field (LOPS2). 

The decision on whether to stay longer than the 2 years foreseen baseline on LOPS2 will be made by the ESA SWT after the first in flight science data have been analysed. Whether we move to the north after 2 years or later, the pointing coordinates of LOPN1 are unlikely to change significantly, because the large field of PLATO puts hard constraints. Potential SOP target fields, for the end of the baseline or for extended mission phases, have not been studied in detail yet. We note that significant work to prepare the respective input catalogues is needed before final decisions on such fields can be taken. Therefore, preparations for SOP fields have to start well in advance before observations can start. At the time of writing this manuscript, the SWT considers it unlikely that SOPs will be part of the nominal 4 year mission baseline. They are, however, kept as an option for mission extensions.

\section{The PLATO Science Ground Segment: Data Products and Releases }
\label{sec:SGS}



\subsection{Generation of the Data Products} 
\label{sec:genDP}

\begin{figure}
\includegraphics[width=\linewidth]{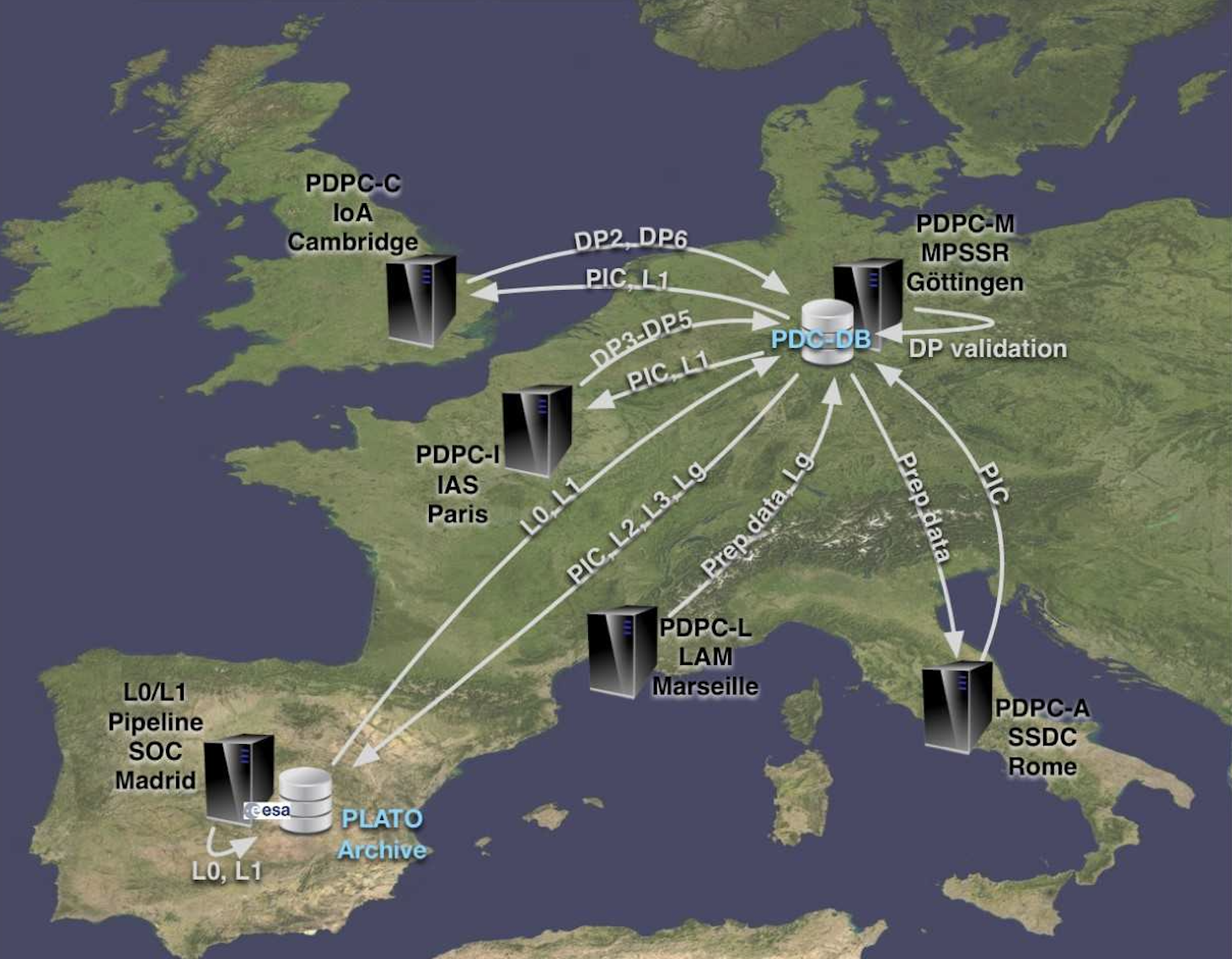}
\caption{Data flow within the Science Ground Segment for the processing of the PLATO science data products.}
\label{Fig:PDC}
\end{figure}

The PLATO Science Ground Segment (SGS) is in charge of the validation, calibration, and processing of the PLATO observations to generate the PLATO science data products. The SGS encompasses the PLATO Science Operations Centre (SOC) under ESA and the following entities within the PMC: the PLATO Data Centre (PDC), PLATO Science Management (PSM), and the PMC Calibration/Operation Team (PCOT). 

The scientific specifications for the on-ground software for generating the data products are provided by the PSM, which benefits from extensive expertise from the European scientific community. 
These specifications account for the lessons learned from CoRoT, {\it Kepler\/}, K2, CHEOPS, and TESS and build on the latest advances in stellar and exoplanet sciences. 
During operations, the PSM will scientifically review the data processing chain and provide updated scientific specifications for algorithms and tools where needed. For each target field, PSM provides the requirements for the PIC. 
 PSM also includes the Ground-Based Observing Programme (GOP), which will provide the Lg Data (ground-based follow-up) under PSM responsibility, including the observations needed for the confirmation of the planets and the radial velocities for the determination of their masses.
 The PSM will also scientifically validate the final list of planets and their characteristics. 
The PCOT  will support the calibration activities and operations, e.g. by providing input to the procedures needed for payload operation and for scientific mission planning.


The PDC responsibilities include the definition and implementation of the L1 and Calibration (CPDS) pipeline (to be run by SOC at ESAC, Spain) and the development of the software for processing the L2 and L3 data products (L3 combines L2 and Lg). See Section \ref{sec:DataProds} for a detailed description of PLATO´s L0-L3 and Lg data products. 
The generation of the data products will be performed within two main pipelines, the Exoplanet Analysis Pipeline (EAS) and the Stellar Analysis Pipelines (SAS). These pipelines do  not work independently, and several intermediate products (e.g. stellar rotation periods; transit detections) will be interchanged between them. A more detailed description of these pipelines will be provided elsewhere after finalising their design. 

The PDC will include a main database system (PDC-DB, see Fig. \ref{Fig:PDC}) that will comprise the PLATO data products, the input catalogue, and all the preparatory data on the PLATO targets that are required for the processing of the Level 2 and Level 3 data products, in particular specifically acquired ground-based follow-up data. The PDC will generate the validated PLATO input catalogue and manage the preparatory and follow-up data. Computing resources will be distributed among five Data Processing Centers: PDPC-C 
for the exoplanet analysis system, PDPC-I 
for the stellar analysis system, PDPC-A 
for the PLATO Input Catalogue, PDPC-L 
for the preparatory and follow-up database management, and PDPC-M 
for running the data analysis support tools.



The ESA SOC will be responsible to carry out the scientific mission planning based on the delivered PLATO Input Catalogue and the approved Guest Observer targets and will send the payload telecommands to the ESA Mission Operations Centre (MOC) in Darmstadt, Germany. Once executed on board the spacecraft, the resulting science data dumped to ground will be provided via the MOC to the SOC. 

The SOC will run the Level 0 pipeline to reconstruct the data as originally generated by the cameras on-board and will apply both a time correlation and a coordinate transformation to the data. The resulting Level 0 products are then fed into the Level 1 pipeline which uses calibration data provided from the CPDS pipeline. The L0 and L1 products will then be provided to the PDC for further processing to produce the L2 and L3 and Lg products. Both the SOC and the PDC will perform validation of the Level 0 and Level 1 data before releasing them to the science community via the PLATO Archive (PAX - hosted at the SOC) (see also Fig.~\ref{fig:dataRelase}).

\subsection{Data Products}
\label{sec:DataProds}

Here we explain which data products will be available to PLATO users for analysis and science. For each target object, the following will be accessible via the ESA archive center:
\begin{itemize}
    \item Level-0: Depending on target, Level-0 data include unprocessed imagettes or on-board pre-processed light curves and centroid curves. Level-0 data are downloaded for the respective targets from all cameras. The concept for the optimal aperture used for the in-flight photometric extraction is given in~\citep{marchiori2019A&A...627A..71M}. \\
    
    \item Level-1: Calibrated light curves and centroid curves produced on-ground for all targets and cameras. Level-1 data include, e.g., corrections for instrumental effects and the light curves and centroid curves derived from imagettes. In addition, for the normal cameras, camera-averaged Level-1 light curves and centroid curves are provided for each target star. For the fast cameras, no average light curves are produced, so that the colour information is preserved.  \\

    \item Level-2: First scientific data products. Concerning exoplanets, Level-2 data include the transit candidates and their parameters, planetary ephemeris of the system, depth and duration of the transit, estimated radius, and their corresponding uncertainties. For planets with detected Transit Time Variations (TTVs), the resulting list of TTV planet parameters will be provided.\\

    For the target stars, Level-2 data include the results of the asteroseismological analysis, and their corresponding uncertainties. When possible, the stellar rotation periods and stellar activity properties inferred from activity-related periodicities in the light curves are provided, as well as the seismically-determined stellar masses, radii and ages of stars, obtained from stellar model fits to the frequencies of oscillation.\\
    
    \item Level-3:  The final "PLATO Catalogue", including the list of confirmed planetary systems, fully characterised by combining information from the planetary transits, analysis of the planet-hosting stars, and the results of ground-based observations. The confirmation of the planets relies on the Lg ground-based follow-up data.\\
    
    \item Lg, ground-based observations data: These data include the results of the PLATO ground-based campaigns, e.g., observations for filtering false planet transit detections and spectroscopy to determine planetary masses.
\end{itemize}

In addition, housekeeping and auxiliary data, like, e.g., pointing information and quality control data will be available with all data product levels.\\

\subsection{Data Releases}
\label{sec:DataReleases}

\begin{figure}
    \centering
    \includegraphics[width=\linewidth]{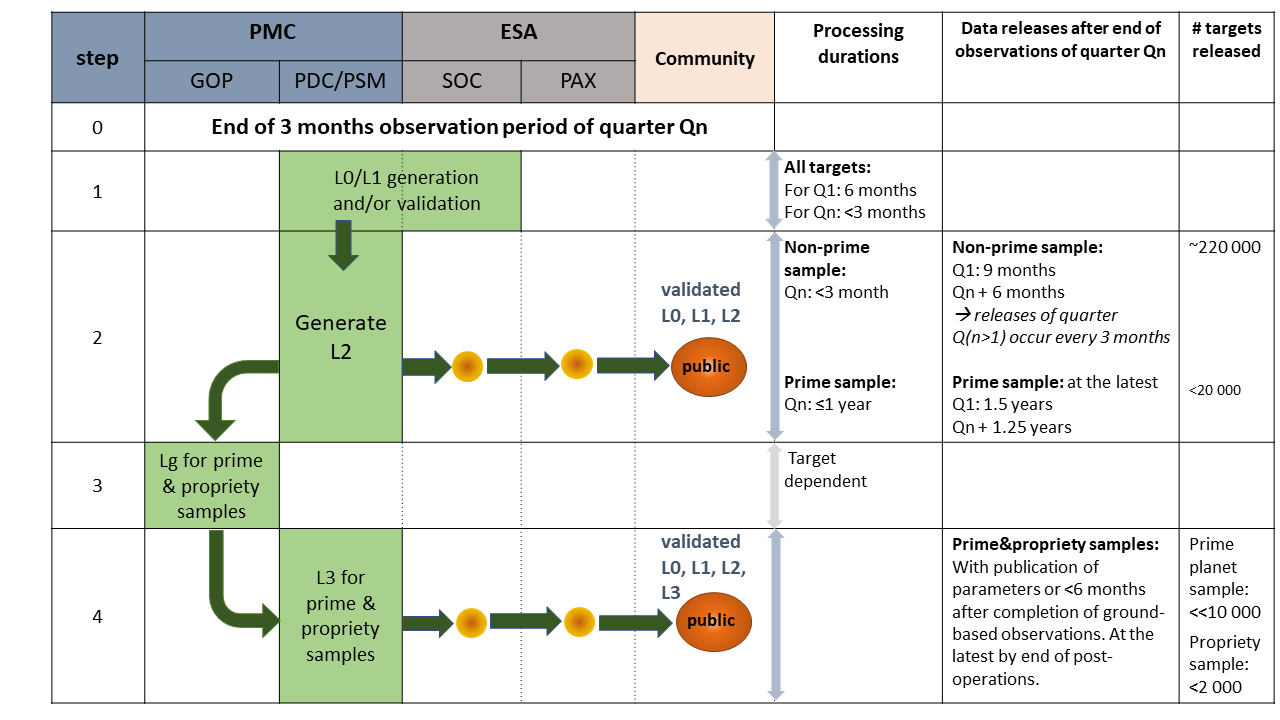}
    \caption{Illustration of the data release strategy of PLATO. Number of targets are estimates and refer to a baseline observing strategy of 2 long field pointings.}
    \label{fig:dataRelase}
\end{figure}

The PLATO data releases have been defined as a function of the product levels described in the previous section and of four observing target groups. The following target groups have been considered: (i) Prime sample (see definition in Section \ref{sec:StellarSamples});
(ii) non-Prime sample; (iii) PMC proprietary targets; (iv) targets proposed by the Guest Observers. Figure \ref{fig:dataRelase} illustrates the release of PLATO data for groups (i) to (iii) as described below. The data rights and releases corresponding to the guest observer’s programme will be outlined in Section \ref{sec:GOProgram}. 

The timeline for the releases of Level-0, Level-1, and Level-2 products has been specified with a cadence of three months. This is the time interval between two 90$\degree$ rotations of the spacecraft, which are necessary to keep the adequate orientation with respect to the Sun. After each three months data acquisition period, three months will be required for the validation of the Level-1 products by the Science Operations Centre and by the PMC. Exceptionally, six months will be needed for Level-1 product validation of the first three months of nominal mission data. This longer validation period is justified because the data pipelines are expected to undergo the most important updates, resultant from the system calibration and characterisation with the first on-board data. Note also that reprocessing of the data will take place as needed due to algorithm improvements, and at the end of each pointing when the full data set is available.
 
For the $~$20,000 prime sample targets, Level-0, Level-1 and Level-2 products for each observing quarter will be publicly released via the ESA PLATO Archive (PAX) as soon as possible after the Level-2 product scientific validation, but no later than one year after the three month Level-1 product validation period. This will allow for the consolidation of the planet candidate list and the initiation of the ground-based observations by the GOP Team(s).

For the $\approx$240,000 non-prime sample targets, Level-0, Level-1 and Level-2 products for each observing quarter will be publicly released within three months of the corresponding Level-1 product validation period.

The mission Science Management Plan establishes that the Level-3 and Lg products of the prime sample targets are a deliverable to the community. In this regard, it is required that the Level-3 and Lg data of these targets be publicly released immediately after the publication of the planetary parameters, or as soon as possible, but no later than six months after the completion of the ground-based observations. Ground-based observations data for prime sample targets that are not confirmed to be planets, will also be made publicly available in the PAX as soon as possible, but no later than six months after the ground-based observations for each target have been performed.

 The proprietary period for each proprietary target allocated to the PMC will end six months after the completion of the ground-based observations for the confirmation and characterisation of the associated planet. The proprietary period will finish in any case at the end of the mission post-operations phase.



\section{The Ground-based Observations Programme}
\label{sec:GOP}

The Ground-based Observation Programme (GOP) within the PMC plays an important role for the PLATO mission. It is in charge (for the "prime" and "proprity" samples) of organising and performing the ground-based observations needed to confirm the planetary nature of PLATO transit candidates and determine planet and stellar parameters at the required precision to reach the goals of the mission. The PMC will establish and manage the GOP team with contributions from international European partners taking part in the follow-up. The GOP Team will respond to calls issued by ground-based facilities (e.g., ESO) to grant observing time for the confirmation and characterisation of the candidates in the “prime sample” (see Sect.\,6). 

A key planet parameter for the confirmation and characterisation of planets is their mass, which can be determined by means of high-resolution spectroscopic observations providing precise radial velocities (RV)  as e.g. the HARPS or Carmenes spectrographs. For terrestrial planets in the habitable zone of their stars, a precision at the level of a few 10s of cm/s is required that can only be achieved by the new generation of high-resolution super-stable spectrographs as e.g. ESPRESSO on the ESO VLT. In addition to RV measurements, photometric, imaging and lower-resolution spectroscopy facilities will be very useful as well to discard false positive scenarios. In this context, PLATO measurements of stellar rotation and activity cycles (see Sect.\,4.6) will be useful to mitigate the impact of stellar activity on the radial velocity follow-up and eliminate false positive detection of non-transiting planets arising from activity. 

 The PLATO consortium is following a strategy first established by the CoRoT mission by setting up a coordinated approach to ground-based follow-up managed by the mission consortium in order to secure sufficient resources and efficient planning of ground-based activities. As mentioned above, this will involve extensive ground-based observations of different types and on a variety of facilities, including the largest telescopes with instruments capable of providing the most precise RV measurements. Dedicated smaller facilities taking part, e.g., in the filtering process of planet candidates will also contribute to PLATO. Observations have therefore to be scheduled in an optimal way, making a proper match between facilities and the observation needs, avoiding unnecessary duplications while ensuring a coherent use of the facilities registered in the GOP to obtain data at the required level of quality. Quality checks of the data  will be performed to ensure that only data of sufficient quality is used in the decision making process and eventually in the system modelling, before possibly conducting additional observations when needed.  Existing data in available archives will be used as well, and reprocessed when needed. The GOP data products (new or reprocessed) are delivered to the PDC where PLATO light curves are processed together with the ground-based data and where the stellar and planetary parameters are derived.

The GOP is also responsible for gathering any preparatory observations being attempted (e.g. multiplex spectroscopic observations of stars in the PLATO field). For target samples beyond the propriety and prime samples, the GOP will perform and support preparatory and follow-up observations on a best effort basis taking advantage of momentarily available unused facilities and the GOP operational infrastructure developed for the mission. It is also foreseen to use data obtained on PLATO targets by the general community  and generously provided to the PMC. 

Although the GOP organisation and subsystems are designed to work in an automatic way, human intervention is still necessary in occasional cases, in order to have a smooth flow of information and data between participating facilities and the PDC, with a proper allocation of the needed resources. Teams are foreseen to be established to address GOP-related matters at the scientific and operational levels.

The participation to the GOP activities is fully open to the astronomers from ESA member states who will actively involve themselves in the project. They usually become full consortium members with the same rights and obligations as the other PMC members (strictly following the PLATO data access and publication policies). The participation of colleagues from outside of ESA is considered on a case by case basis. In order to allow for an optimal though flexible programming and use of facilities, it will also be possible to have actual observers at the telescopes who are not PLATO members but are active in time sharing coordination of projects run on specific facilities.
Facilities taking part in the GOP are registered with the relevant associated information in a dedicated database maintained by the GOP. By facility we mean telescope+instrument (including an efficient data reduction software). The main a priori constraints they will have to fulfil are i) to be accessible by GOP members for extended periods of time, and ii) to have demonstrated their performance accuracy (through published results or benchmark tests), so they can be deployed for observations in the most efficient way.

\section{The Guest Observer Programme} 
\label{sec:GOProgram}


The PLATO Guest Observer (GO) programme will allow all interested groups and individuals to submit proposals to observe targets which are not included in the PLATO prime sample (see Section \ref{sec:DataProds}) and have science goals which are not covered by the PLATO core science objectives.  A detailed description of the scope of the programme and of its boundaries with respect to the core science will be outlined in the GO programme policies and procedures that are currently in preparation. A few examples
of complementary science cases which could be addressed via the GO programme are discussed in Section \ref{sec:ComplScience}. An average of 8\% of the science data rate (excluding calibration data), averaged over the mission lifetime, will be allocated to the GO programme. This will allow for an extended complementary science programme, while preserving the resources for an optimal observation of the core science targets. The total number of GO objects that may be observed with this allocation will range from thousands to tens of thousands, depending on the sampling times and whether imagettes or light curves are requested. ESA will appoint an independent Time Allocation Committee (TAC) for the evaluation and selection of the proposals based on scientific merit. 

GO targets would have to lie within the pre-defined PLATO sky fields. 
The first call for GO proposals will be issued by the ESA SOC nine months before launch and after the publication of the PIC and selection of the prime sample. The call will include tools which will allow potential users to predict the signal-to-noise for a range of spectral type and class of object. More calls are expected to be issued during the mission (once per year, to be confirmed). 

GO programmes can contain targets that are part of the PIC, but not of the prime sample. For targets in common with the PIC, access to the associated Level-0  and Level-1 products will be granted with the condition that the observations are exclusively used in relation with the science objectives of the proposal. Exploitation for complementary science of non-public PIC target data will only be carried out through approved GO programmes. 


GO programmes can also include on a best effort basis the observation of Targets of Opportunity (ToO). These may contain objects which can be identified in advance but which undergo unpredictable changes (e.g., recurrent novae), as well as objects that can only be identified in advance as a class (e.g., novae, supernovae, tidal disruption events, gamma ray bursts). The identification of ToOs shall come from the GO programme. For ToOs
that cannot be anticipated, proposals may be submitted at any time during the mission. The TAC will prioritise ToOs with respect to on-going GO programmes, to permit the interruption of lower priority targets observations if required by the ToO. ToOs can be observed with the constraint that the reaction time between triggering the ToO and the start of the observation will be in line with the ongoing mission planning cycle, with execution of ToOs being performed on a best effort basis. The baseline planning cycle during nominal mission foresees interaction once per month.

Level-0 and Level-1 products for all PLATO observed targets, including the GO programme observations, will be generated by the ESA SOC. The proprietary period of the targets selected through the GO programme call will be one year, starting at the time of the delivery to the Guest Observer of the last portion of the relative Level-1 data. During the execution of the observations, the SOC will deliver Level-0 and Level-1 products to the PI of the GO programme observer every three months. ESA will require the GOs to provide their data and results, to make them accessible through the PLATO Archive.

\section{Expected Instrument Performance} 
\label{sec:InstrPerformance}

 The design of the PLATO mission is driven by the need to obtain a core sample which is dominated by stellar flux white noise and where instrumental effects play only a minor role. 
 Although the nominal mission operation time is 4.25 years at this point, the payload is designed to maintain a level of performance that allows the science objectives to be achieved up to an extended mission end-of-life (EOL) of 6.5 years. 
 The EOL scenario assumes that up to two cameras could be lost during the mission due to technical failures (i.e. 22 operating N-CAMs instead of the 24 N-CAMs available beginning-of-life, BOL). 
 This number results from a reliability analysis performed during the design phase.
 Along the text, we are using the EOL as conservative worst-case scenario for mission performance, unless stated differently.
 For example, the targets of the PLATO Input Catalogue are chosen according to their BOL performance. 
 This is the most informative choice for the scientific community and representative of the performance that will be found at the beginning of the mission.
 However, the planet yield and stellar characterization performances are computed with EOL performance, to make sure that the scientific goals of the project are reached in a realistic worst-case scenario.

 \begin{figure}
     \centering
     \includegraphics[width=\linewidth]{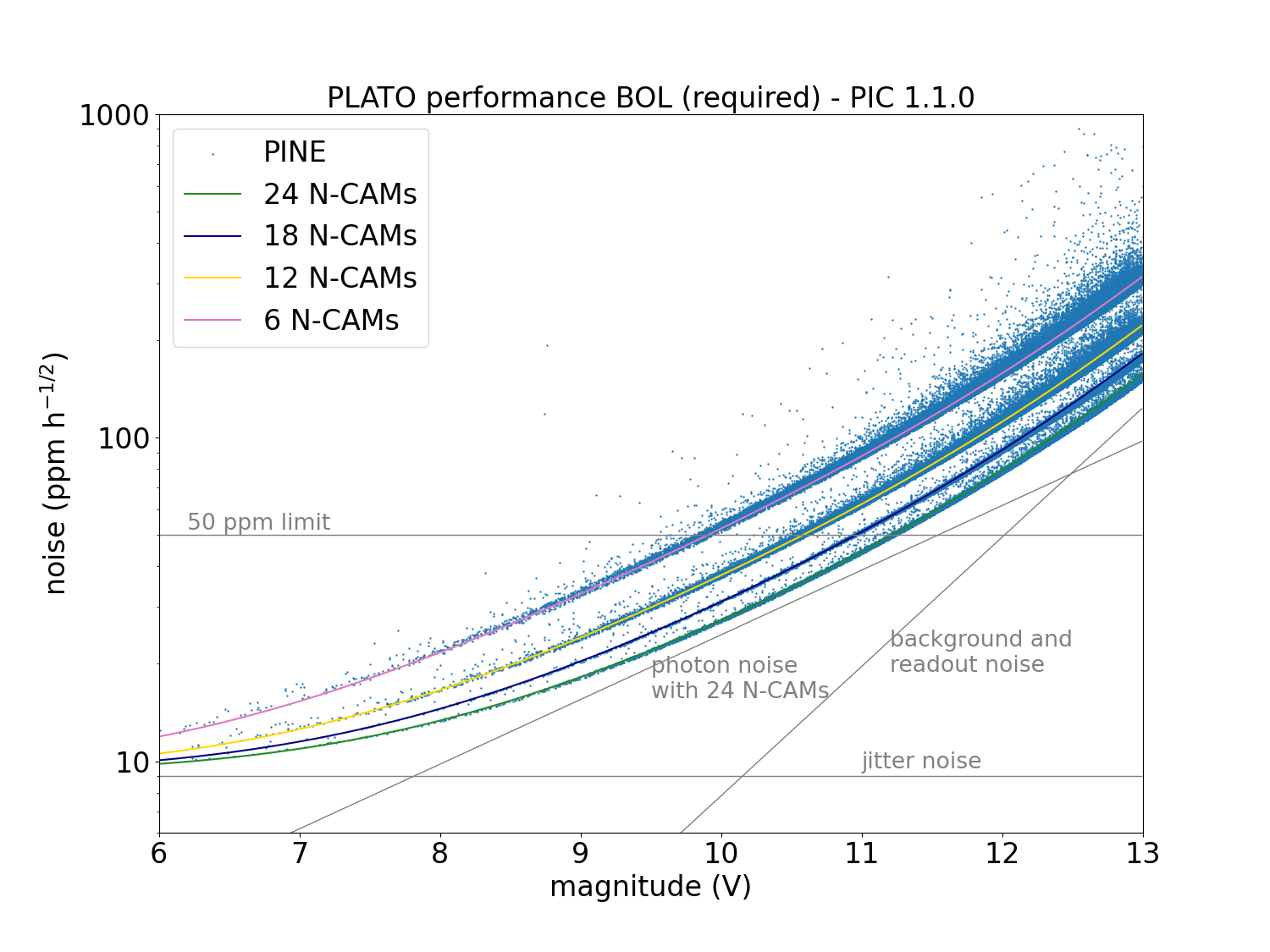}
     \caption{Expected PLATO noise performance calculated with the PLATO Instrument Noise Estimator~\citep[PINE,][]{boerner2024}. The values refer to beginning of life scenario (BOL, i.e. 24 N-CAMs, see Section~\ref{sec:InstrPerformance}). The stellar counts and properties are taken from the PIC 1.1.0. A simple noise model (including jitter, photon noise, and readout and background noise) provides a good approximation to the expected in-flight performance of the mission. The 50 ppm level is the limiting value for the P1 and P2 samples.
      For stars 8 < V < 12, the noise is dominated by photon noise. }
     \label{fig:noise-magnitude}
 \end{figure}

In order to derive realistic instrument performance estimates, a full simulation of the PLATO payload and satellite behaviour has to be made. A number of simulators have been developed for this purpose each focusing on various aspects of the mission. The consortium’s primary simulator is PlatoSim~\citep{jannsen2024A&A...681A..18J}. 
It is an end-to-end camera simulator at pixel level taking into account detailed instrumental noise properties as well as realistic stellar variability, producing time series of imagettes as well as light curves. Complementary to PlatoSim,  a light curve simulator~\citep{samadi2019} has been developed, which models the variability in the Fourier domain, and then converts it to a light curve.

  At the time of writing this paper, instrument requirements for PLATO have been settled and a number of simulations of the expected instrument performance have been performed. Figure \ref{fig:noise-magnitude} illustrates the importance of different instrumental noise sources for PLATO (see ~\citet{boerner2024}). Throughout the main target magnitude range of PLATO (8 mag to 11 mag) photon noise dominates the signal over instrumental effects. At the very bright end, jitter noise from the satellite dominates. Below about 12.5 mag electronic read-out noise starts to dominate.

Stars brighter than magnitude 8 will saturate with the N-CAMs (brighter than 4.5 with the F-CAMs). 
However, it shall be possible to extract precise photometry from saturated targets. 
For slightly saturated stars (around magnitude 8) no extension of the nominal window of 6x6 pixels will be necessary.
For moderately saturated targets (between magnitudes 4 and 8) saturated star descriptors are needed with the N-CAMs while smearing photometry will be possible with the F-CAMs.
Finally, for the few highly saturated targets (expected 10 per camera) smearing photometry is technically possible, but alternatives are being considered~\citep[e.g.][]{white2017MNRAS.471.2882W}.

\begin{figure}
    \centering
    \includegraphics[width=\linewidth]{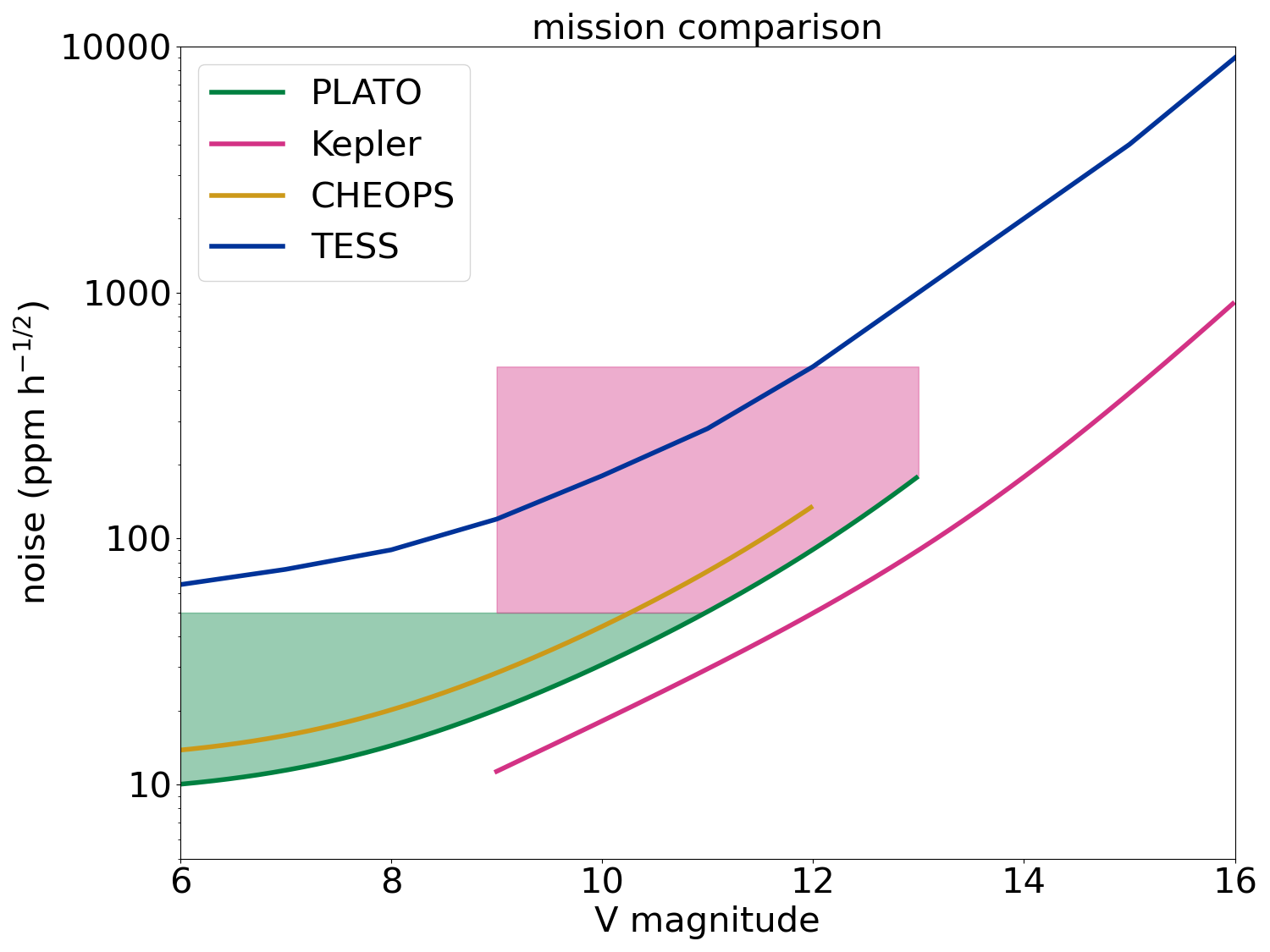}
    \caption{Comparison of the noise levels expected for PLATO with other missions. 
    The approximate parameter space for PLATO samples (green: P1, red: P5) is indicated for illustrative purposes. 
    For PLATO we assume 24 N-CAMs and BOL performance (see Section~\ref{sec:InstrPerformance}).
    The values for {\it Kepler} are taken from~\citet{keplerInstrumentHandbook}, 
    the values for TESS from~\citet{sullivan2015}~\citep[but see also][]{kunimoto2022}, 
    and the values for CHEOPS from~\citet{2024A&A...687A.302F}. 
    They lines shown should be understood as reasonable approximations of the performance of each instrument and not used as reliable estimators for an individual target. Please, refer to the different publications where the values are extracted from.}
    \label{fig:mission-comp}
\end{figure}

The noise-to-signal simulator PINE (~\citet{boerner2024}) is using the current best knowledge of the instrument design. PINE values were used, e.g., for the development of the PIC. The detailed simulation of stellar noise-to-signal ratios (NSR)s across the PLATO FoV is shown in Fig.~\ref{fig:noise-magnitude} versus stellar magnitude. 
As can be seen, the required  P1 noise performance of 50 ppm in 1 hour is met in the center of the F0V (24 cameras) for stars brighter than 11 mag. 
At the corners (6 cameras) the same performance is reached for stars brighter than 10 mag. 
Overall, when applied to real sky fields, the noise performance of PLATO is sufficient to provide the required number of stars in the respective P1 to P5 samples.

Of course, the particular design on PLATO with 24 overlapping cameras asks for a more detailed simulation taking into account the real geometry of the PLATO FoV. The PLATO stellar noise budget depends not only on the position of target stars in the field, hence the number of cameras, but also on the variations of the point-spread-function (PSF) across the FoV of each camera. These effects will be considered in future studies.

To put PLATO into perspective with other transit detection missions, Figure \ref{fig:mission-comp} shows the expected noise level of PLATO with respect to CHEOPS, TESS and {\it Kepler\/}/K2. For illustration purposes the main planet detection ranges for PLATO P1 and P5 samples are indicated as well as the main planet detection magnitude range of {\it Kepler\/}/K2. PLATO´s sensitivity is better than TESS and CHEOPS, but below {\it Kepler\/}/K2 as expected from the respective aperture sizes of these missions. However, PLATO will sample on average brighter stars than {\it Kepler\/}/K2 and will hence detect the majority of its planets around brighter stars, thereby facilitating asteroseismology and ground-based follow-up observations.

\section{PLATO spacecraft and mission configuration} 
\label{sec:mission}

The requirements that drive the design of the PLATO spacecraft and the mission configuration are the transit detection and characterisation of small exoplanets in long-period orbits ($\approx$1 year) around solar like stars, and the observation of stellar oscillations in planet host stars. To achieve these science goals, PLATO will perform high-precision photometric observations for periods of 2 to 4 years, that should be as much as possible undisturbed by the space environment and the platform operations. Consequently, maintenance of the thermal and pointing stabilities, while keeping the interruptions to a minimum, are key aspects of the spacecraft design and of the mission configuration. 

\subsection{PLATO spacecraft}
\begin{figure}
\includegraphics[width=11cm]{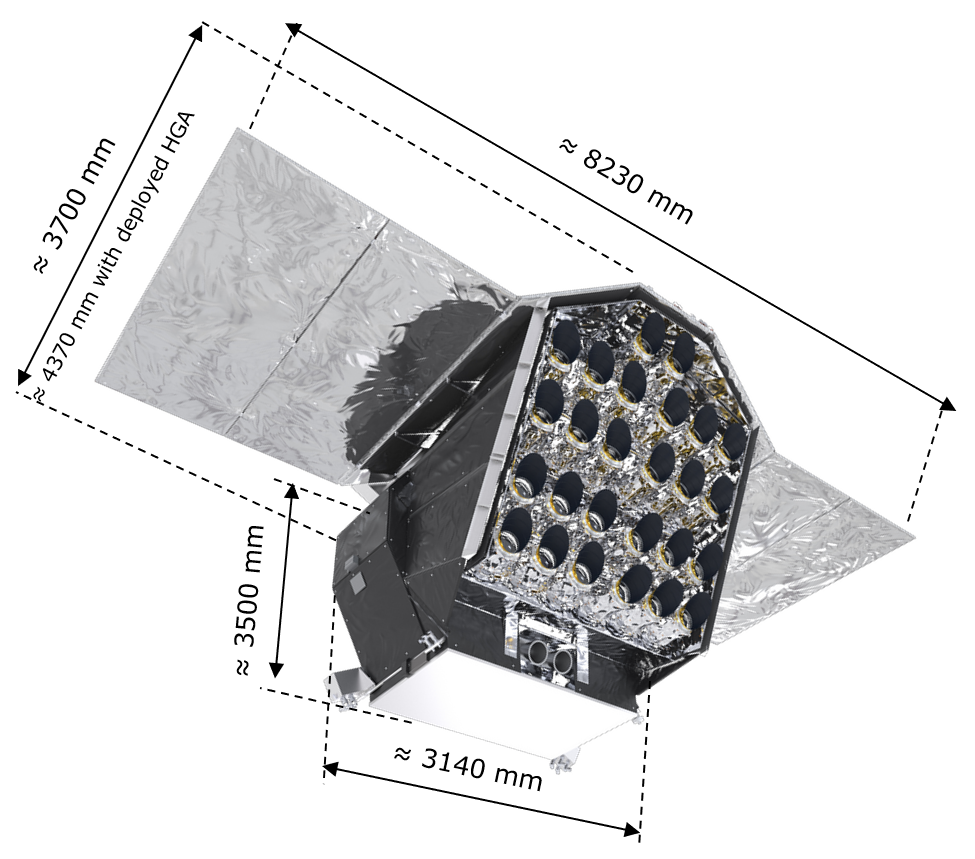}  
\caption{Artist impression of the PLATO Spacecraft with its dimensions. Credits: ESA/ATG Medialab.}
\label{fig:satellite}       
\end{figure}

The spacecraft is a 3-axis stabilised system with a launch mass of approximately 2500 kg, including consumables, a size of about 3.5 m (x) × 3.6 m (y) × 3.7 m (z) in stowed configuration, and a deployed wingspan of approximately 9 m (as shown in Figure \ref{fig:satellite}). The spacecraft accommodates a payload of 26 cameras and its design ensures an adequate protection from exposure to the Sun.

Based on a modular architecture design, the spacecraft consists of a Payload Module and a Service Module. The Payload Module, which accommodates the cameras, is thermally and mechanically decoupled as much as possible from the rest of the spacecraft by means of a truss structure with flex joints. The module is mostly made of carbon fibre to increase stiffness and reduce thermo-elastic deformations. The Payload Module contains the Optical Bench where the 24 Normal Cameras and the two Fast Cameras are accommodated. It is integrated with the Service Module via three bipods. Additional thermal decoupling is achieved by wrapping the Optical Bench in multi-layer insulator towards the sunshield/solar array and the Service Module. This design leads to a very stable environment by clearly separating the scientific payload from housekeeping activities and other variable influences.

The Service Module (see Fig.~\ref{fig:satellite2}) contains all the systems necessary to operate the spacecraft in the designated orbit, such as, shielding, power, propulsion, attitude control, thermal control, communication, commanding, and data management. The structure, based on a carbon fibre central tube and shear panels, provides the interface to the launcher as well as a stiff base for the payload module. The Service Module electronic units are accommodated on the bottom panels that act as radiators. A sunshield protects the cameras from the Sun, while guaranteeing an unobstructed view to space. It also carries three solar panels mounted on the spacecraft body and four deployable panels on two wings to provide the 3000 W of required power to the satellite. The satellite attitude control system is based on reaction wheels, thrusters, gyros, Sun sensors and star trackers.  The thermal control system is based on heaters, radiators, and multi-layer insulation that provide the required temperatures and thermal stability to all units. This system facilitates a radiative environment to cool down the front-end-electronics units and provides the conditions to keep the low telescope temperatures. The antenna sub-system of the spacecraft includes two fixed low gain antennas operating in X-band that guarantee hemispherical coverage when the spacecraft inertial attitude is not available from the star trackers. They are complemented with a steerable low-gain antenna (as part of the high gain antenna assembly (HGAA)) for the high X-band data rates. The downloading of the science data and the main satellite commanding is carried out via a dual X/K-band high gain antenna assembly. This enables communication for satellite control using the X-band, while the downloading to the ground of the daily 435 Gb of data uses the K-band with a data rate of up to 72 Mbps. 

The PLATO spacecraft is designed and built by ESA, having as Prime contractor an industrial core team comprising OHB, ThalesAlenia and Beyond Gravity, that lead a pool of European companies.

\subsection{Launch and operations}
\label{sec:launch}
PLATO is planned to be launched with an Ariane 62 rocket by the end of 2026. Since the PLATO payload's line of sight is up-oriented towards the zenith, and since blinding and illumination of the cameras by the Sun must be avoided, the strategy selected for launch and transfer includes an intermediate circular parking Low-Earth orbit. In addition, the upper stage of Ariane 62 provides a barbeque mode, such that, it can rotate around its longitudinal axis during its coasting phase, avoiding illumination of the payload. By adopting this strategy, launch is possible all year round with some exclusion windows to avoid Moon and Earth eclipses during transfer. PLATO will be injected from the parking orbit into an eclipse-free Lissajous orbit around the Earth-Sun Lagrangian point 2 (L2), located $\approx$1.5 million km from Earth. During the transfer phase, which will take about 30 days, the commissioning phase will start and will run until the check-out and calibration of the spacecraft and its payload are completed, at the latest three months after launch. At that time the routine science operations will begin with a nominal duration of 4 years. Extension of the scientific operations will be possible as the satellite is being built and verified for an in-orbit lifetime of 6.5 years and will accommodate consumables for at least 8.5 years. 

The shape of the halo orbit around L2 depends on the exact date and time of launch. When PLATO is orbiting L2, the payload will be protected from the solar light by reorienting the sunshield every three months with a 90° spacecraft rotation around its line-of-sight. The orbit around L2 is maintained by regular station-keeping manoeuvres planned by the Mission Operations Centre (MOC) approximately every 28 days. During science operations, a communication session with the nominal ground station will be established several days per week. The scheduling of mission operations is strongly constrained by the required science duty cycle, which must be above 93\%. This is critical to minimise the probability of missing transits of long period planets, and for the detectability of stellar oscillation modes. Furthermore, periodic observation gaps that could generate disturbing peaks in the frequencies of interest in the power spectrum of star oscillations must be avoided.

\section{Payload design overview} 
\label{sec:payload}


\subsection{General payload description}
\label{sec:generalPayloadDescription}

The PLATO payload has two major science drivers constraining its design. Compared to previous missions like {\it Kepler\/}, PLATO has been optimised to observe stars bright enough to allow radial velocity measurements on ground to complement the photometric measurements in space. Since such bright stars are relatively scarce on the sky, a very large total field of view of 2132 deg$^2$ was required. Furthermore, to be able to observe in a given target field the brightest as well as fainter stars, the instrument photon collecting area is split into 26 individual cameras. The 24 cameras which are used exclusively for science operation are termed "normal" cameras whereas the two additional cameras, used also for fine-pointing on top of science operation, are termed "fast" cameras.
With this configuration PLATO will be able to obtain very stable, high precision, long-time photometric data of a large number of bright stars. Combining the information provided by the planetary transits, with the stellar data derived by asteroseismology techniques and the radial velocity measurements obtained from the ground, PLATO will provide a high-precision characterisation of the radius, mass and age of exoplanetary systems.

\begin{figure}
\includegraphics[width=\linewidth]{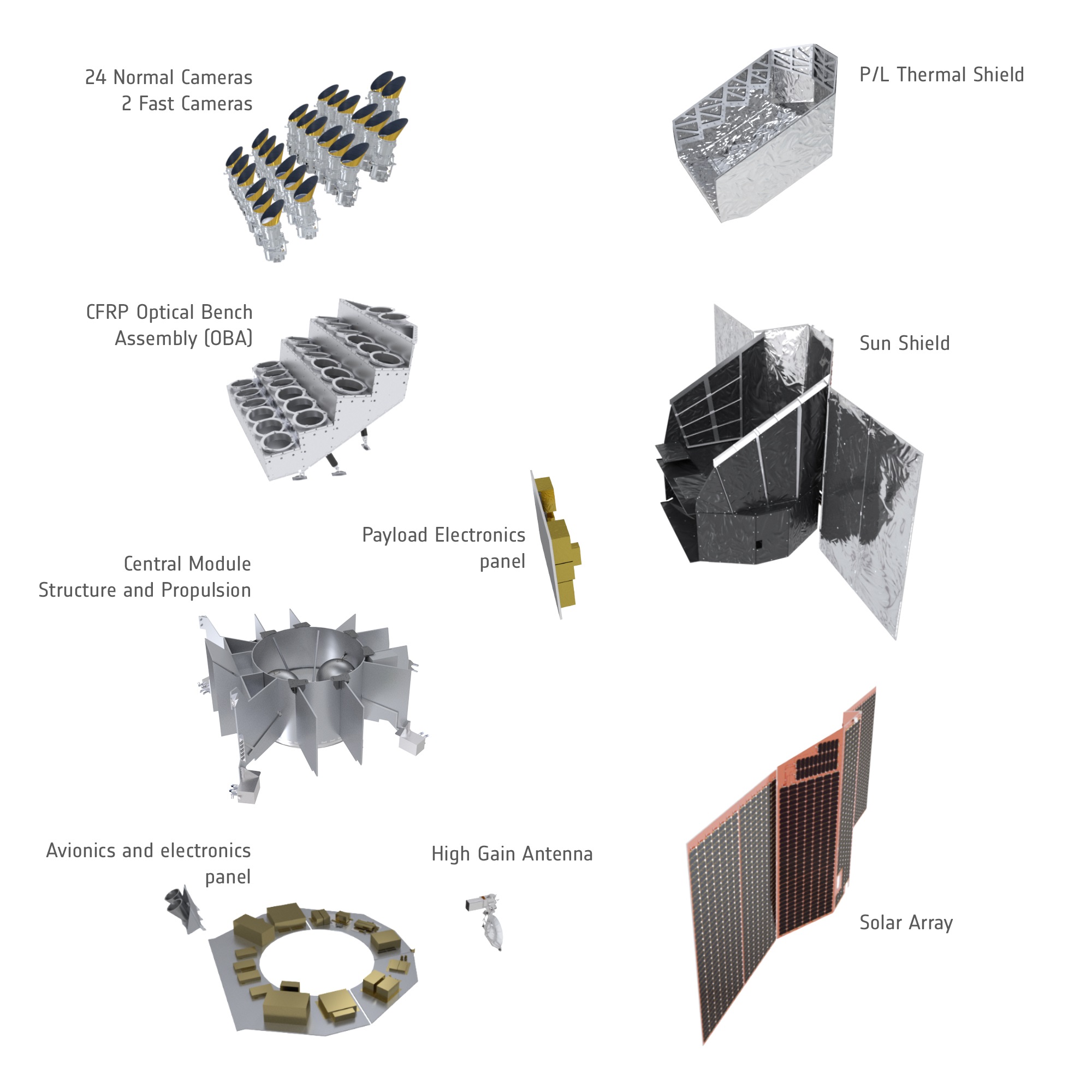}  
\caption{Expanded artist impression of the PLATO Spacecraft showing its different modules and elements, with its 26 cameras seating on the Payload Module, and the Data Processing Units fixed below directly on the radiator. Credits: ESA/ATG Medialab.}
\label{fig:satellite2}       
\end{figure}

The data read by the CCDs of each camera are first processed by a Data Processing Unit (DPU) handling two cameras at the same time. The processed data are then transferred to the Instrument Control Unit (ICU) that finally compresses them before sending them to the spacecraft on-board mass memory for downloading to ground. The different sub-systems composing the PLATO spacecraft and payload can be seen in Fig.~\ref{fig:satellite2}.

\begin{figure}
\includegraphics[width=5.5cm]{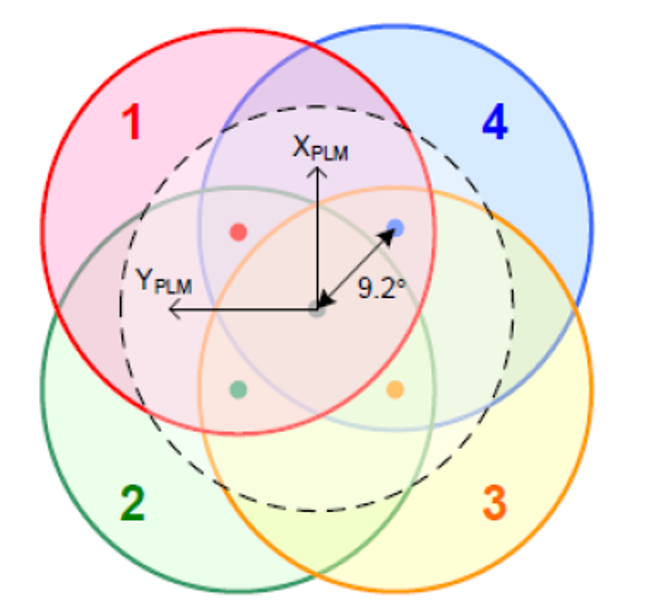}    
\includegraphics[width=5.5cm]{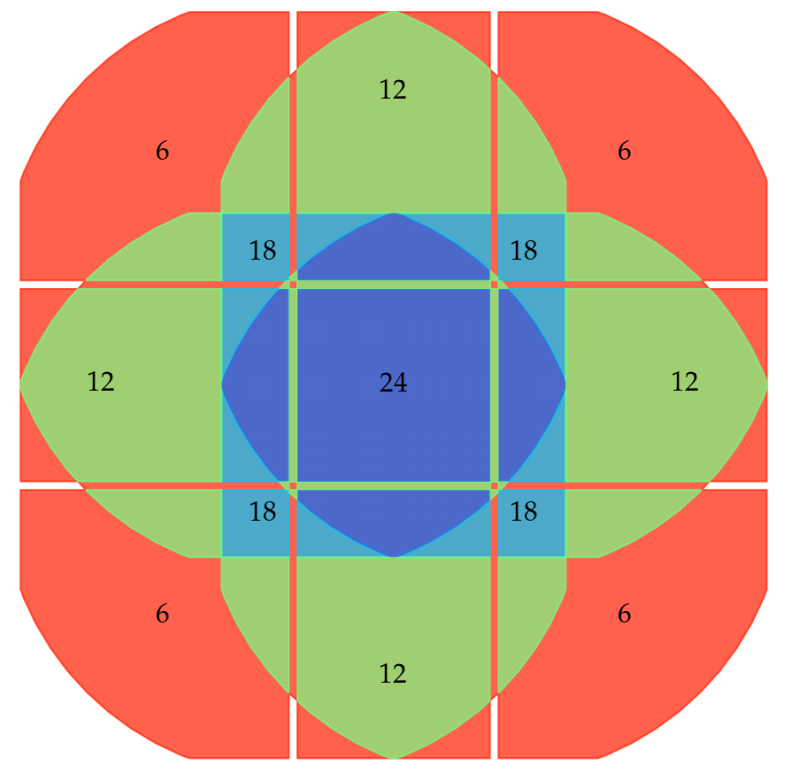}

\caption{Illustration of the PLATO field coverage. Left: position on the sky of the optical axis of the four groups of 6 coaligned "normal" cameras, and of the two "fast" cameras (dashed black line). Right: Number of "normal" cameras observing simultaneously the different sectors of the overall PLATO field of view. The center of the field (dark blue) is observed by all 24 cameras. At the edges (orange) only 6 cameras observe simultaneously the same field on the sky.}
\label{fig:FoV}       
\end{figure}

\begin{figure}
  \centering
\includegraphics[width=10cm]{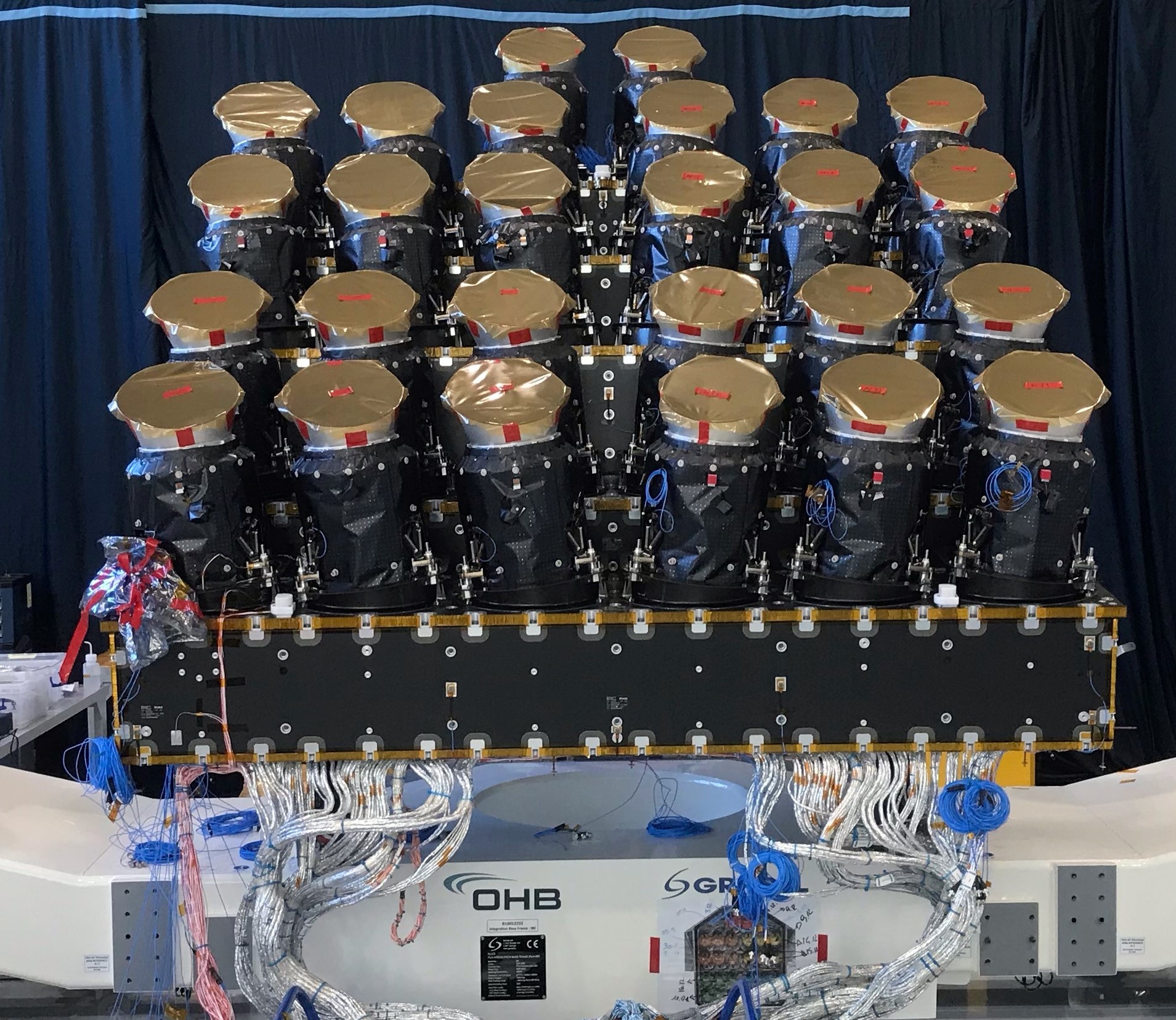} 
\caption{The PLATO Payload Module Structural and Thermal Model (STM) during integration at the OHB facilities. Note the relative pointing of each of the 4 groups of 6 normal cameras, and the two fast cameras located at the top of the PLM. Credit: OHB AG. }
\label{fig:PLM_STM}       
\end{figure}

Table \ref{tab:instrument} summarises the main data of the instrument. Each 20~cm class camera (12~cm entrance pupil) has a total field of view (FoV) of 1037 square degrees ~\citep[see][]{pertenais21b}. The 24 normal cameras are organised on the optical bench of the spacecraft in 4 groups of 6 co-aligned cameras each. The sky fields of the four camera groups are offset with respect to a central line of sight by 9.2 degrees, pointing towards the four corners of a square of 13.0 deg, in order to widen the total overlapping FoV of the instrument. With this approach the number of brightest stars within SNR requirements is augmented with respect to a full coalignment of the whole set of cameras. Figure \ref{fig:FoV} illustrates how many cameras observe the same part of the target field for a pointing. The total instrument field therefore is about 2132 deg² (with rounded edges). Figure \ref{fig:PLM_STM} shows the Structural and Thermal Model (STM) of the payload during integration at OHB, ready for the qualification activities of the Payload Module. 

The two fast cameras are optimised to observe the brightest stars (4 - 8.2 mag) and are part of the fine-pointing loop of the satellite. They point towards the center of the PLATO payload field of view.
These two cameras provide full redundancy to secure the fine-pointing capabilities of the mission. They differ however in their entrance filter, one with a red and one with a blue spectral filter allowing for colour information on the bright targets. Thanks to a very fast readout of 2.5~s, they provide very high quality data for the Fine Guidance System (FGS) algorithm as shown in  \cite{griessbach21}. This algorithm allows the satellite to reach the excellent pointing stability required for the different science cases detailed above: 

\begin{itemize}
    \item FGS noise equivalent angle:  $<$ 0.025 arcsec around X and Y, and 0.1 arcsec around Z FGS measurement reference frame axes.
    \item FGS measurement bias stability: $<$ 0.010 arcsec around X and Y, and 0.040 arcsec around Z FGS measurement reference frame axes.
\end{itemize}

\begin{table}
\caption{PLATO Instrument parameters.}
\label{tab:instrument}
\begin{tabular}{ wl{3.7cm} wl{7.5cm}}
\hline\noalign{\smallskip}
PLATO Instrument &    \\
\noalign{\smallskip}\hline\noalign{\smallskip}
Total FoV                & 2132 deg² \\
FoV by 24 CAMs          & 294 deg² \\
FoV by 18 CAMs          & 171 deg² \\
FoV by 12 CAMs          & 796 deg² \\
FoV by 6 CAMs           & 871 deg² \\
Number of cameras      & 26 total: 24 normal, 2 fast cameras\\
Normal camera groups       & 4, with 6 normal cameras each\\
Angle between camera groups & Arranged on four corners of a 13.0 degrees side square \\
PDE & 3 arcsec HCA over 3 months in the directional direction\\
    & 6 arcsec HCA over 3 months in the rotational direction \\
PRE & 3 arcsec HCA in the directional direction \\
    & 6 arcsec HCA in the rotational direction \\
RPE & 1 arcsec HCA  over 25s in the directional direction\\
    & 2 arcsec HCA over 25s in the rotational direction \\
Photometric range   & 4 - 16 magV \\
Instrument mass     & $<$ 623.15 kg \\
Peak power consumption     & $<$ 890~W\\
Daily data generation       & 435 Gbit/day \\
Normal electronic chain     & 12  DPUs: 1 for 2 normal cameras \\
                            & 2 normal AEUs: 1 for 12 normal cameras \\
Fast electronic chain       & 2 fast-DPUs: 1 per fast camera   \\
                            & 1 fast AEU: 1 for both fast cameras \\
\noalign{\smallskip}\hline
\end{tabular}
PDE: Attitude Pointing Drift Error; PRE: Attitude Pointing Repeatability Error; RPE: Relative Attitude Pointing  Error.\\
FoV: Total Field of View; HCA: half cone angle; DPU: Data Processing Unit; AEU: Ancillary Electronic Unit.
\end{table}


\subsection{PLATO Cameras}
\label{sec:cameras}

\begin{figure}
\includegraphics[width=7cm]{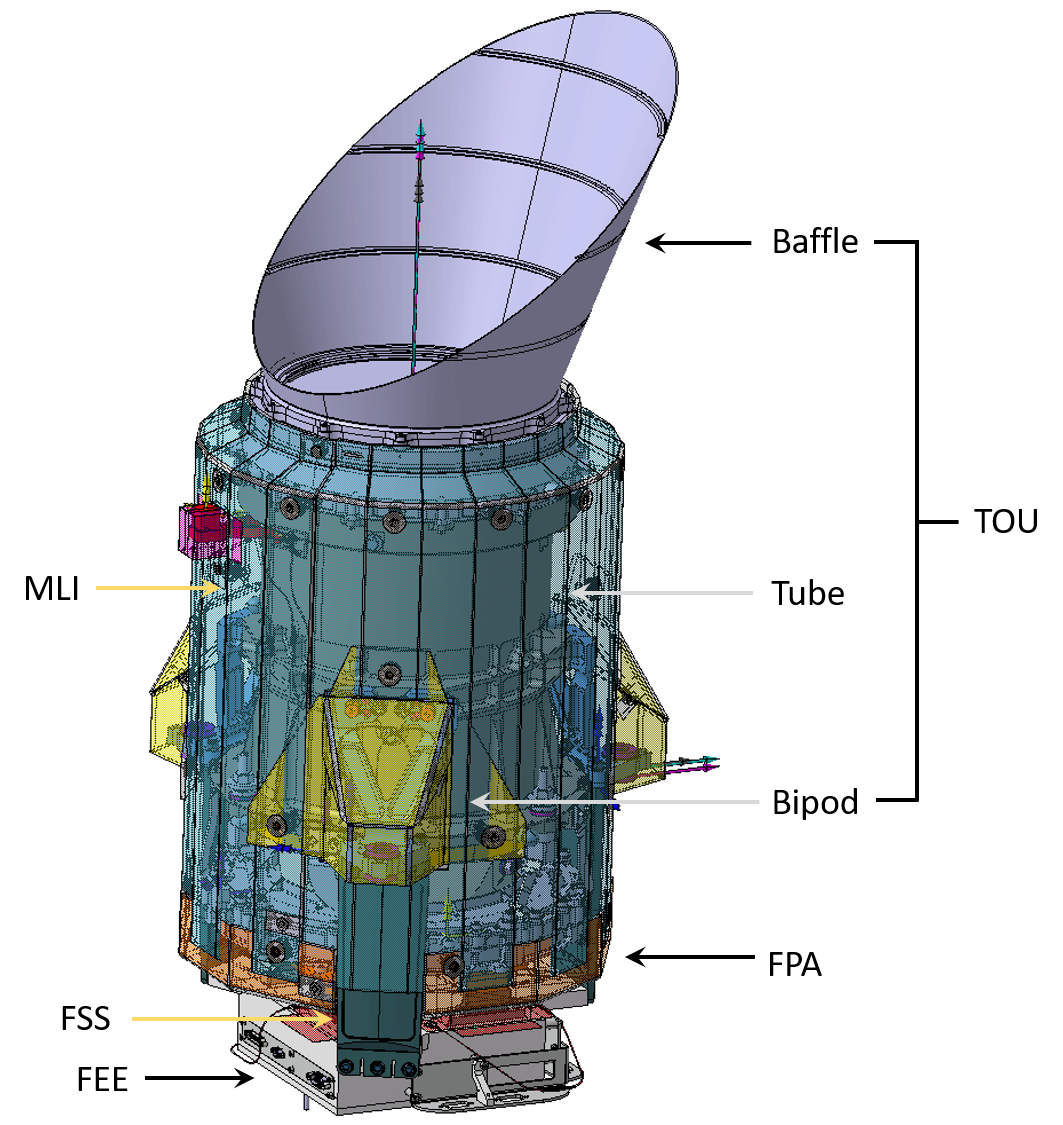}
\includegraphics[width=5cm]{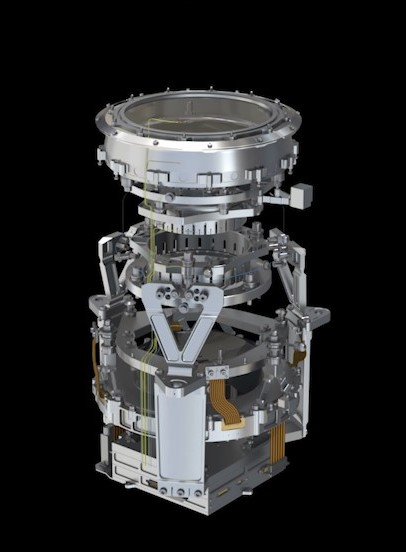}
\caption{CAD views of Camera with all its sub-systems. Left: Complete camera; right: Internal structure (credits: ESA/ATG Medialab).  }
\label{fig:CAM}       
\end{figure}

All the cameras are externally identical (except for the baffles) and their interfaces to the spacecraft optical bench are the same, with three main bipods, see Fig.~\ref{fig:CAM}. Mechanically, their central part is the tube of the Telescope Optical Unit (TOU), made of AlBeMet. The six lenses of the TOU are all mounted inside this tube. This tube is also the support for the baffle that acts both as straylight baffle and as a radiator for the whole camera. While the Focal Plane Assembly (FPA) with the CCDs is fixed to the TOU tube by three bipods, the Front End Electronics of the Normal Cameras (N-FEE) are mounted directly to the main interfaces of the cameras to the optical bench (the main bipods) with their own Support Structure (FSS). This allows for conductive coupling to evacuate the high-power dissipation. The Fast Front End Electronic units (F-FEE), much heavier and with a higher thermal dissipation, are however directly mounted to the optical bench of the spacecraft and have no thermo-mechanical links to the cameras (except for the CCD flexi connections).

Thermally, each camera is individually controlled by a Thermal Control System (TCS) that sets its temperature thanks to three heaters placed on the tube of the TOU. Using a Proportional-Integral (PI) controller algorithm, the temperature of each camera can be held very stable, up to $\pm10$mK over 14~hours using 600~s sliding window averages. Thanks to the athermal design of the camera and the AlBeMet tube properties, this temperature stability around the heaters is very well propagated to all the camera sub-systems and especially the CCDs. The heat dissipated by the CCDs is conductively transferred directly to the baffle that can then dissipate it into cold space. This TCS is also used to fine tune the focus of each camera in space and ensure that each camera is operated in its best possible focus position~\citep{pertenais21a,clermont}. By changing the temperature of the camera, the characteristics of the optical system are changing leading to a slight shift of the focus. This thermal refocusing is used on-ground and then during commissioning to calibrate each camera individually to its best focus temperature. Testing under representative thermal vacuum conditions on 18 flight units up to now has shown that the best focus will be between $-77\,\degree$C and $-86\,\degree$C, with an average value around $-80\,\degree$C, well within the operational limits between  $-70\,\degree$C and $-90\,\degree$C.
The precise value for each camera will be determined during the commissioning phase once in space.


The optical design of each camera \citep{magrin2010} consists of a six lens system with a central Calcium Fluoride lens \citep{2012SPIE.8442E..49B} close to the stop, and with a first aspheric lens protected, for thermal and radiation hardness purposes, by a Suprasil window. All the optical materials have been the subject of specific radhard studies \citep{2018OExpr..2633841C} and the whole design is a fine tuning balance between radiation, thermal and mass budgets \citep{2016SPIE.9904E..30M}.
To obtain more information on the current status of the point spread function analysis and the camera focusing activities, the interesting reader can refer to~\citep{2022SPIE12180E..1DB} and~\citep{2022SPIE12180E..4NP}.

\begin{figure}
\includegraphics[width=6cm]{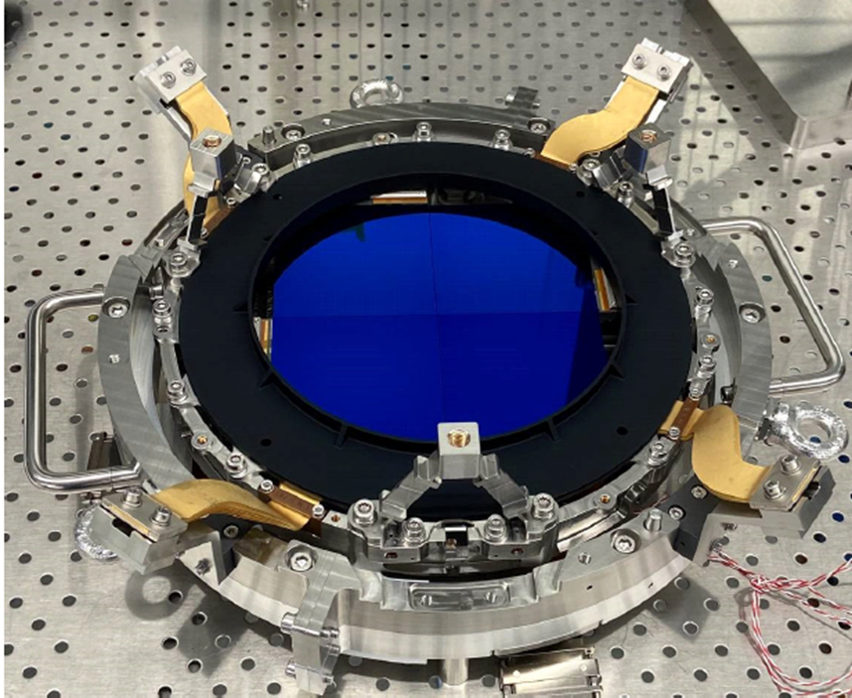}
\includegraphics[width=6cm]{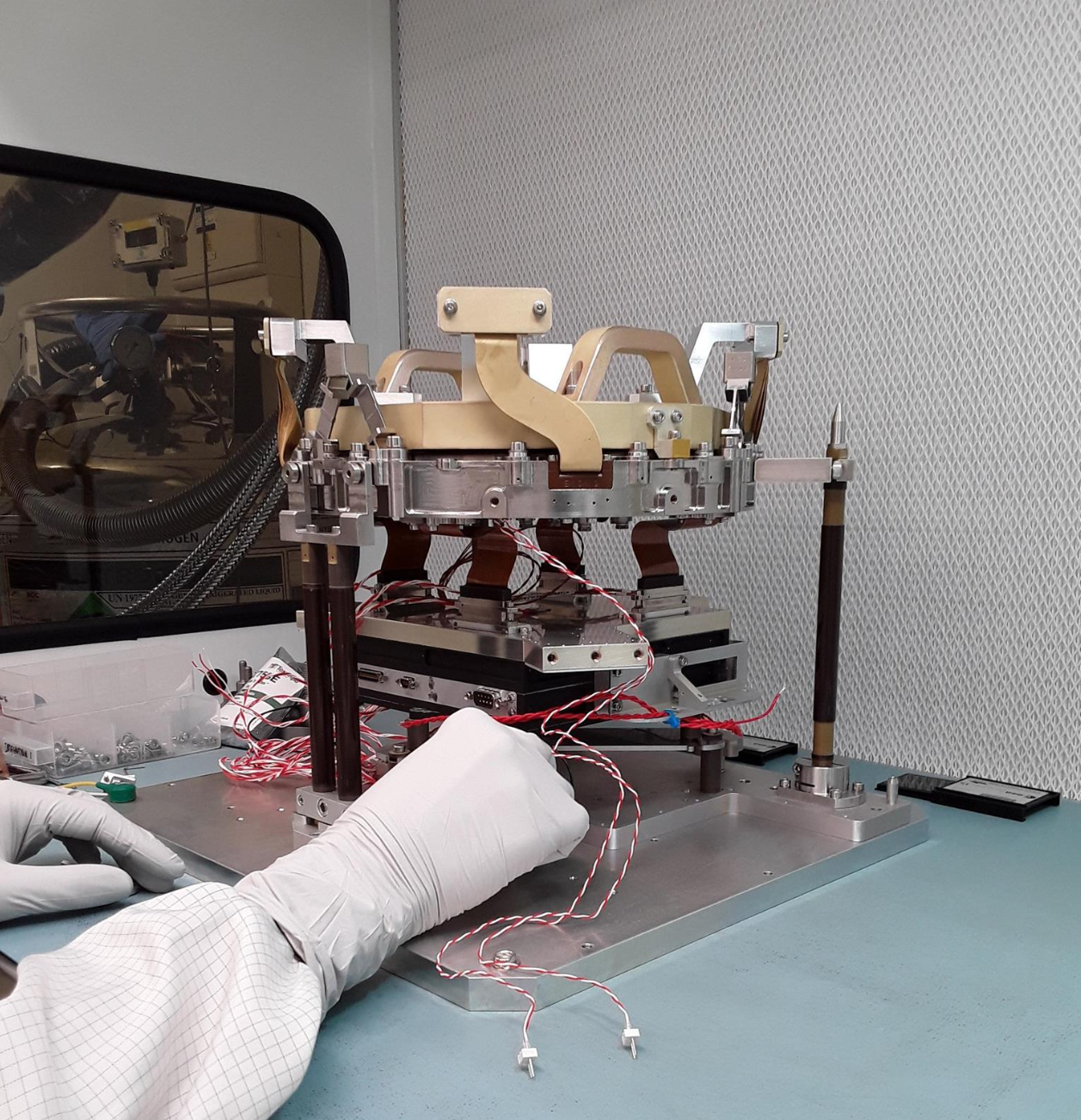}
\caption {Left: Engineering model of a complete Focal Plane Assembly populated with full--frame CCDs. Right: Engineering models of the FPA and the Normal Front End Electronics integrated for functional testing at MSSL. }
\label{fig:FPA}       
\end{figure}

An anti-reflection coating is placed on every optical surface to maximise the optical transmission in the spectral range [500-1000]~nm. Each 12~cm aperture camera has a FoV of 1037 square degrees as shown in \cite{pertenais21b}. This wide FoV is achieved thanks to a circular optical FoV of the TOU of around 18.88° radius that illuminates the 4 full-frame CCDs of 81.18~$\times$~81.18~mm² each on the Focal Plane Assemblies (FPA). As mentioned above, each FPA is electrically connected to a Front End Electronics (either "normal" or "fast" ones), completing the PLATO Camera. One of the FPA models used for qualification is shown in Fig.~\ref{fig:FPA}, with 4 full-frame (i.e. "normal" CCDs).  More details on the FPA can be found in \citet{moreno}. Table \ref{tab:cameras} summarises the the main parameters of the individual cameras.
The transmission and quantum efficiency curves can be consulted in~\citep{boerner2024}.

\begin{table}
\caption{PLATO Camera parameters.}
\label{tab:cameras}
\begin{tabular}{lll}
\hline\noalign{\smallskip}
Normal cameras  &   \\
\noalign{\smallskip}\hline\noalign{\smallskip}
entrance pupil diameter & 120 mm (the frontal lens is 200 mm wide) \\
focal length        & 247.5 mm\\
camera FoV          & 1037 deg² \\
photometric range   & 4-16 magV\\
transmission        & 500-1000 nm\\
no. CCDs/camera     & 4  \\
read-out mode       & full frame \\
read-out cadence    & 25 sec \\
number of effective pixels    & 4510$\times$4510 \\
pixel size          & 18$\times$18 $\mu$m\\
pixel scale         &  15 arcsec/pixel \\
nominal working temperature         &  -80°C \\
\noalign{\smallskip}\hline
\hline\noalign{\smallskip}
Fast cameras, where different from above  &   \\
\noalign{\smallskip}\hline\noalign{\smallskip}
camera FoV          & 610 deg² \\
photometric range   & 4 - 8.2 magV  \\
transmission        & 505-700 nm\\
                    & 665-1000 nm \\
read-out mode       & frame transfer\\
read-out cadence    & 2.5 sec \\
number of effective pixels    & 4490$\times$2245 \\
\noalign{\smallskip}\hline

\end{tabular}
\end{table}

\subsection{PLATO Data Processing System}
\label{sec:DPS}

The PLATO Data Processing System (DPS) is made up of an on-board segment and a ground segment. The on-ground data processing approach is presented in Section \ref{sec:SGS}. Concerning the on-board segment, with 24 normal cameras working at the cadence of 25 seconds and two fast cameras working at the cadence of 2.5 seconds, the amount of raw data produced each day is over 100 Terabit. This volume must be compared to the hundreds of Gigabit which can be actually downloaded each day to the ground. It is clearly not possible to transmit the whole amount of raw data. The role of the on-board processing is to reduce by a factor of more than 1000 the flow rate by downlinking star intensities, centroid curves and imagettes at the cadence required by the science applications. 

\begin{figure}
\includegraphics[width=\linewidth]{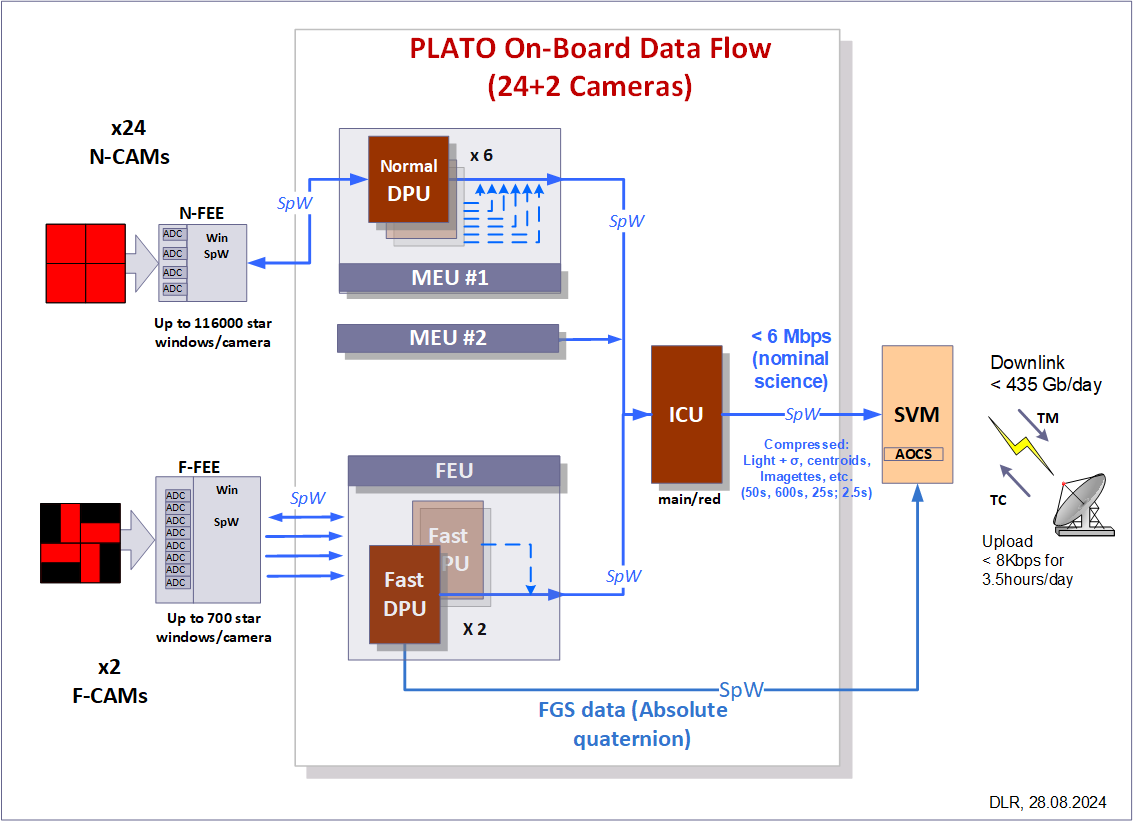}    
\caption{Simplified overview of the PLATO on-board data flow.}
\label{fig:dps}       
\end{figure}

\begin{figure}
\includegraphics[width=\linewidth]{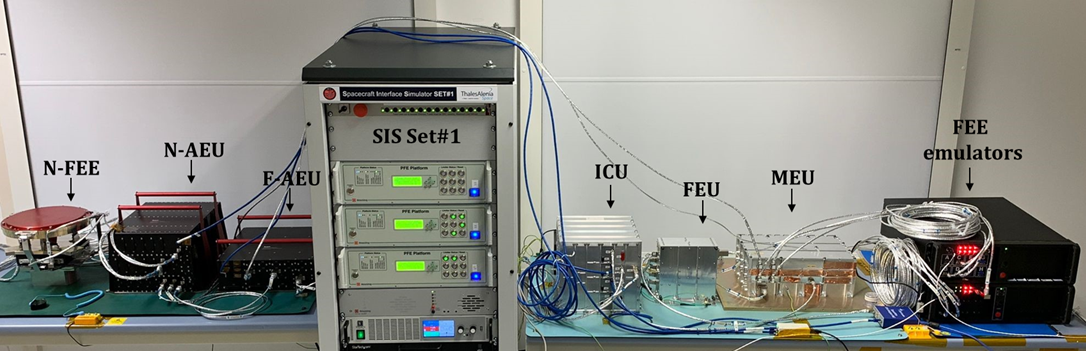}    
\caption{DPS engineering models test bench at DLR.}
\label{fig:dps_test_bench}       
\end{figure}

The PLATO payload on-board data processing system is made up of an Instrument Control Unit (ICU), two Main Electronic Units (MEU) and a Fast Electronic Unit (FEU). 
Each MEU contains 6 normal Data Processing Units (N-DPUs) for processing data from the normal cameras. Each N-DPU is dealing with two normal cameras in terms of managing the N-FEE and acquiring data. The nominal processing cadence for the N-DPUs is 25 sec.
There are two F-DPUs gathered in the FEU. Each F-DPU is responsible for processing the data of one fast camera. The processing cadence for F-DPUs is 2.5 sec. The F-DPUs main purpose is providing attitude data for the Fine Guidance System (FGS) directly to the Service Module (SVM) AOCS and to the ICU. Furthermore, it manages the F-FEE and handles scientific data of bright stars.
There are two ICU channels which work in cold redundancy. The ICU is responsible for the management of the payload, the communication with the SVM and the compression of scientific data before transmitting them as telemetry to the SVM.
Data is routed through a SpaceWire network from FEU and MEU to the ICU, and then from the ICU to the SVM. For the FGS data, dedicated SpaceWire links between FEU and spacecraft AOCS are used.
Figure~\ref{fig:dps} gives an overview of the PLATO data processing system architecture and of the data flow rates while Fig.~\ref{fig:dps_test_bench} shows the engineering models during  tests. It focuses on the sharing of the main functions and the data flows. It is a simplified view of the hardware architecture. 

An on-board software is present in the ICU and both normal and fast DPUs.
The ICU software is in charge of collecting all science data and housekeeping. It manages the PLATO payload sub-systems (DPS and FEE).
In observation mode the N-DPU software must first assemble windows from the images-data-stream, received from the FEE. 
Then the software performs an initial treatment, which consists of computing the background noise and some correction parameters like smearing and offset. 
A part of the image data will be sent to the ICU as imagettes without further processing. On a selection of stars, the N-DPU calculates the centroid and the received flux.
Flux and centroid measurements are computed on-board using optimal binary mask as explained in~\citet{marchiori2019A&A...627A..71M}. 
The N-DPU is able to stack fluxes and centroids from each camera over periods of 50 seconds and 600 seconds and send it to the ICU.
Before averaging the values of fluxes and centroids, an outlier rejection algorithm is used on-board. 
These data are then compressed by a factor of at least 2.5 for transmission to the SVM mass memory from where they are downloaded to the ground. 
The star catalogue defined on-ground will be uploaded in the ICU and then forwarded to the DPUs.

The main purpose of the F-DPU is however to perform the attitude calculation of the F-CAMs based on the positions of the stars on the CCD. This Fine Guidance System (FGS) data processing using the knowledge of the camera geometrical model and the differential aberration is providing quaternions of the CAM attitude to the AOCS system of the spacecraft. This is a key component needed to reach the pointing performance presented in Table \ref{tab:instrument}.

\section{Synergies with other Missions} 
\label{sec:synergy-missions}

Exoplanet missions form a significant part of ESA and NASA's ongoing and future space science programs 
(\cite{stapelfeldt2021} for an overview). PLATO will start its science operations in 2027. By this time, the now ongoing extended operations periods for  CHEOPS and TESS will have ended,  but further prolongations could extend their lifetimes until and beyond PLATO launch. Prolongation of James-Webb-Space-Telescope (JWST) operations beyond 2026 will allow it to observe PLATO targets. 
Two years after PLATO, the ARIEL mission is set to launch and plans to spectroscopically observe transiting exoplanets. Here we outline the key aspects of synergies with PLATO for these and other missions:

{\it TESS:} A discussion of the planet yield of PLATO versus TESS \citep{ricker2015} can be found in Section \ref{tab:planet-yield}. The majority of planets detected by TESS have orbital periods $<$10 days~\citep{Guerrero2021}, while PLATO planets focus on orbital periods $>$27 days, including a significant fraction on long-periods $>$100 days. Most of the planet detections around solar-like stars  from PLATO will have $<$4 \rearth, outnumbering the numbers expected from TESS for this type of stars. There is a clear complementarity of TESS and PLATO. We note that if TESS enters a third extended mission period from 2026-2029, it will overlap with PLATO. This offers the opportunity for dedicated complementary observations where beneficial. This could include, for example, a pre-characterization of the PLATO field for short-period planets to prolonge TTV measurement baselines.

{\it CHEOPS:} The 30\,cm-aperture space telescope is a follow-up mission carrying out high-precision photometry of known bright exoplanetary systems \citep{benz2020, fortier2024A&A...687A.302F}. CHEOPS has been used primarily to measure precise planetary radii, revising the composition of small planets \citep[e.g,][]{bonfati2021A&A...646A.157B, lacedelli2022MNRAS.511.4551L, egger2024A&A...688A.223E}, to confirm transiting long-period planets \citep[e.g.,][]{delrez2021NatAs...5..775D, osborn2022A&A...664A.156O}, characterise multiplanet systems \citep[e.g,][]{leleu2021A&A...649A..26L, luque2023Natur.623..932L} and has also been used to characterise the emissive and reflective properties of planetary atmospheres \citep[e.g,][]{lendl2020A&A...643A..94L, brandeker2022A&A...659L...4B}.

A major benefit to PLATO from CHEOPS results from improvements made in the planetary composition and structure modelling that inform the interpretation of precise planet radii and masses in terms of, e.g, planet formation, evolution and interior-atmosphere coupling. While many of the related science questions can only be fully answered when a larger sample of accurately characterised planets from PLATO becomes available, the ongoing activities within CHEOPS pave the way for PLATO data to come. With large parts of the PLATO LOPS2 field accessible to CHEOPS, a cross-calibration of CHEOPS and PLATO data of the same transits will benchmark the calibration of PLATO data (e.g. in terms of contamination removal). Preparatory observations of known transiting planets in the PLATO LOPS2 filed by CHEOPS have started and will be beneficial to maximising science return on these objects, crucially extending the baseline of transit observations for TTV studies.

{\it Ariel:} will provide spectra of the atmospheres of a large number (around 1000) of transiting planets \citep{tinetti2018}.  These planets will mostly be hot and warm gaseous exoplanets (Jupiters, Saturns, Neptunes) as well as of super-Earths/sub-Neptunes orbiting bright stars of various types, with FG stars being the predominant host. Several surveys are already identifying potential Ariel targets that meet its mission objectives. For instance, TESS is expected to discover over 4500 planets around bright stars and more than 10000 giant planets around fainter stars (see Section \ref{tab:planet-yield}).  Other surveys,  including PLATO, are expected to provide thousands of aditional candidates suitable for study with Ariel. Notably, about 20-40 targets are expected to come from a PLATO LOPs (see also Nascimbeni et al. 2024, submitted). Moreover, PLATO will likely uncover new small, warm planets around bright stars, providing an exciting pool of additional candidate targets for Ariel. PLATO will play a crucial role by delivering high-precision planetary radii, extended TTV datasets, and refined ephemerides for planets within its fields. Data from PLATO’s N-CAM (white-light phase curves) and F-CAMs (two-colour information) will further aid in prioritizing Ariel candidates, particularly for hot, nearby targets (see Section \ref{sec_atmospheres}). The combined insights on radii, masses, and stellar parameters - especially stellar ages - will significantly enhance the scientific output of Ariel. For example, precise ages from PLATO will help break degeneracies in the study of warm Jupiters, enabling detailed investigations of planetary evolution, including contraction, atmospheric loss, and tidal interactions. In addition, PLATO’s pre-characterization of stellar activity (e.g., rotational modulation, activity cycles, flare and CME frequencies, granulation) will be invaluable. These data will be particularly useful for targets observed simultaneously by both missions.

{\it JWST:} The main synergy of PLATO and JWST \citep{gardner2006}  concerns newly detected warm and/or small planets which can be followed up by JWST to investigate their atmospheres, phase curves and albedos. PLATO will form a target finder for JWST on these highly interesting targets in future and provide relevant complementary planet and stellar activity data (see discussion on ARIEL above).

{\it GAIA:} Through its astrometric survey of the Milky Way, Gaia \citep{gaia2016A&A...595A...1G} will discover a large population of massive  planets (expected to be super-Neptunes and more massive), many of which will be at long periods.
Most of these will be found from their trigonometric signatures but some may also be detectable through their transits. A significant harvest of new long period ($>$ 5 years), massive exoplanets is expected in the upcoming Gaia data releases (e.g. Gaia DR4). For those with sufficient orbital information, inclinations could be accurate enough to predict potential transit times and if these fall with the PLATO observation fields monitoring observations will be possible at low cost. These long period, transiting planets, would be extremely valuable. While many of these systems may only transit once during the PLATO observations, combined modelling of the astrometric and transit observations will produce accurate orbital and planetary parameters. 
For PLATO itself, the GAIA observations not only form the basis of the PLATO input catalogue, but allow contamination with the PLATO targets/pixels to be estimated. Furthermore, while GAIA produces limited temporal coverage it may reveal the presence of line of sight background eclipsing binaries as has been demonstrated for TESS candidates by \cite{panahi2022A&A...667A..14P}. This vetting capability will be extremely important in keeping the PLATO ground based followup tractable.

{\it Euclid and Roman Space telescope:} The Nancy Grace Roman Space telescope \citep{spergel2015arXiv150303757S} is expected to detect exoplanets from microlensing, from transits and from coronographic imaging, covering a wide range of planet system parameters. ESA's Euclid mission also has some capability to detect extrasolar planets, and studies demonstrating the joint detection capabilities of the two missions have been performed (e.g. \citealt{bachelet2022A&A...664A.136B,kerins2023magnifying}). 
In particular the microlensing technique used by these missions has the potential to detect planets as small as Earth, and smaller, at orbital distances of 1 AU, although not with accurate planet parameters as anticipated by PLATO. Nevertheless, PLATO together with Roman and Euclid will be help solidify our understanding of the frequency of earth-sized planets in earth-like orbits. Both Euclid and the Roman Space Telescope will study a relatively distant stellar and hence planetary population which will contrast with the nearby PLATO planets. The large number of planets detected by these surveys will enable biases in their planet catch to be understood, hence leading to a better understanding of the planetary demographics in these different environments.

For all the missions discussed here, a combined analysis of their planetary frequencies will enable the most comprehensive picture of planetary demographics to be determined, allowing us to deduce how planets are born and evolve (through modelling).

\bmhead{Acknowledgments}
This work presents results from the European Space Agency (ESA) space mission PLATO. The PLATO payload, the PLATO Ground Segment and PLATO data processing are joint developments of ESA and the PLATO Mission Consortium (PMC). Funding for the PMC is provided at national levels, in particular by countries participating in the PLATO Multilateral Agreement (Austria, Belgium, Czech Republic, Denmark, France, Germany, Italy, Netherlands, Portugal, Spain, Sweden, Switzerland, Norway, and United Kingdom) and institutions from Brazil. Members of the PLATO Consortium can be found at https://platomission.com/. The ESA PLATO mission website is https://www.cosmos.esa.int/plato. We thank the teams working for PLATO for all their work. \\

The authors gratefully thank David Ciardi and an unknown referee for carefully reading this publication and providing many helpful comments for clarifying the current status of PLATO to a large readership. 

Specific acknowledgements: \\
The German PMC, PDC, DPS, and F-FEE team members are supported by the Geraman Aerospace Agency (Deutsches Zentrum f\"ur Luft- und Raumfahrt, e.V., DLR) grant numbers 50OO1401, 50OP2001, 50OP2103, 50OP2104, 50OP2101, 50OP1902 and 50OP2102.\\
The DLR team members acknowledge the funding by the Research and Development Department of the German Aerospace Center (Deutsches Zentrum f\"ur Luft- und Raumfahrt, e.V., DLR).\\
M.L. acknowledges support of the Swiss National Science Foundation under grant number PCEFP2\_194576. The contribution of M.L. has been carried out within the framework of the NCCR PlanetS supported by the Swiss National Science Foundation under grants 51NF40\_182901 and 51NF40\_205606. \\
The research and results presented in this paper has received funding from the BELgian federal Science Policy Office (BELSPO) through various PRODEX grants for PLATO development and from the  KU Leuven Research Council (grant C16/18/005: PARADISE).\\
This project was supported by the KKP-137523 "SeismoLab" \'Elvonal grant of the Hungarian Research, Development and Innovation Office (NKFIH) and by the Lend\"ulet Program  of the Hungarian Academy of Sciences under project No. LP2018-7.\\
Project no. C1746651 has been implemented with the support provided by the Ministry of Culture and Innovation of Hungary from the National Research, Development and Innovation Fund, financed under the NVKDP-2021 funding scheme. \\
This work was supported by the Hungarian National Research, Development and Innovation Office grants OTKA K131508 and KH-130526, and the \'Elvonal grant KKP-143986. Authors acknowledge the financial support of the Austrian-Hungarian Action Foundation (101\"ou13, 112\"ou1). \\
This work was supported by Funda\c c\~ao para a Ci\^encia e a Tecnologia (FCT) through research grants UIDB/04434/2020 and UIDP/04434/2020.\\
This work was supported by FCT - Funda\c c\~ao para a Ci\^encia e a Tecnologia through national funds and by FEDER through COMPETE2020 - Programa Operacional Competitividade e Internacionaliza\c c\~ao by these grants: UIDB/04434/2020; UIDP/04434/2020 and co-funded by the European Union (ERC, FIERCE, 101052347). Views and opinions expressed are however those of the author(s) only and do not necessarily reflect those of the European Union or the European Research Council. Neither the European Union nor the granting authority can be held responsible for them. \\
This work was (partially) supported by the Spanish MICIN/AEI/10.13039/501100011033 and by "ERDF A way of making Europe" by the “European Union” through grant PID2021-122842OB-C21, and the Institute of Cosmos Sciences University of Barcelona (ICCUB, Unidad de Excelencia ’Mar\'{\i}a de Maeztu’) through grant CEX2019-000918-M.\\
This work is part of the project Advanced technologies for the exploration of the Universe and its components, of the area of Astrophysics and High Energy Physics, within the frame of the R\&D\&I Complementary Plans of the Spanish Government that are part component 17 of the Recovery and Resilience Mechanism. This contract is funded by the European Union - NextGenerationEU (MICIN/PRTR funds) and by Generalitat de Catalunya. \\
We acknowledge financial support from the Agencia Estatal de Investigaci\'on of the Ministerio de Ciencia e Innovaci\'on MCIN/AEI/10.13039/501100011033 and the ERDF “A way of making Europe” through project PID2021-125627OB-C31, from the Centre of Excellence “Mar\'{\i}a de Maeztu” award to the Institut de Ci\`encies de l’Espai (CEX2020-001058-M) and from the Generalitat de Catalunya/CERCA programme. \\
Funding for the Stellar Astrophysics Centre was provided by The Danish National Research Foundation (Grant DNRF106).\\
Support from PLATO ASI-INAF agreements n. 2022-28-HH.0.\\
Giampaolo Piotto, Marco Montalto, Luca Malavolta, Valerio Nascimbeni, Luca Borsato, Giacomo Mantovan, Valentina Granata: Support from PLATO ASI-INAF agreements n. 2022-28-HH.0.\\
This work on the PLATO space mission was supported by CNES.\\
This study is supported by the Research Council of Norway through its Centres of Excellence funding scheme, project number 223272 (CEED) and 332523 (PHAB).\\
Brazilian participation on the PLATO mission is funded by Funda\c{c}\~{a}o de Amparo \`{a} Pesquisa do Estado de S\~{a}o Paulo (FAPESP) under grant 2016/13750-6.\\
The team at IAC acknowledges support from the Spanish Research Agency of the Ministry of Science and Innovation (AEI-MICINN) under grant 'Contribution of the IAC to the PLATO Space Mission' with reference PID2019-107061GB-C66, DOI: 10.13039/501100011033.\\
U.C.Kolb and C.A. Haswell were supported by grant ST/T000295/1 from STFC. C.A. Haswell was supported by grant ST/X001164/1 from STFC.\\
NW, SA, PB, GB, FDA, DWE, DF, DLH, SH, JI, NM, MR, GR, LS, MW have been fully or partly supported by grant funding from the UK Space Agency through grants ST/R004838/1 and ST/X001571/1. \\
The PDPC-C hardware is largely provided through the UK STFC IRIS digital research infrastructure (https://www.iris.ac.uk) project. \\
D.M.B. gratefully acknowledges a senior postdoctoral fellowship from the Research Foundation Flanders (FWO; grant number [1286521N]), the Engineering and Physical Sciences Research Council (EPSRC) of UK Research and Innovation (UKRI) in the form of a Frontier Research grant under the UK government’s ERC Horizon Europe funding guarantee (SYMPHONY; grant number [EP/Y031059/1]), and a Royal Society University Research Fellowship (grant number: URF\textbackslash R1\textbackslash 231631)\\
PK and MK would like to acknowledge the funding of the CZ contribution to PLATO mission from ESA PRODEX under PEA-4000127913 contract. \\
MS and RK would like to acknowledge the funding from the LTT-20015 grant of the MEYS.\\
UH, OK, TO acknowledge support from the Swedish National Space Agency (SNSA/Rymdstyrelsen).\\
I.L. and A.B. extend their gratitude to the Funda\c c~ao para a Ci\^encia e Tecnologia (FCT, Portugal) for the financial support provided to the Center for Astrophysics and Gravitation (CENTRA/IST/ULisboa) under Grant Project No. UIDB/00099/2020.\\
The work of the Porto team was supported by FCT - Fundação para a Ciência e a Tecnologia through national funds by grants: UIDB/04434/2020; UIDP/04434/2020.\\
The Portuguese team thanks the Portuguese Space Agency for the provision of financial support in the framework of the PRODEX Programme of the European Space Agency (ESA) under contracts number 4000133026, 4000140773, and 4000124770.\\
CAB and INTA authors are funded by Spanish MCIN/AEI/10.13039/501100011033 grants PID2019-107061GB -C61 and -C62. \\
JPG, MLM, JRR, ARB and RGH acknowledge financial support from project PID2019-107061GB-C63 from the ‘Programas Estatales de Generación de Conocimiento y Fortalecimiento Científico y Tecnológico del Sistema de I+D+i y de I+D+i Orientada a los Retos de la Sociedad.\\
E.A.’s work has been carried out within the framework of the NCCR PlanetS supported by the Swiss National Science Foundation under grants 51NF40$\_$182901 and 51NF40$\_$205606.\\
RA acknowledges funding from the Science \& Technology Facilities Council (STFC) through Consolidated Grant ST/W000857/1.\\
Paul Beck acknowledges support by the Spanish Ministry of Science and Innovation with the \textit{Ram{\'o}n\,y\,Cajal} fellowship number RYC-2021-033137-I and the number MRR4032204.\\
S.M.~acknowledges support from the Spanish Ministry of Science and Innovation (MICINN) with the Ram\'on y Cajal fellowship no.~RYC-2015-17697, the grant no.~PID2019-107187GB-I00, and through AEI under the Severo Ochoa Centres of Excellence Programme 2020--2023 (CEX2019-000920-S).\\
JMMH is funded by Spanish MCIN/AEI/10.13039/501100011033 grant PID2019-107061GB-C61.\\
M.L.M. acknowledge financial support from the Severo Ochoa grant CEX2021-001131-S funded by MCIN/AEI/10.13039/50110001103.\\
J.P.G. acknowledge financial support from the Severo Ochoa grant CEX2021-001131-S funded by MCIN/AEI/10.13039/50110001103.\\
C.P.M. acknowledge financial support from the Severo Ochoa grant CEX2021-001131-S funded by MCIN/AEI/10.13039/50110001103.\\
F.J.P. acknowledge financial support from the Severo Ochoa grant CEX2021-001131-S funded by MCIN/AEI/10.13039/50110001103.\\
J.R.G. acknowledge financial support from the Severo Ochoa grant CEX2021-001131-S funded by MCIN/AEI/10.13039/50110001103.\\
M.A.S.C. acknowledge financial support from the Severo Ochoa grant CEX2021-001131-S funded by MCIN/AEI/10.13039/50110001103.\\
R.S.M. acknowledge financial support from the Severo Ochoa grant CEX2021-001131-S funded by MCIN/AEI/10.13039/50110001103.\\
M.V.A. acknowledge financial support from the Severo Ochoa grant CEX2021-001131-S funded by MCIN/AEI/10.13039/50110001103.\\
B.A.M. acknowledge financial support from the Severo Ochoa grant CEX2021-001131-S funded by MCIN/AEI/10.13039/50110001103.\\
A.C. acknowledge financial support from the Severo Ochoa grant CEX2021-001131-S funded by MCIN/AEI/10.13039/50110001103.\\
J.M.G.L. acknowledge financial support from the Severo Ochoa grant CEX2021-001131-S funded by MCIN/AEI/10.13039/50110001103.\\
Juan Carlos Morales acknowledge financial support from the Agencia Estatal de Investigaci\'on of the Ministerio de Ciencia e Innovaci\'on MCIN/AEI/10.13039/501100011033 and the ERDF “A way of making Europe” through project PID2021-125627OB-C31, from the Centre of Excellence “Mar\'{\i}a de Maeztu” award to the Institut de Ci\`encies de l’Espai (CEX2020-001058-M) and from the Generalitat de Catalunya/CERCA programme.\\
Ignasi Ribas acknowledge financial support from the Agencia Estatal de Investigaci\'on of the Ministerio de Ciencia e Innovaci\'on MCIN/AEI/10.13039/501100011033 and the ERDF “A way of making Europe” through project PID2021-125627OB-C31, from the Centre of Excellence “Mar\'{\i}a de Maeztu” award to the Institut de Ci\`encies de l’Espai (CEX2020-001058-M) and from the Generalitat de Catalunya/CERCA programme.\\
Aldo Serenelli acknowledge financial support from the Agencia Estatal de Investigaci\'on of the Ministerio de Ciencia e Innovaci\'on MCIN/AEI/10.13039/501100011033 and the ERDF “A way of making Europe” through project PID2021-125627OB-C31, from the Centre of Excellence “Mar\'{\i}a de Maeztu” award to the Institut de Ci\`encies de l’Espai (CEX2020-001058-M) and from the Generalitat de Catalunya/CERCA programme.\\
DJA is supported by UKRI through the STFC (ST/R00384X/1) and EPSRC (EP/X027562/1).\\
SLC acknowledges funding from the Science \& Technology Facilities Council (STFC) through an Ernest Rutherford Research Fellowship ST/R003726/1.\\
PPA acknowledges the support of Funda\c c~ao para a Ci\^encia e Tecnologia FCT/MCTES, Portugal, through national funds by the following grants UIDB/04434/2020, UIDP/04434/2020.FCT, 2022.03993.PTDC.\\
Tiago Campante is supported by FCT in the form of a work contract (CEECIND/00476/2018).\\
MC acknowledges the support of Funda\c c~ao para a Ci\^encia e Tecnologia FCT/MCTES, Portugal, through national funds by these grants UIDB/04434/2020, UIDP/04434/2020.FCT, 2022.06962.PTDC.FCT, 2022.03993.PTDC and CEECIND/02619/2017.
ZsB acknowledges the support by the J\'anos Bolyai Research Scholarship of the Hungarian Academy of Sciences.\\
Gy. M. Szab\'o acknowledges support from the PRODEX Experiment Agreement No. 4000137122.\\
R\'obert Szab\'o: Lend\"ulet Program  of the Hungarian Academy of Sciences, project No. LP2018-7/2022.\\
R\'obert Szab\'o: KKP-137523 "SeismoLab" \'Elvonal grant of the Hungarian Research, Development and Innovation Office (NKFIH).\\
R\'obert Szab\'o: MW-Gaia COST Action (CA18104).\\
KV was supported by the Bolyai J\'anos Research Scholarship of the Hungarian Academy of Sciences, and by the Bolyai+ grant \'UNKP-22-5-ELTE-1093. \\
D.M.B. gratefully acknowledges funding from the Research Foundation Flanders (FWO) by means of a senior postdoctoral fellowship (grant agreement No. 1286521N).\\
Thibault Merle is granted by the BELSPO Belgian federal research program FED-tWIN under the research profile Prf-2020-033$\_$BISTRO.\\
Thierry Morel acknowledges financial support from Belspo for contract PRODEX PLATO mission development.\\
S.N.B acknowledges support from PLATO ASI-INAF agreement n.~2015-019-R.1-2018.\\
DLB acknowledges support from NASA through the Astrophysics Science Smallsat Studies program (80NSSC20K1246) and the TESS GI Program (80NSSC21K0334).\\
DBdeF acknowledges financial support from the Brazilian agency CNPq-PQ2 (Grant No. 305566/2021-0). Research activities of STELLAR TEAM of Federal University of Ceará are supported by continuous grants from the Brazilian agency CNPq.\\
J.~K. gratefully acknowledges the support of the Swedish National Space Agency (SNSA; DNR 2020-00104) and of the Swedish Research Council  (VR: Etableringsbidrag 2017-04945).\\
AJM acknowledges support from the Swedish Research Council (grant 2017-04945) and the Swedish National Space Agency (grant 120/19C).\\
Support for MC is provided by ANID grants ICN12\textunderscore009 (Millennium Institute of Astrophysics, FB10003 (Basal-CATA2), and 1231637 (FONDECYT).\\
CD acknowledges SNSF under grant TMSGI2$\_$211313.\\
MK acknowledges the support from ESA-PRODEX PEA4000127913.\\
TL was supported by a grant from the Branco Weiss Foundation.\\
PM acknowledges support from STFC research grant number ST/M001040/1.\\
F.J.P. acknowledges financial support from the grant CEX2021-001131-S funded by MCIN/AEI/ 10.13039/501100011033.\\
BR-A acknowledges funding support from FONDECYT Iniciación grant 11181295 and ANID Basal project FB210003.\\
A.R.G.S. acknowledges the support by FCT through national funds and by FEDER through COMPETE2020 by these grants: UIDB/04434/2020 \& UIDP/04434/2020. A.R.G.S. is supported by FCT through the work contract No. 2020.02480.CEECIND/CP1631/CT0001.\\
CdB acknowledges support from a Beatriz Galindo senior fellowship (BG22/00166) from the Spanish Ministry of Science, Innovation and Universities.\\

\section{Appendix A}
\label{app:planetary radius estimation}

We outline briefly the simple analytic model of the estimation of planet radius accuracy. All calculations below are valid for beginning of life (BOL). We start from the following form of the conversion of $V$ magnitudes to fluxes measurable by PLATO
\begin{equation}
    f = 180 000 \mathrm{e}^{-}\mathrm{/s} \cdot 10^{-0.4(V-11)}
\end{equation}
The noise level at flux $f$ can be estimated (Cabrera et al., in prep.) as:
\begin{equation}
    N = \sqrt{(9\cdot10^{-6})^2 + \frac{1/f + (150/f)^2}{N_{\mathrm{camera}} \cdot T/t_\mathrm{exp}}}
\end{equation}
Here $N$ is the noise level for $T=1$ hour time-scale and $N_\mathrm{camera} = 24$ was assumed. $t_{exp} = 25$s is the exposure cadence. The first constant term is the jitter-noise, the last term is the readout noise. It is assumed that the magnitude zero-point is the same for all cameras and the readout-noise is also independent on which camera we consider.

The radius ratio precision can be derived from the equation
\begin{equation}
    \delta = \left( \frac{R_\mathrm{planet}}{R_\mathrm{star}} \right)^2 L_D
\end{equation}
where $L_D$ describes the effect of limb darkening. 
The current knowledge of limb darkening for low magnetic field, quiet, solar-type stars is known sufficiently well that the contribution to the uncertainty in the radius ratio is negligible (c.f. Section 4.4. of ~\citealt{maxted2023MNRAS.519.3723M} as well as ~\citealt{ludwig2023A&A...679A..65L}). 
However, numerical simulations show that about over 800 Gauss surface magnetic field the theory predicts higher limb darkening than that was observed. 
In the same way, some simulations result in limb brightening effects difficult to understand at this point (see~\citealt{ludwig2023A&A...679A..65L}).
So, the very active stars can have different properties and more work is needed to calibrate properly their limb darkening behaviour (c.f. ~\citealt{csizmadia2013a,espinoza2015MNRAS.450.1879E,agol2020AJ....159..123A,patel2022AJ....163..228P}).
The results are not changing if we consider different limb darkening laws; therefore, for sake of simplicity, we used linear limb darkening with a constant linear limb darkening coefficent $u=0.6$ which is a solar-like value ($L_D = 1/(1-u/3)$). Then we have by differentiation that
\begin{equation}
    \frac{\Delta k}{k} = \frac{\sqrt{2}}{2} \frac{N(V)}{\delta} \frac{1}{\sqrt{N_\mathrm{transit} D}}
\end{equation}
abbreviating the planet-to-star radius ratio $k = R_\mathrm{planet} / R_\mathrm{star}$. 
The factor $\sqrt{2}$ is valid if the baseline is determined by observing only one full transit duration length before and after the transit.
However, the factor will decrease to 1 if the baseline length is much longer than the transit duration (in the limit to infinity).
Since the noise level is calculated for 1 hour in Cabrera et al. (in prep), the transit duration $D$ plays a role here. Despite the simplicity of this analytic model, the results are in quite good agreement with \citet{barros2020,morris2020}; and \citet{2023A&A...675A.106C}.

\bibliography{bibliography_file}   


\end{document}